# Mechanics of hierarchical twisted and coiled polymer artificial muscles: Decoupling force from kinematic limits

Ye Xiao [a,b], Zhao Luo [a,b], Falin Tian [c], Xinghao Hu [d,*], Dabiao Liu [e,*], Chun Li [a,b,*]

[a] *Department of Engineering Mechanics, Northwestern Polytechnical University, Xi'an 710072, People's Republic of China*

[b] *Shenzhen Research Institute of Northwestern Polytechnical University, Shenzhen, 518057, China*

[c] *Laboratory of Theoretical and Computational Nanoscience, National Center for Nanoscience and Technology, Chinese Academy of Sciences, Beijing 100190, China*

[d] *School of Mechanical Engineering, Jiangsu University, Zhenjiang, 212013, People's Republic of China*

[e] *Department of Engineering Mechanics, School of Aerospace Engineering, Huazhong University of Science and Technology, Wuhan 430074, People's Republic of China*

[*] Corresponding author:
huxh@ujs.edu.cn (X. Hu), dbliu@hust.edu.cn (D. Liu), lichun@nwpu.edu.cn (C. Li).

## Abstract

Thermally actuated twisted and coiled polymer (TCP) artificial muscles exhibit exceptional specific work capacities but are limited by an inherent competition between load-bearing capacity and actuation stroke. To address this limitation, we investigate a hierarchical helical structure designed to decouple force generation from kinematic limits. We propose a coupled thermo-mechanical model incorporating inter-filamentary contact mechanics and geometric nonlinearities to predict the assembly's equilibrium response. The results indicate that this hierarchical topology significantly amplifies isometric actuation stress compared to monofilament baselines, while maintaining a biological-like contraction stroke of approximately 22%. A critical topological threshold governed by the balance between cooperative load-sharing and geometric confinement is identified. Beyond an optimal bundle complexity, the geometric jamming dominates, as excessive inter-filamentary friction hinders actuation. Furthermore, we elucidate a stiffness-stroke synergy in homochiral configurations, where high helical angles amplify the thermal untwisting torque to overcome increased structural rigidity. Crucially, the volumetric energy density exhibits scale invariance regarding the hierarchical radius, implying that absolute force output can be linearly scaled through geometric upsizing without compromising efficiency. These findings provide a mechanics-based rationale for the structural programming, demonstrating that soft actuator performance limits are dictated by topological order rather than intrinsic material properties.

## 1. Introduction

Artificial muscles, inspired by biological muscle systems, originated from early explorations in electroactive polymers and shape-memory alloys during the mid-20th century, with significant advancements accelerating in the 2010s through materials like conducting polymers and carbon nanotubes (Mirvakili and Hunter, 2018). These actuators hold substantial significance in mimicking natural muscle contraction, offering significant advantages, including high power-to-weight ratios (often exceeding 1 kW/kg), silent operation, inherent compliance for safe human interaction, and large strains up to 50%, surpassing traditional electric motors in flexibility and energy efficiency (Zhang et al., 2019). Their primary applications span robotics for soft grippers and exoskeletons, biomedical devices including prosthetics and drug delivery systems, and aerospace for adaptive structures, enabling more efficient and biomimetic engineering solutions (Zhu et al., 2022; Gao, e al. 2025).

Owing to the ability to amplify small material expansions into substantial macroscopic motions through geometric leveraging, twisted and coiled structures predominate in artificial muscles, especially for thermally actuated polymer systems. In these configurations, thermal actuations induce untwisting through anisotropic expansion in various materials. For example, Lima et al. (2012) demonstrated hybrid carbon nanotube (CNT) yarns that achieved tensile strokes through thermal or electrochemical heating, highlighting the potential for rapid, reversible actuation in twisted forms. Evolving to twisted and coiled structures, these enhance performance by incorporating helical coiling, which converts torsional changes into axial contraction or extension. Haines et al. (2014) pioneered coiled nylon fibers from fishing line, producing contractions up to 49% under thermal stimuli, setting a benchmark for low-cost, high-strain actuators. Meanwhile, much effort has been directed toward the synthesis of helical fibers using several different materials. Foroughi et al (2011, 2021) developed coiled CNT bundles that enabled torsional actuation with thermal inputs, offering high torque densities. Mu et al. (2015) prepared graphene-oxide composites to enhance the durability and self-folding capabilities under heat. He et al. (2022) employed polylactic acid fibers to demonstrate biodegradable coiled actuators with strains exceeding 40% for biomedical applications. In addition, Leng et al (2021) proposed twisted-fiber muscles incorporating guest materials as paraffin in CNTs, where thermal expansion drove volume changes and improved actuation speed. These studies collectively illustrate the versatility of twisted and coiled designs across carbon-based, polymeric, and composite materials, enabling thermal actuation with strokes often above 30% and applications in morphing structures and soft grippers, while emphasizing global efforts to integrate sustainable materials for reduced environmental impact (Cherubini et al., 2015; Aziz et al., 2020; Higueras-Ruiz et al., 2021; Singh

et al., 2024).

In recent years, considerable attention has been paid to twisted and coiled polymer (TCP) artificial muscles. Experimental advancements in thermally driven variants have progressed steadily since Haines et al. (2014) fabricated actuators from commercial nylon and sewing thread, achieving 5.3 kW/kg power density and 49% contraction by Joule or external heating, thus establishing polymers as accessible alternatives to exotic materials. Furthermore, Cherubini et al. (2015) optimized coiling parameters in polyethylene fibers, enhancing cyclic stability to over 1 million cycles under thermal loads, which improved reliability for long-term applications. Mu et al. (2018) explored multi-ply twisted nylon, which increased load capacity while sustaining 20% strain, and demonstrated molecular-channel effects for guest-driven actuation. Saharan and Tadesse (2018) designed a robotic hand with locking mechanism using TCP muscles for extreme environments, integrating sensors for real-time feedback. Semochkin (2016) developed a device for producing artificial muscles from nylon fishing line with heater wire, facilitating humidity-triggered actuation. Swartz et al. (2018) presented experimental methods for acquiring untwisted monofilament thermal properties and thermal actuation data of straight twisted polymer actuators. Sutton et al. (2016) designed an assistive wrist orthosis using conductive nylon actuators, confirming potential in biomedical orthoses. Zhang et al. (2024) proposed a compound

TCP actuator with payload-insensitive untwisting characteristics, improving force output in multi-strand configurations. Tsai et al. (2025) examined high cycle performance of TCP actuators, optimizing endurance for prolonged operation. Zheng et al. (2025) measured performance of TCP with bionic wrist mechanisms via temperature self-sensing, showing fatigue tolerance in robotic applications. The above contributions have merged experimental fabrication with performance optimization, emphasizing thermal efficiency and material scalability for cross-disciplinary uses like wearable robotics (Meng et al, 2020; Sun et al, 2025).

Theoretical research on TCP has evolved to complement experiments, providing mechanistic insights into the mechanical properties of artificial muscles upon thermal actuation. For example, Xiao and Tian (2015) proposed microstructural constitutive models with helical and entropic chains for intrinsic actuation prediction. Karami and Tadesse (2017) modeled TCP muscles using a phenomenological approach, demonstrating enhanced load capacity in silver-coated variants. Lamuta et al. (2018) developed a theory for tensile actuation of fiber-reinforced coiled muscles using energy minimization and formulated three-dimensional strain modeling through an energetic approach, incorporating viscoelastic effects to predict rotation and capture hysteresis, which improved accuracy over initial quasi-static assumptions. Because of the complex three-dimensional stress-strain field induced by the twisting process, TCP artificial muscles are prone to forming intricate coupling mechanisms and microstructure evolution during actuation. Therefore, some fine theoretical model for TCP actuation were emerged successively. For example, Yang and Li (2016) proposed a top-down multi-scale modeling for actuation response of polymeric artificial muscles, introducing material anisotropy to couple microscopic mechanisms with macroscopic responses, thereby yielding more accurate predictions than single-scale methods. Wang et al. (2022) proposed discrete elastic rod simulations to model the microstructure evolution of artificial muscle upon thermal actuation, estimating the energy-conversion efficiency and power output of TCP actuation. Xiao et al. (2022) analyzed the synergistic effect of axial- torsional-radial deformation of TCP artificial muscles by a bottom-up theoretical model, predicting the dependence of mechanical properties on the microstructural parameters. Sun et al. (2022) presented a physics-based modeling of twisted-and-coiled actuators using Cosserat rod theory, forecasting stroke and force. Liu et al. (2024) proposed a new thermo-mechanical actuation model for twisted and coiled actuators with initial curvature based on finite strain theory, enhancing predictions for dynamic responses. Wang et al. (2024) analyzed the mechanics and physics of TCP actuators, scaling to large-diameter bundles for high-power outputs. Gao et al. (2024) conducted driving response analysis of TCP actuators considering viscoelastic precursor fibers, validating underwater performance.

Thermally actuated single-fiber TCP artificial muscles exhibit substantial axial strokes, frequently surpassing 20-30% strain under modest loads, which can be attributed to the anisotropic thermal expansion and helical microstructure untwisting (Xiao et al., 2023). Nevertheless, the load-bearing capacity TCP actuation is limited, generally confined to stresses of 1-10 MPa, induced by intrinsic material constraints such as polymer chain slippage and reduced torsional stiffness in single fibers, which impede efficient force transfer (Yuan et al., 2019; Chen et al., 2024). This results in notably low energy conversion efficiencies, typically 0.5-2% in nylon-based variants, stemming primarily from pronounced thermal losses, frictional dissipation during untwisting, and operation within narrow temperature differentials (e.g., 50-150°C) that limit Carnot efficiencies to below 30% (Wang et al., 2022; Wang et al., 2024). By comparison, traditional systems like internal combustion engines attain practical efficiencies of 20-40%, highlighting the disparity that confines artificial muscles to specialized prototypes (Weissman et al., 2025). Enhancing load capacity through approaches such as multi-fiber bundling or hierarchical helical structures, without compromising axial stroke, is thus critical for advancing their integration into engineering domains.

Hierarchical helical structures, drawing from biological motifs such as plant tendrils and DNA, feature nested twists or multi-strand coils to optimize load distribution and enhance performance in the research object.

These designs have been widely adopted in engineering applications, notably in wire ropes for infrastructure projects. Costello (1997) quantified strand-level static behavior and mapped it through an upscaling step to the global properties of the wire rope. This strategy sidesteps the intricate geometric complexities of multi-level helices of mechanical analysis and has informed numerous subsequent studies. Zhao et al. (2014) established a bottom-up theoretical model to predict the mechanical responses of carbon nanotube ropes with multi-level helical chirality, revealing enhanced deformation ability and tunable elastic properties. Zheng et al. (2018) investigated the influence of fiber migration on mechanical properties of yarns with helical structures, highlighting improved strength via hierarchical arrangements. Gao et al. (2021) formulated a microstructure-based crack-bridging model for helical fiber-reinforced biocomposites, identifying an optimal helical angle that maximizes toughness and providing a scaling law for macroscopic fracture properties. Meng et al. (2022) developed a three-dimensional crack twisted-bridging model for biological materials with hierarchical Bouligand structures, revealing high, direction-independent fracture toughness influenced by nanofiber pitch angle, length, and interfacial properties. Han et al. (2024) proposed a multi-scale mechanical model for multilevel helical structures with filament damage, elucidating the interplay between helical characteristics and material failure. Xu et al. (2023,2024) prepared water/humidity-responsive actuators from multi-strand carboxyl methyl cellulose fibers with helical structures, experimentally showing superior actuation and mechanical endurance. Espinosa et al. (2012) conducted multiscale experimental mechanics on hierarchical carbon-based materials, confirming theoretical predictions of high strength and toughness through in situ testing. Transitioning to hierarchical helical structures in TCP actuation, Peng et al. (2021) modeled multi-level coiling for enhanced force and scalability in nylon-based systems. Gotti et al. (2020) and Lang et al. (2024) reviewed fibrous hierarchies for TCP actuators, highlighting structural advantages of hierarchies over films and noting improved fatigue resistance via geometric tuning. These progresses underscore hierarchical designs' role in bridging micro-to-macro scales, enhancing tunability, stiffness, and multi-stimuli response for applications in smart textiles and biomimetic robotics.

Despite substantial axial strokes in thermally driven single-fiber TCP artificial muscles, their load-bearing capacity remains modest, yielding energy conversion efficiencies typically below 1%, thus limiting practical engineering deployment (Kongahage et al., 2021). Although existing reinforcements like guest materials or multi-ply twists improve force output, they often reduce stroke length or introduce hysteresis and thermal lag, compromising overall performance (Liu et al., 2024). Consequently, a critical challenge persists: how to elevate load capacity without diminishing axial stroke, a prerequisite for transitioning these actuators to high-impact applications such as load-bearing exoskeletons and industrial robotics. In the present paper, we introduce a novel twisted and coiled artificial muscle fabricated from multi-strand nylon fibers via hierarchical helical principles, achieving a significant elevation in load-bearing capacity to several-fold improvements in force output, while preserving the axial stroke comparable to single-fiber counterparts. The TCP actuator is constructed by some bundles helically wound around its centerline, and each bundle consists of a certain number of filaments helically wound around its helical centerline. The fabrication process involves controlled twisting and coiling of bundled filaments to form multi-level helices, optimizing inter-fiber interactions for enhanced torque transfer and structural stability under thermal stimulation. Meanwhile, a comprehensive mechanical model of the hierarchical helical structure is derived to predict deformation behaviors, systematically analyzing the effect of geometric parameters, such as helical angle, radius, hierarchy of chirality, and hierarchy levels, alongside microstructure evolutions on key performance metrics, including effective stiffness and stress in individual filaments with hierarchical helical structures.

The present paper is outlined as follows. Section 2 details the fabrication and thermomechanical characterization of the hierarchical helical structures. Section 3 formulates the theoretical framework, deriving the equilibrium equations and effective constitutive relations through contact mechanics, stiffness homogenization, and recursive structural analysis. Section 4 outlines the finite element implementation for model verification. Section 5 presents the experimental validation and conducts a systematic parametric investigation to elucidate the effects of hierarchy, chirality, and geometry on actuation performance. Section 6 discusses the underlying physical mechanisms, focusing on stiffness synergy, topological limits, and scaling laws. Finally, Section 7 summarizes the main conclusions.

## 2. Materials and methods

Nylon 6 fibers served as the primary material in this paper. To investigate the thermal properties, two samples were prepared for comparison: Sample 1 was a straight precursor nylon fiber with a diameter of 0.3 mm (Precursor fiber), and Sample 2 consisted of twisted nylon fibers of the same diameter (bias angle approximately $30\ ℃$), heat-set in an environmental chamber at 160 °C for 180 minutes.

### 2.1 Thermomechanical characterizations

Differential scanning calorimetry (DSC) was performed on the two samples using a DSC instrument, with nitrogen as both purge gas (40 mL/min) and protective gas (50 mL/min). Two thermal cycles were conducted. In the first cycle, samples were cooled from $25\ ℃$ to $-50\ ℃$ and then heated to $300\ ℃$ at $10\ ℃/\mathrm{min}$. The second cycle repeated the cooling to -50 °C, followed by heating to $300\ ℃$ at $10\ ℃/\mathrm{min}$.

Figure 1 compares the DSC curves for both samples across the two cycles. The nylon fiber morphology comprises two phases: (I) a soft amorphous phase, indicated by a broad glass transition region from approximately $20\ ^\circ\mathrm{C}$ to $80\ ^\circ\mathrm{C}$; (II) a hard crystalline phase, represented by a melting peak at $180\ ^\circ\mathrm{C}$. Although endothermic peak heights varied, Sample 2 exhibited largely unchanged key thermal properties compared to Sample 1, including the glass transition temperature ($T_g \approx 50\ ^\circ\mathrm{C}$), melting peak position, and overall transition trends, indicating that twisting minimally interrupted fiber morphology and thermal behavior.

In the first cycle, Sample 2 displayed reduced heat-flow values relative to Sample 1 within the $50 - 150\ ^\circ\mathrm{C}$ range, a reduction attributed to the elimination of precursor-fiber-induced residual stresses through annealing. For the second-cycle curves, both samples stabilized and aligned closely, indicating that the initial cycle (heating to $300\ ^\circ\mathrm{C}$) erased the fibers' work history, resulting in consistent thermal behavior.

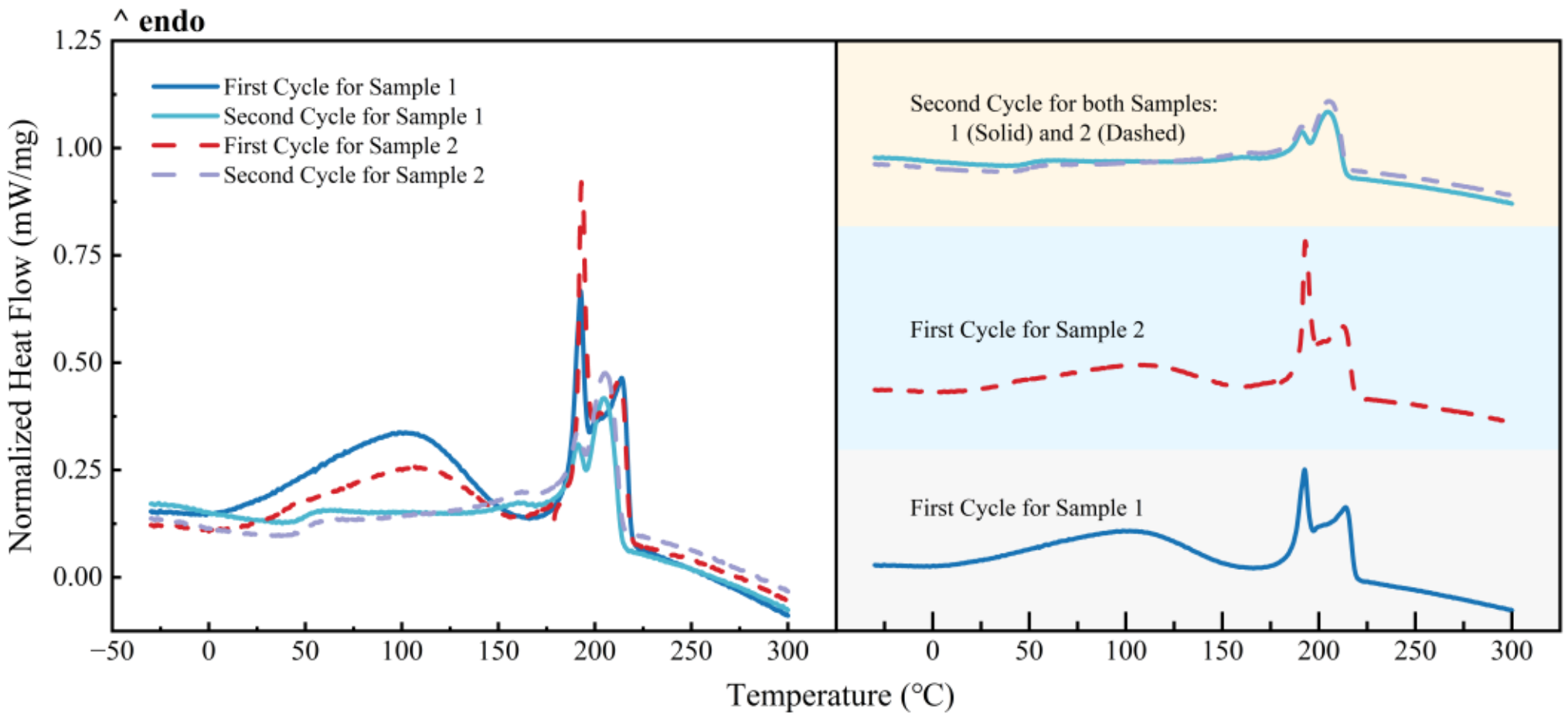

Fig. 1 The results for Sample 1 and Sample 2 by DSC.

## 2.2 Sample preparation with hierarchical helical structure

Hierarchical artificial muscles were fabricated from nylon fibers by a stepwise helical-winding process combined with heat-setting. Initially, multiple precursor nylon fibers were aligned, fixed at one end, and twisted using a stepper motor connected to the other end, forming a secondary helical structure (bundle-filament structure) through multi-strand winding. Twist density, which governs the helix or bias angle as a critical geometric parameter, was precisely controlled by the number of rotations.

This secondary structure was then heat-set in an environmental chamber at 160 ℃ for 180 minutes to eliminate residual internal stresses by annealing, thereby stabilizing the helical configuration (as validated by prior DSC tests). Next, multiple heat-set secondary structures were aligned and subjected to the same twisting process to form a tertiary helical structure (rope-bundle-filament structure). Finally, this tertiary structure underwent another heat-setting at 160 ℃ for 180 minutes. This graded helical winding and heat-setting process yielded stable hierarchical nylon fiber artificial muscles.

## 2.3 Fabrication of twisted and coiled structures

The previously fabricated hierarchical helical structure was treated as a single filament for this process. With one end fully constrained, the other end was connected to a rotary stepper motor to insert twist. Upon reaching a preset twist density—allowing for the regulation of the fiber bias angle— the fiber was wound around a cylindrical mandrel (rod). The diameter of the mandrel determined the helical radius of the coil, while the pitch was controlled during the winding process to adjust the inclination angle. Following formation, the samples were heat-set in a constant temperature oven at 160°C for 180 minutes to release residual stresses and stabilize the coiled geometry, resulting in dimensionally controlled and morphologically stable samples

2.4 Mechanical testing methodology

To characterize the axial-torsional coupled mechanical properties of the hierarchical nylon fiber artificial muscles at room temperature, a tension-torsion tester was employed. Tested samples encompassed various tertiary helical configurations with different geometric parameters, including two precursor fiber radii (0.15 mm and 0.085 mm), chirality (homochiral or heterochiral), structural arrangements (2×2, 2×3, 3×3, 3×4, 4×4), and helix angles (10°, 15°, 20°), with initial lengths ranging from 154 to 183 mm.

Samples were precisely mounted in fiber clamps to ensure alignment, pre-stretched to a taut state, and subjected to uniaxial tension at a constant axial displacement rate of 2 mm/min on one end. The tester recorded axial force and torque arising from the helical structure's resistance to untwisting in real time. For reliability, five replicates per configuration were tested, and results were averaged.

## 2.5 Thermal actuation performance testing

To characterize the actuation performance of the helical fiber artificial muscles under thermal stimuli, a custom thermal actuation testing system was established at room temperature. The artificial muscle was suspended vertically from a rigid bracket, while the lower end was attached to variable weights to apply a constant external load, simulating isotonic operating conditions.

To minimize the effects of external air disturbances and uneven heat dissipation, the muscle was housed within a ceramic tube acting as an insulated heating chamber. A thermocouple attached to the inner wall of the ceramic tube monitored the temperature in real-time to ensure the thermal input approximated the sample environment. A controllable power supply regulated the heating elements to achieve stable temperature ramps from room temperature to the target value.

Actuation displacement was measured using a non-contact infrared displacement sensor aligned with the bottom reference surface of the sample. The testing protocol involved cyclic thermal loading between room temperature and 120°C. During these heating-cooling cycles, the system synchronously recorded the axial contraction displacement, temperature, and time, allowing for the evaluation of the actuation stroke and the recovery behavior during the cooling phase.

## 3. Theoretical models

To elucidate the underlying mechanisms governing the axial-torsional coupling observed in these TCP artificial muscles with hierarchical helical structures, a bottom-up theoretical framework was proposed to investigate the mechanical response and geometric evolution under deformation, providing quantitative predictions that align with experimental results. Unlike phenomenological models, this framework explicitly accounts for the finite deformation kinematics and the recursive geometric constraints inherent in the multi-level helical assembly.

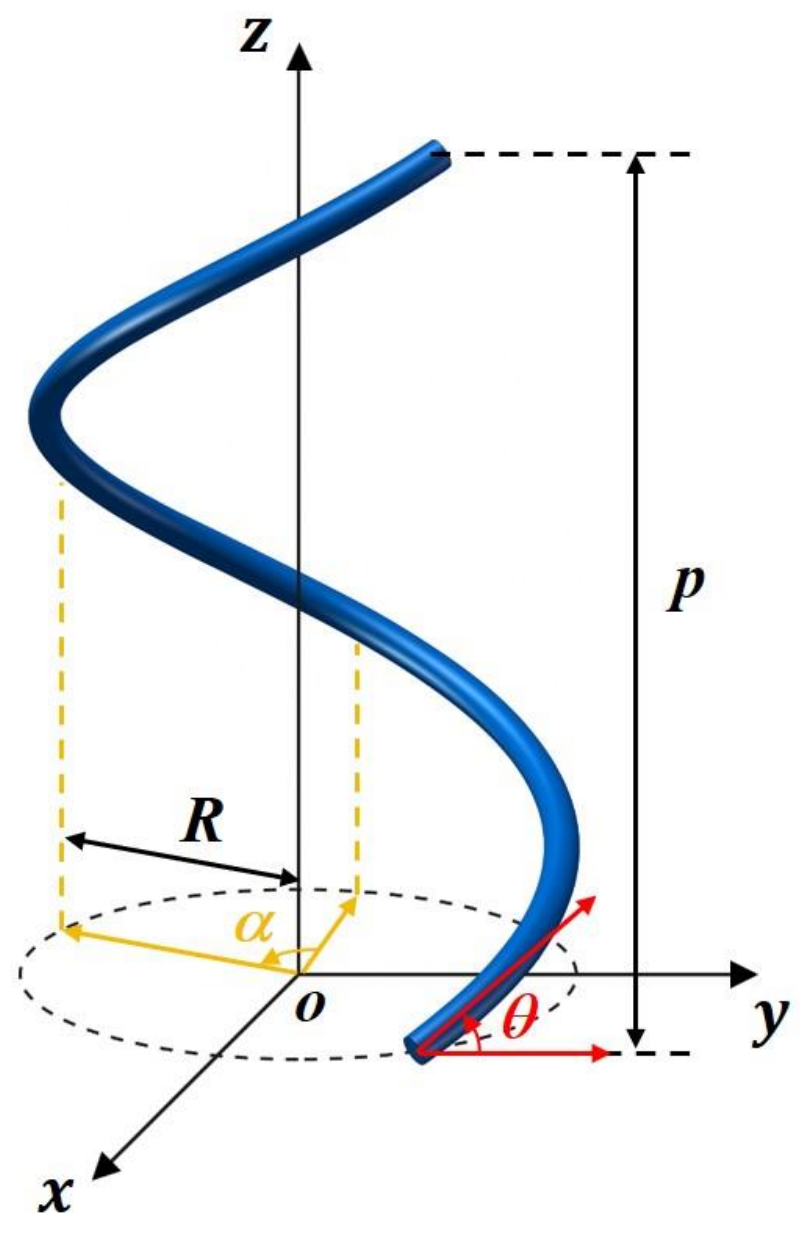

Fig. 2 Geometrical descriptions of a helical curve.

### 3.1 Geometric description of a helical structure

The geometry of a helical curve can be characterized by the helical radius $R$, helical angle $\theta$, helical pitch $p$, and phase angle of the helix $\alpha$, as shown in Fig. 2. In the Cartesian basis $\{\mathbf{e}_1, \mathbf{e}_2, \mathbf{e}_3\}$, an arbitrary point in the curve can be defined as follows,

$$\begin{cases} x = R\cos\alpha, \\ y = R\sin\alpha, \\ z = p\dfrac{\alpha}{2\pi}. \end{cases} \tag{1}$$

Then, the position vector on the helix coiled around the z-axis is expressed as,

$$\mathbf{R}(s) = R\cos\alpha\mathbf{e}_1 + R\sin\alpha\mathbf{e}_2 + \frac{p\alpha}{2\pi}\mathbf{e}_3. \tag{2}$$

where $s$ is the arc length. Based on the Frenet-Serret frame, the orthonormal unit basis vectors along the tangential (**T**), normal (**N**), and binormal (**B**) directions of the curve are derived as,

$$\mathbf{T}(s) = \frac{d\mathbf{R}}{ds} = \frac{1}{\zeta}\left(-R\sin\alpha\mathbf{e}_1 + R\cos\alpha\mathbf{e}_2 + \frac{p}{2\pi}\mathbf{e}_3\right), \tag{3-a}$$

$$\mathbf{N}(s) = \frac{d\mathbf{T}/ds}{|d\mathbf{T}/ds|} = -\cos\alpha\mathbf{e}_1 - \sin\alpha\mathbf{e}_2, \tag{3-b}$$

$$\mathbf{B}(s) = \mathbf{T}(s)\times\mathbf{N}(s) = \frac{1}{\zeta}\left(\frac{p}{2\pi}\sin\alpha\mathbf{e}_1 - \frac{p}{2\pi}\cos\alpha\mathbf{e}_2 + R\mathbf{e}_3\right). \tag{3-c}$$

Here, the metric factor is $\zeta = \left|\frac{d\mathbf{R}}{d\alpha}\right| = \frac{\sqrt{4\pi^2R^2+P^2}}{2\pi}$. Consequently, the curvature and torsion are,

$$\kappa(s) = \frac{R}{\zeta^2}, \qquad \tau(s) = \frac{p}{2\pi\zeta^2}. \tag{4}$$

3.2 Kinematics of the hierarchical helix

As mentioned in our earlier work (Xiao et al, 2022), the Darboux vector $\omega_i(s)$ can be introduced to illustrate the rotation of the Frenet frame in a curve,

$$\omega_i(s) = \kappa(s)\mathbf{B}(s) + \tau(s)\mathbf{T}(s). \quad (5)$$

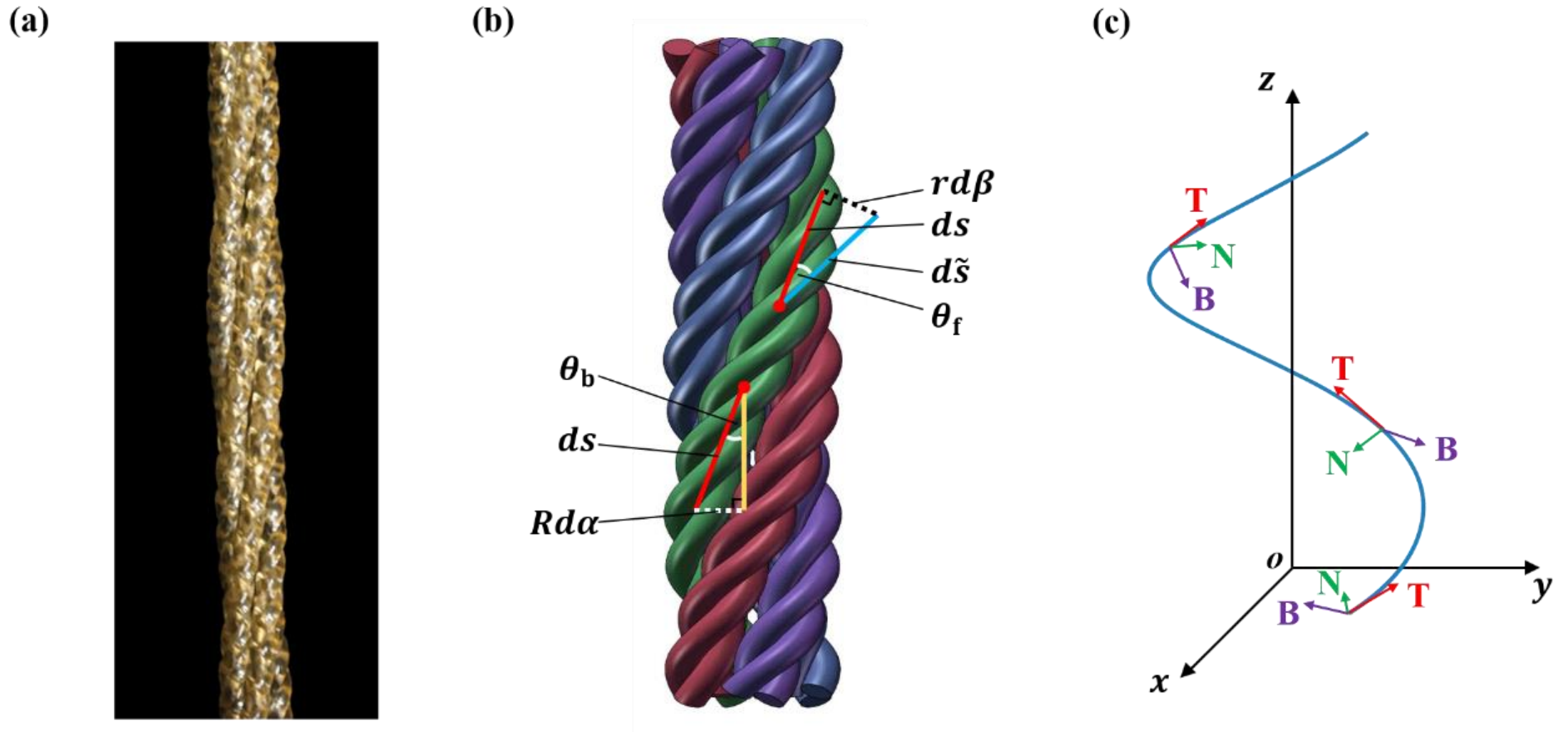

Fig. 3 Optical image (a) and geometrical descriptions (b) of a hierarchical fiber in artificial muscles with hierarchical helical structure (3×4); (c) local right-handed Frenet coordinate frame.

To construct the hierarchical helical morphology, a recursive geometric generation method is employed. As illustrated in Fig. 3, the centerline of a secondary-level structure $\mathbf{r}(\tilde{s})$ is generated by superimposing a secondary helical modulation onto the moving Frenet frame of the primary filament:

$$\mathbf{r}(\tilde{s}) = \mathbf{R}(s) + r\cos\beta\mathbf{N}(s) + r\sin\beta\mathbf{B}(s). \quad (6)$$

where $r$ and $\beta$ denote the radius and phase angle of the secondary helix, respectively. For the geometric compatibility between the hierarchical structure, we relate the differential arc length of the filament $d\tilde{s}$ to that of the bundle $ds$. Differentiating the position vector $\mathbf{r}(\tilde{s})$ with respect to the phase angle $\alpha$ yields:

$$\frac{d\mathbf{r}}{d\alpha} = \frac{dR}{d\alpha} + r\left(-\sin\beta\frac{d\beta}{d\alpha}\mathbf{N} + \cos\beta\frac{d\mathbf{N}}{d\alpha}\right) + r\left(\cos\beta\frac{d\beta}{d\alpha}\mathbf{B} + \sin\beta\frac{d\mathbf{B}}{d\alpha}\right), \quad (7)$$

Substituting the relations $d\mathbf{N}/d\alpha = \zeta(-\kappa\mathbf{T} + \tau\mathbf{B})$ and $d\mathbf{B}/d\alpha =$

$-\zeta\tau\mathbf{N}$, and simplifying the vector components, we derive the squared metric tensor components. The Jacobian of the transformation, $J_{\text{Jaco}} = d\tilde{s}/ds$, is explicitly derived as:

$$\frac{d\tilde{s}}{ds} = \frac{1}{\zeta}\left|\frac{d\mathbf{r}}{d\alpha}\right| = \left[\left(1-\kappa r\cos\beta\right)^2 + \left(\tau r + \frac{d\beta}{ds}\zeta\right)^2\right]^{\frac{1}{2}}, \tag{8}$$

The equations above serve as the fundamental bridge between hierarchical levels, rigorously relating the primary filament strain to the secondary bundle deformation and twist.

### 3.3 Equilibrium equations of a helical filament in the hierarchy

On the basis of the concise geometrical descriptions of a two-level helical structure, the equilibrium governing equations of a filament in the artificial muscle with the hierarchical helical structure are achieved. Due to the fabrication process of the artificial muscles, these helical filaments of carbon nanotubes of polymers are often discontinuous, yet linked by the guest materials, such as paraffin and nylon. Therefore, all the filaments of artificial muscles are assumed to be long enough with circular cross-section in the undeformed configuration.

In this section, a straight bundle (secondary structure), containing *m* filaments winded into helices, is presented with three loading conditions as axial tension, torsion and bending.

#### 3.3.1 Orthogonality of inter-filament contact forces

The bundle made of filaments is treated as a model system, in which the adjacent filaments are identified contact without friction demonstrated by Costello (1997). Referring to the Frenet frame {**T**, **N**, **B**}, the differential equilibrium equations of a single filament in the bundle can be expressed as (Xiao et al, 2014),

$$\frac{dF_i}{ds}+\psi_{ijk}\omega_j F_k - f_i = 0, \qquad \frac{dM_i}{ds}+\psi_{ijk}\omega_j M_k - \psi_{ij3}F_j = 0. \tag{9}$$

Here $F_i$ and $M_i$ are the internal forces and moments on the infinitesimal filament along the tangent, normal, and binormal directions, respectively. $f_i$ is designated as the contact pressures on the infinitesimal filament along the three directions induced by the adjacent filaments, as show in Fig. 4. $\omega_i$ denotes the Darboux vector, $\psi_{ijk}$ is the permutation sign, and $s$ is the arc-length parameter.

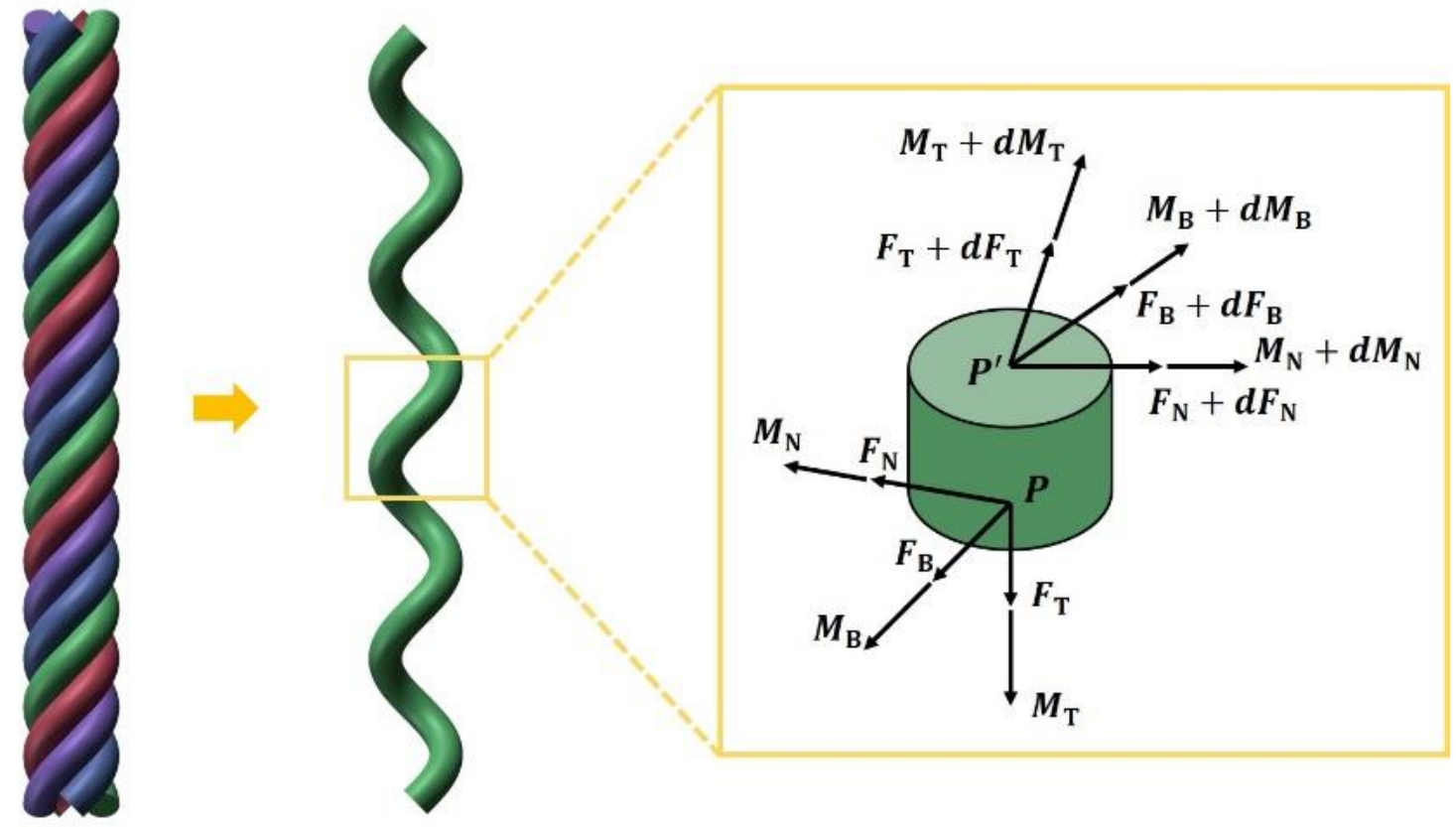

Fig. 4 Model of helical structures. (a) a straight bundle consisting of m filaments and (b) internal forces of a helical filament.

To simplify the Darboux vector, let $i = \mathbf{T},\ \mathbf{N},\ \mathbf{B}$ and the geometrical relations hold for a helical curve:

$$\kappa_{\mathbf{N}}^{\mathrm{f}}=0, \quad \kappa_{\mathbf{B}}^{\mathrm{f}}=\frac{\cos^2\theta_{\mathrm{f}}}{r_{\mathrm{b}}}, \quad \tau^{\mathrm{f}}=\frac{\eta_{\mathrm{f}}\sin\theta_{\mathrm{f}}\cos\theta_{\mathrm{f}}}{r_{\mathrm{b}}}. \tag{10}$$

Similarly, the initial curvature and torsion of filament in the undeformed

configuration are

$$\kappa_{\mathbf{N}}^{\mathrm{f}(0)}=0,\ \ \kappa_{\mathbf{B}}^{\mathrm{f}(0)}=\frac{\cos^2\theta_{\mathrm{f}}^0}{r_{\mathrm{b}}^0},\ \ \tau^{\mathrm{f}(0)}=\frac{\eta_{\mathrm{f}}\sin\theta_{\mathrm{f}}^0\cos\theta_{\mathrm{f}}^0}{r_{\mathrm{b}}^0}. \tag{11}$$

where $\tau^{\mathrm{f}}$, $\kappa_{\mathbf{N}}^{\mathrm{f}}$ and $\kappa_{\mathbf{B}}^{\mathrm{f}}$ are the components of $\omega_i$ along with the tangential, normal and binormal directions, the superscript (0) or 0 is designated to represent the undeformed configuration of the hierarchical structure, $\eta_{\mathrm{f}}$ denotes the helical chirality of filaments. $r_{\mathrm{b}}$ denotes radius of the bundle and is defined as a function of the axial strain $\varepsilon_{\mathrm{f}}$ and filament's radius $r_{\mathrm{f}}$,

$$r_{\mathrm{b}}=r_{\mathrm{b}}^0-r_{\mathrm{f}}^0\nu\varepsilon_{\mathrm{f}}\csc\frac{\pi}{m}. \tag{12}$$

$$r_{\mathrm{f}}=r_{\mathrm{f}}^0\left(1-\nu\varepsilon_{\mathrm{f}}\right). \tag{13}$$

here, $\upsilon$ is Poisson's ratio.

As mentioned in the literatures (Liu et al, 2018; Chu et al, 2021; Xiao, et al, 2022), since a straight bundle subjected to axial tension and torsion deforms uniformly along the length direction, all the internal forces in the helical structure remain with the arc-length $s$, *i.e.* $\frac{dF_{\mathbf{T}}}{ds}=\frac{dF_{\mathbf{N}}}{ds}=\frac{dF_{\mathbf{B}}}{ds}=\frac{dM_{\mathbf{T}}}{ds}=\frac{dM_{\mathbf{N}}}{ds}=\frac{dM_{\mathbf{B}}}{ds}=0$. Thus, Eq. (9) can be reduced to a series of algebraic equations,

$$F_{\mathbf{T}}=EA\varepsilon_{\mathrm{f}}, \tag{14a}$$

$$M_{\mathbf{B}}=EI(\kappa_{\mathbf{B}}-\kappa_{\mathbf{B}}^0), \tag{14b}$$

$$M_{\mathbf{T}}=GJ(\tau-\tau^0), \tag{14c}$$

$$M_{\mathbf{N}}=EI(\kappa_{\mathbf{N}}-\kappa_{\mathbf{N}}^0). \tag{14d}$$

where $\varepsilon_{\mathrm{f}}$ and $A$ are the axial strain and cross-sectional area of the filament.

$E$ and $G$ are Young's modulus and shear modulus, respectively. $I$ is the moment of inertia and $J$ is the polar moment of inertia.

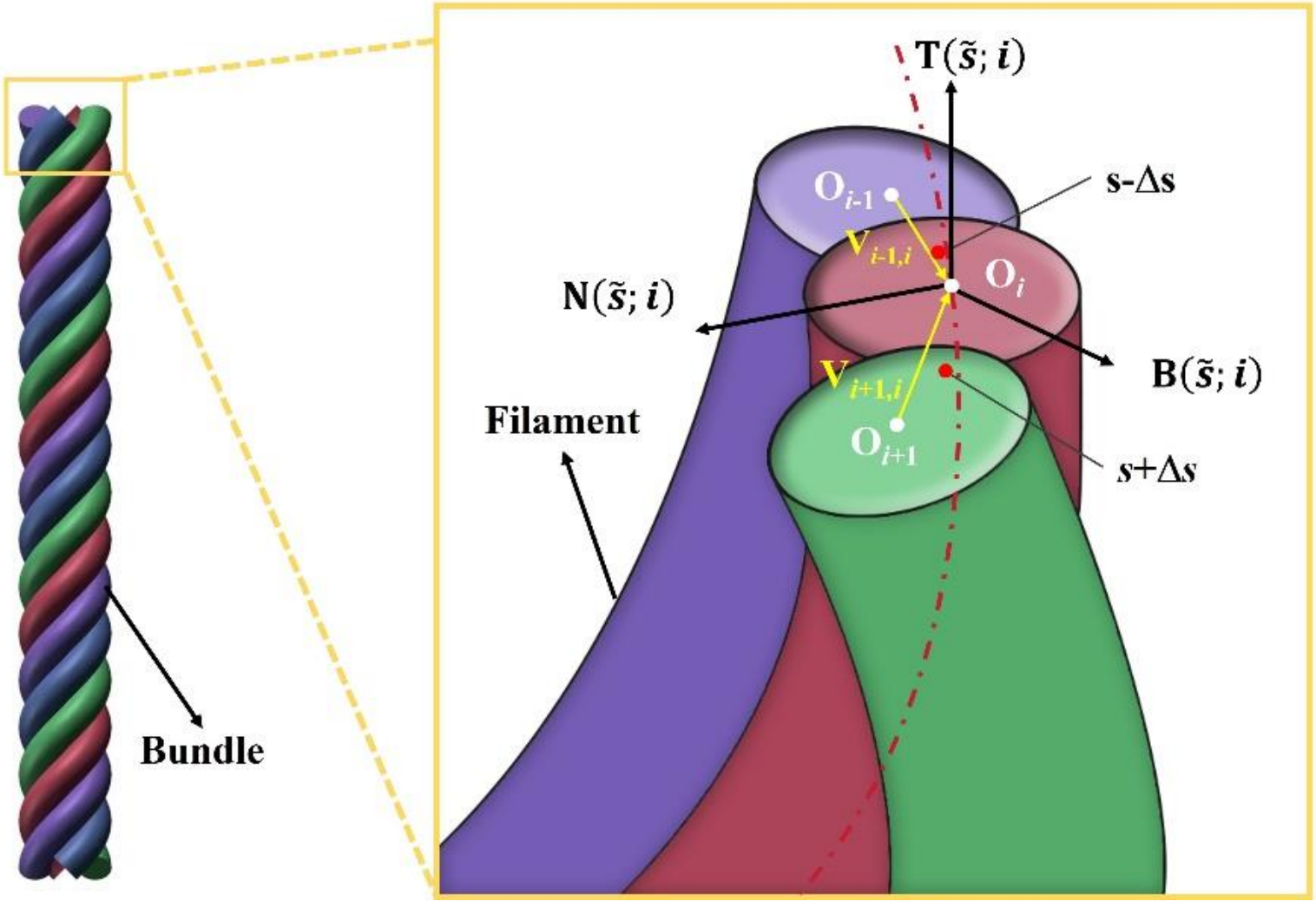

Fig. 5 The force analysis of the three adjacent filaments in a straight bundle.

Meanwhile, a critical aspect of the bundle mechanics is the vector nature of the interaction forces. Let $O_i$ be the centroid of the $i$-th filament, as shown in Fig. 5. The interaction vector $\mathbf{v}_{i-1,i+1}$ connecting the contact points with neighbors $i-1$ and $i+1$ is derived from the difference in their position vectors:

$$\mathbf{v}_{i-1,i+1} = \mathbf{R}(s+\Delta s, i-1) - \mathbf{R}(s-\Delta s, i+1), \tag{15}$$

where the angular offset for the $k$-th filament is $\alpha_k = \alpha + 2\pi k/m$. Upon substituting the parametric helical equations and applying trigonometric expansions to the planar components, the following expressions are obtained:

$$\mathbf{v}_x = R\left[\cos(\alpha + \Delta\alpha + \frac{2\pi(i-1)}{m}) - \cos(\alpha - \Delta\alpha + \frac{2\pi(i+1)}{m})\right], \tag{16}$$

$$\mathbf{v}_y = R\left[\sin(\alpha + \Delta\alpha + \frac{2\pi(i-1)}{m}) - \sin(\alpha - \Delta\alpha + \frac{2\pi(i+1)}{m})\right], \tag{17}$$

Owing to the symmetry condition $\Delta\alpha \approx 2\pi/m$, the dot product of this vector with the normal vector $\mathbf{N}(s,i) = [-\cos\alpha_i, -\sin\alpha_i, 0]$ is evaluated. Through algebraic cancellation, it is found that:

$$\mathbf{v}_{i-1,i+1} \cdot \mathbf{N}(s,i) \equiv 0. \tag{18}$$

This orthogonality condition implies that the line connecting the neighbors is tangent to the curvature circle. Since the resultant contact force $f$ from neighbors $i-1$ and $i+1$ must be symmetric with respect to the bisecting plane (the normal plane of filament $i$), the tangential and binormal components vanish. Thus, it is demonstrated that the contact pressure is purely normal:

$$f_{\mathbf{T}} = 0, \tag{19a}$$

$$f_{\mathbf{N}} = F_{\mathbf{T}}\kappa_{\mathbf{B}} - F_{\mathbf{B}}\tau, \tag{19b}$$

$$f_{\mathbf{B}} = 0. \tag{19c}$$

Substituting Eqs. (19a-c) into Eq. (10), the shear force relation to be decoupled as:

$$F_{\mathbf{B}} = M_{\mathbf{T}}\kappa_{\mathbf{B}} - M_{\mathbf{B}}\tau. \tag{20}$$

Therefore, the shear force $F_B$ is entirely determined by the coupling between torsion and bending, uninfluenced by friction or tangential contact.

### 3.3.2 Kinematic compatibility and strain analysis

The actuation of the TCP muscle is driven by anisotropic thermal expansion, manifesting as a coupling between radial expansion and axial-torsional deformation. This coupling is governed by strict kinematic compatibility constraints.

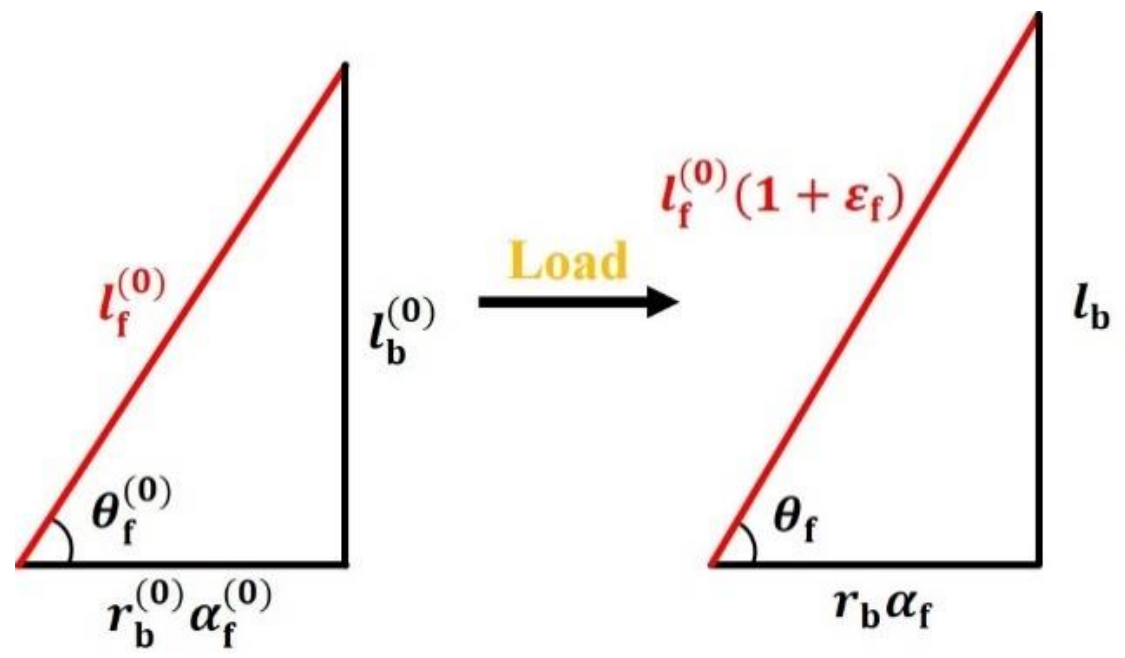

Fig. 6 The geometric relations between the helical filament and straight bundle in the undeformed and the deformed configurations.

By considering the "unrolled" geometry of the helical filament on the surface of the bundle cylinder, the filament forms the hypotenuse of a right triangle, where the legs represent the bundle's axial length $l_b$ and circumference $2\pi r_b$, as shown in Fig. 6. In the undeformed configuration, the geometric relations are:

$$l_f^0 = \frac{l_b^0}{\sin\theta_f^0}, \tag{21}$$

$$2\pi r_b^0 N_{turn} = l_b^0 \cot\theta_f^0, \tag{22}$$

In the deformed configuration, introducing the bundle axial strain $\varepsilon_b$ and torsional strain $\gamma_b = \frac{r_b \Delta\alpha}{l_b}$, Eqs. (21-22) evolve to:

$$l_f = l_b / \sin\theta_f, \tag{23}$$

$$2\pi r_b N_{turn} + l_b \gamma_b = l_b \cot\theta_f, \tag{24}$$

As illustrated in Fig. 6, the axial strain of the filament $\varepsilon_f$ is derived by comparing the hypotenuse lengths:

$$\varepsilon_f = \frac{l_f - l_f^0}{l_f^0} = \frac{l_b / \sin\theta_f}{l_b^0 / \sin\theta_f^0} - 1 = (1+\varepsilon_b)\frac{\sin\theta_f^0}{\sin\theta_f} - 1. \tag{25}$$

Simultaneously, torsional compatibility requires linking the change in radius to the one in angle. Incorporating the Poisson effect, the radius contracts as $r_\mathrm{b} = r_\mathrm{b}^0 - w\varepsilon_\mathrm{f}$ (by Eq. 10), where $w = r_\mathrm{f}^0 \upsilon \csc\frac{\pi}{m}$ is a geometric factor dependent on the Poisson's ratio and number of filaments in a bundle. Substituting this into the cotangent relation yields the implicit consistency equation:

$$\frac{r_\mathrm{b}^0 - w\varepsilon_\mathrm{f}}{r_\mathrm{b}^0}\cot\theta_\mathrm{f} = \frac{\cot\theta_\mathrm{f}^0 + \gamma_\mathrm{b}}{1+\varepsilon_\mathrm{b}}, \tag{26}$$

To determine the unknown helical angle $\theta_\mathrm{f}$, the expression for $\varepsilon_\mathrm{f}$ is substituted back into Eq. (26). Rearranging terms to isolate $\sin\theta_\mathrm{f}$ and $\cos\theta_\mathrm{f}$ yields a linear trigonometric equation of the form $A\sin\theta_\mathrm{f} + B\cos\theta_\mathrm{f} = C$ as:

$$(r_\mathrm{b}^0 + w)\sin\theta_\mathrm{f} - \left[\frac{r_\mathrm{b}^0(1+\varepsilon_\mathrm{b})}{\gamma_\mathrm{b} + \cot\theta_\mathrm{f}^0}\right]\cos\theta_\mathrm{f} = w(1+\varepsilon_\mathrm{b})\sin\theta_\mathrm{f}^0 \tag{27}$$

Then, the explicit solution is obtained by the sine addition formula method:

$$\theta_\mathrm{f} = \arcsin\frac{w(1+\varepsilon_\mathrm{b})\sin\theta_\mathrm{f}^0}{\sqrt{\left(r_\mathrm{b}^0 + w\right)^2 + u^2}} + \arctan\frac{u}{\left(r_\mathrm{b}^0 + w\right)}, \tag{28}$$

$$u = \frac{r_\mathrm{b}^0\left(1+\varepsilon_\mathrm{b}\right)}{\gamma_\mathrm{b} + \cot\theta_\mathrm{f}^0}. \tag{29}$$

$$\gamma_\mathrm{b} = \eta_\mathrm{f} r_\mathrm{b}^0 \cdot \phi_\mathrm{b}, \tag{30}$$

$$\phi_\mathrm{b} = \eta_\mathrm{f}\frac{\alpha_\mathrm{f} - \alpha_\mathrm{f}^0}{l_\mathrm{b}^0}. \tag{31}$$

In summary, all the internal forces of the helical filament, in a straight bundle with axial strain $\varepsilon_\mathrm{b}$ and torsional strain $\gamma_\mathrm{b}$, can be calculated by the double-stage mechanical model above.

### 3.3.3 Homogenization of bending stiffness

In this section, the mechanical behavior of a bending bundle consisting by $m$ filaments will be presented. When the bundle undergoes bending with a curvature radius ρ, the helical symmetry is broken, and the cross-section of the bundle becomes non-uniform along the binormal axis, as shown in Fig. 7. Therefore, it is incompatible with present bending rigidity of a bundle that the filament's profile is regarded as a circle or even an ellipse in the analysis of the mechanical properties.

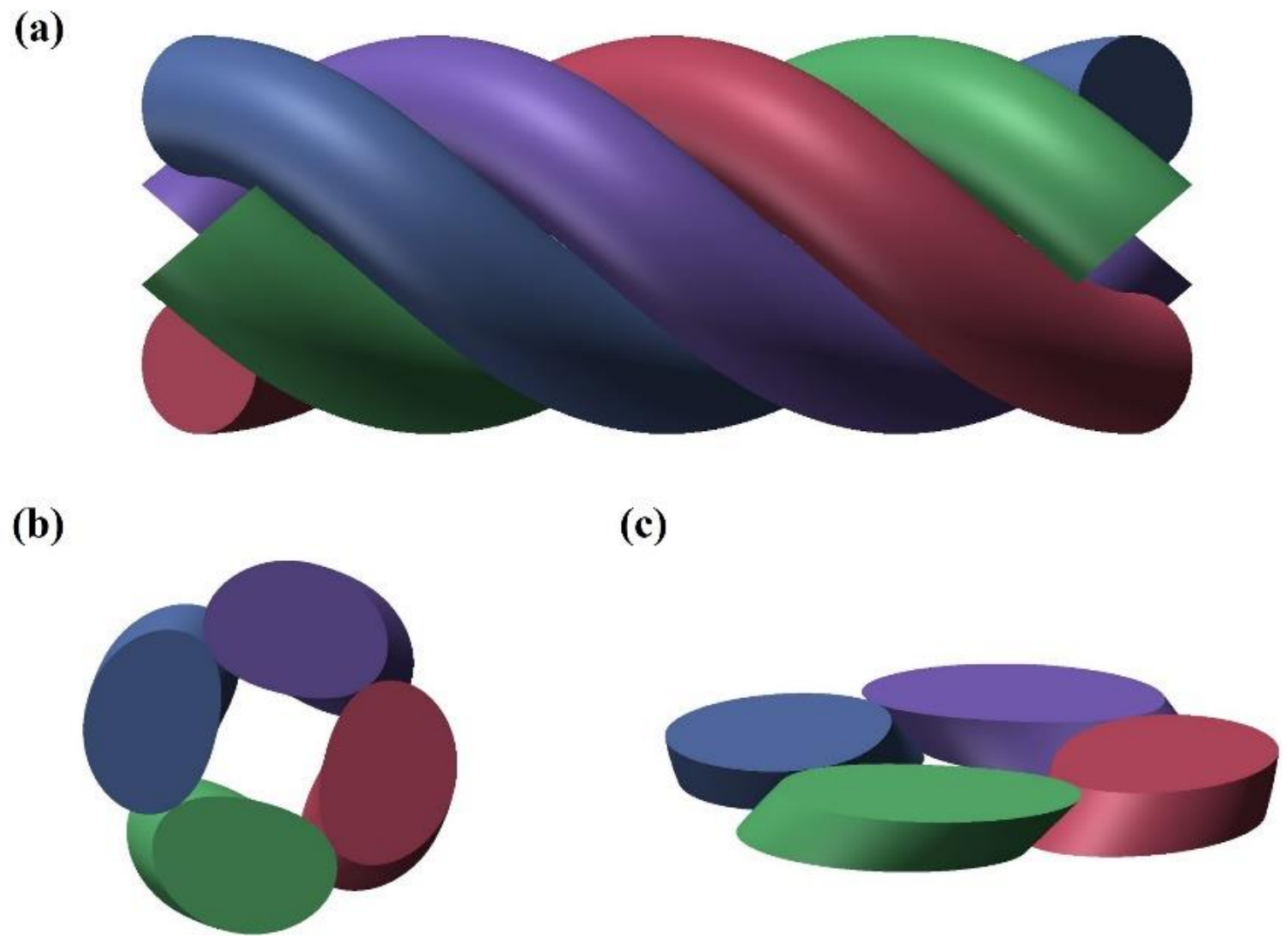

Fig. 7 The lateral and front views of a straight bundle with 4-filaments

The centerline of a filament in a bent bundle is parameterized by superimposing the helical winding onto a toroidal axis $R_{\text{arc}}(s)$:

$$\mathbf{r}(\tilde{s}) = \mathbf{R}_{\text{arc}}(s_{\text{b}}) + r_{\text{b}} \cos\beta_{\text{f}} \mathbf{N}_{\text{arc}}(s_{\text{b}}) + r_{\text{b}} \sin\beta_{\text{f}} \mathbf{B}_{\text{arc}}(s_{\text{b}}), \tag{32}$$

$$\mathbf{R}_{\text{arc}}(s_{\text{b}}) = \rho\cos\alpha_{\text{b}}\mathbf{e}_1 + \rho\sin\alpha_{\text{b}}\mathbf{e}_2, \tag{33}$$

$$\rho = \frac{ds_{\text{b}}}{d\alpha_{\text{b}}}. \tag{34}$$

where $s_b$ is the arc length of bundle's centerline and $\alpha_{\text{b}}$ denotes the phase angle of the bending bundle, as shown in Fig. 8.

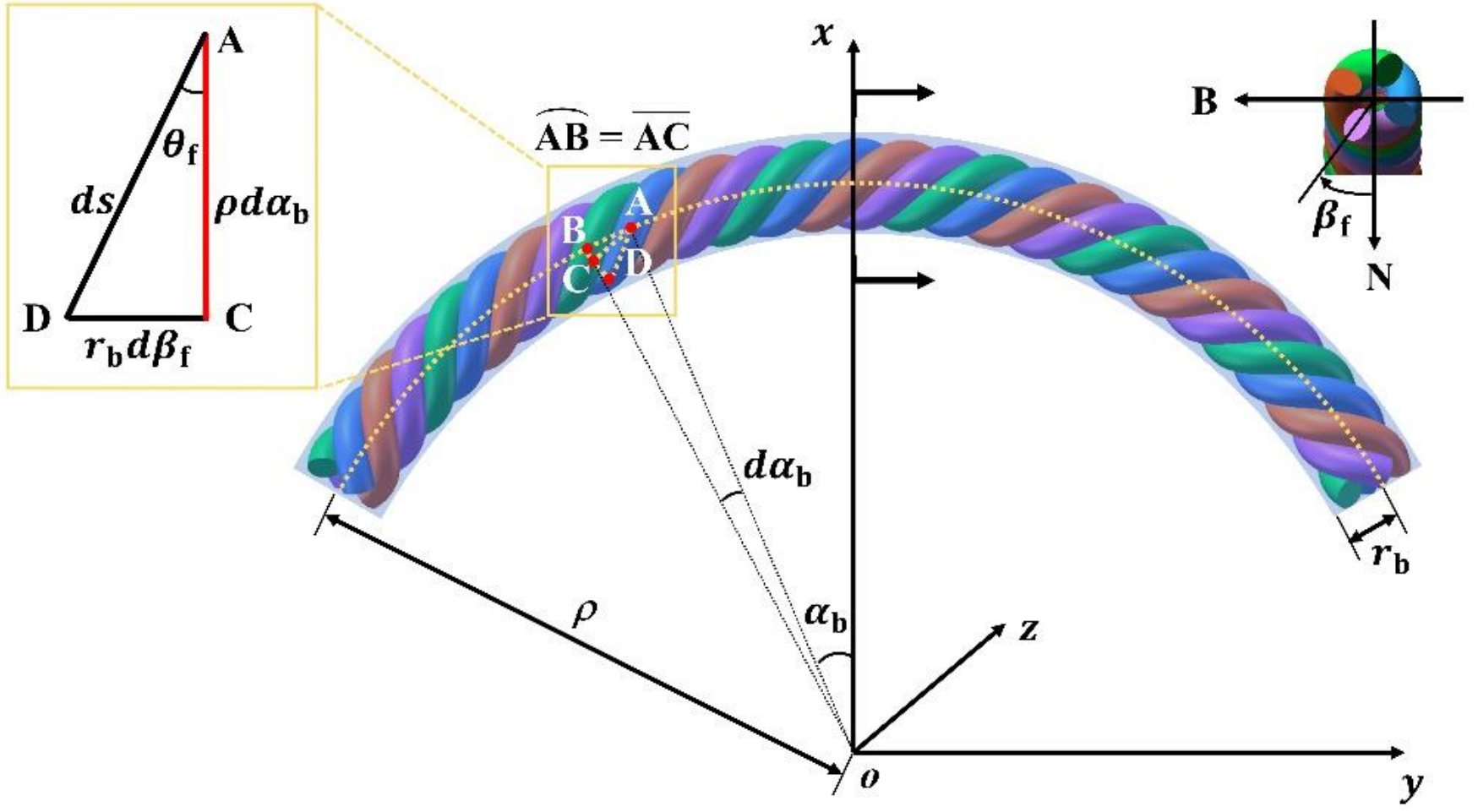

Fig. 8 The schematic diagram of geometric relations for a bending bundle

The local tangent vector $t_i$ is obtained by differentiating $r(s)$ with respect to the arc length, utilizing the Frenet formulas for the toroidal axis as:

$$\mathbf{N}(s_b) = -\cos\alpha_b \mathbf{e}_1 - \sin\alpha_b \mathbf{e}_2, \tag{35}$$

$$\mathbf{B}(s_b) = \mathbf{e}_3. \tag{36}$$

Subsequently, the local curvature corresponding to the cross-section $\kappa_f$ is identified as the second derivative. The effective bending moment is defined as the average projection of the local moments onto the bundle's binormal axis over one full helical period ($2\pi$ variation in $\beta_f$):

$$M_b^z = \frac{m}{2\pi} \int_0^{2\pi} EI_f (\kappa_f \cdot e_{\text{binormal}}) d\beta_f \tag{37}$$

The dot product $\kappa_f \cdot e_{\text{binormal}}$ involves complex geometric terms characterizing the variation of the filament's orientation relative to the bending plane, the detailed derivation is provided in Appendix A. The final expression for the effective stiffness is:

$$(EI)_{\mathrm{b}}^{\mathrm{eff}} = M_{\mathrm{b}}^{z}\rho = \frac{mEI_{\mathrm{f}}\rho}{2\pi}\int_{0}^{2\pi}\frac{\Psi(\beta_{\mathrm{f}},\theta_{\mathrm{f}},\rho,r_{\mathrm{b}})}{\left[(\rho - r_{\mathrm{b}}\cos\beta_{\mathrm{f}})^{2} + \rho^{2}\tan^{2}\theta_{\mathrm{f}}\right]^{\frac{3}{2}}}d\beta_{\mathrm{f}}, \tag{38}$$

where the numerator function $\Psi$ encapsulates the geometric nonlinearity:

$$\Psi = (r_{\mathrm{b}}^{2} - \rho^{2}\tan^{2}\theta_{\mathrm{f}})\cos^{2}\beta_{\mathrm{f}} - \rho\left(r_{\mathrm{b}} + \frac{\rho^{2}\tan^{2}\theta_{\mathrm{f}}}{r_{\mathrm{b}}}\right)\cos\beta_{\mathrm{f}} + \rho^{2}(1 + 2\tan^{2}\theta_{\mathrm{f}}). \tag{39}$$

Eqs. (38-39) accounts for the stiffening effect due to the helical angle and the coupling between the bending curvature and filament torsion.

3.4 Recursive mechanics for tertiary structures

Owing to the similar hierarchical structure, the relation between the bundle and filament can be easily extended to the higher-level structure (the rope). The rope is modeled as a helical assembly of *n* bundles. Based on the principle of hierarchical similarity, the bundles are treated as "effective filaments" possessing the homogenized tensile stiffness (derived from Eq. 28) and bending stiffness (derived from Eq. 38).

The macroscopic kinematics of the rope are defined by the global axial strain $\varepsilon_{\mathrm{r}}$ and torsional strain $\gamma_{\mathrm{r}}$. These global variables drive the deformation of the constituent bundles through an analogous set of compatibility equations, determining the local bundle strain $\varepsilon_{\mathrm{b}}$ and helical angle $\theta_{\mathrm{b}}$:

$$\theta_{\mathrm{b}} = \arcsin\frac{\psi(1+\varepsilon_{\mathrm{r}})\sin\theta_{\mathrm{b}}^{0}}{\sqrt{(r_{\mathrm{r}}^{0}+\psi)^{2} + \chi^{2}}} + \arctan\frac{\chi}{(r_{\mathrm{r}}^{0}+\psi)}. \tag{40}$$

$$\chi = \frac{r_{\mathrm{r}}^{0}(1+\varepsilon_{\mathrm{r}})}{\gamma_{\mathrm{r}} + \cot\theta_{\mathrm{b}}^{0}}, \tag{41}$$

$$\psi = r_{\mathrm{b}}^{0}\upsilon\csc\frac{\pi}{n}, \tag{42}$$

$$r_{\mathrm{r}} = r_{\mathrm{r}}^{0} - r_{\mathrm{b}}^{0}\nu\varepsilon_{\mathrm{b}}\csc\frac{\pi}{n}. \tag{43}$$

Here $r_{\mathrm{r}}$ and $l_{\mathrm{r}}$ denote the radius and axial length of the straight rope. By the explicit solution of $\theta_{\mathrm{b}}$, the curvature and torsion of a bundle in the rope are derived as

$$\kappa_{\mathbf{N}}^{\mathrm{b}}=0,\quad \kappa_{\mathbf{B}}^{\mathrm{b}}=\frac{\cos^2\theta_{\mathrm{b}}}{r_{\mathrm{r}}},\quad \tau^{\mathrm{b}} = \frac{\eta_{\mathrm{b}}\sin\theta_{\mathrm{b}}\cos\theta_{\mathrm{b}}}{r_{\mathrm{r}}}. \tag{44}$$

$$\kappa_{\mathbf{N}}^{\mathrm{b}(0)}=0,\quad \kappa_{\mathbf{B}}^{\mathrm{b}(0)}=\frac{\cos^2\theta_{\mathrm{b}}^{0}}{r_{\mathrm{b}}^{0}},\quad \tau^{\mathrm{b}(0)} = \frac{\eta_{\mathrm{b}}\sin\theta_{\mathrm{b}}^{0}\cos\theta_{\mathrm{b}}^{0}}{r_{\mathrm{b}}^{0}}. \tag{45}$$

Then, the twisting angle per unit length $\phi_{\mathrm{b}}$ is expressed as,

$$\phi_{\mathrm{b}} = \eta_{\mathrm{b}}\left(\frac{\sin\theta_{\mathrm{b}}\cos\theta_{\mathrm{b}}}{r_{\mathrm{r}}} - \frac{\sin\theta_{\mathrm{b}}^{0}\cos\theta_{\mathrm{b}}^{0}}{r_{\mathrm{r}}}\right). \tag{46}$$

and the analytical solution of the torsional strain $\gamma_{\mathrm{b}}$ can be solved by Eqs. (30) and (46).

On the basis of the specified value of bundle's axial strain $\varepsilon_{\mathrm{b}}$ and torsional strain $\gamma_{\mathrm{b}}$, $\varepsilon_{\mathrm{f}}$ and $\theta_{\mathrm{f}}$ can be obtained from Eqs. (25) and (30). Combined with Eqs. (14) and (20), the internal forces of the filament are determined. Then, the total forces and torque in the bundle can be derived by projecting the ones in the filaments along the bundle's centerline,

$$F_{\mathbf{T}}^{\mathrm{b}} = m\left(F_{\mathbf{T}}\sin\theta_{\mathrm{f}} + \eta_{\mathrm{f}}F_{\mathbf{B}}\cos\theta_{\mathrm{f}}\right), \tag{47a}$$

$$M_{\mathbf{T}}^{\mathrm{b}} = m\left(M_{\mathbf{T}}\sin\theta_{\mathrm{f}} + \eta_{\mathrm{f}}M_{\mathbf{B}}\cos\theta_{\mathrm{f}} + \eta_{\mathrm{f}}F_{\mathbf{T}}r_{\mathrm{b}}\cos\theta_{\mathrm{f}} - F_{\mathbf{B}}r_{\mathrm{b}}\sin\theta_{\mathrm{f}}\right), \tag{47b}$$

$$M_{\mathbf{B}}^{\mathrm{b}} = \frac{(EI)_{\mathrm{b}}^{\mathrm{eff}}}{\rho} - \frac{(EI)_{\mathrm{b}}^{\mathrm{eff}(0)}}{\rho}, \tag{47c}$$

Refer to Eq. (44), the curvature radius $\rho_{\mathrm{b}}$ of the bundle is defined as,

$$\rho_{\mathrm{b}} = \frac{r_{\mathrm{r}}}{\cos^2 \theta_{\mathrm{b}}}, \tag{47d}$$

Similarly, the internal force along the binormal direction of the bundle can be represented from Eq. (20),

$$F_{\mathbf{B}}^{\mathrm{b}} = M_{\mathbf{T}}^{\mathrm{b}} \kappa_{\mathbf{B}}^{\mathrm{b}} - M_{\mathbf{B}}^{\mathrm{b}} \tau^{\mathrm{b}} = M_{\mathbf{T}}^{\mathrm{b}} \frac{\cos^2 \theta_{\mathrm{b}}}{r_{\mathrm{r}}} - M_{\mathbf{B}}^{\mathrm{b}} \frac{\eta_{\mathrm{b}} \sin \theta_{\mathrm{b}} \cos \theta_{\mathrm{b}}}{r_{\mathrm{r}}}. \tag{47e}$$

The recursive analytical structure establishes a rigorous, quantitative link between the constituent material properties, the multi-level geometric parameters, and the macroscopic actuation performance of the hierarchical TCP muscle.

### 3.5 Tension and stress analysis of the tertiary structure

Referring to Eqs. (47a) and (47b), the total axial force and torque in the rope can be given by

$$F_{\mathbf{T}}^{\mathrm{r}} = n\left(F_{\mathbf{T}}^{\mathrm{b}} \sin \theta_{\mathrm{b}} + \eta_{\mathrm{b}} F_{\mathbf{B}}^{\mathrm{b}} \cos \theta_{\mathrm{b}}\right), \tag{48a}$$

$$M_{\mathbf{T}}^{\mathrm{r}} = n\left(M_{\mathbf{T}}^{\mathrm{b}} \sin \theta_{\mathrm{b}} + \eta_{\mathrm{b}} M_{\mathbf{B}}^{\mathrm{b}} \cos \theta_{\mathrm{b}} + \eta_{\mathrm{b}} F_{\mathbf{T}}^{\mathrm{b}} r_{\mathrm{r}} \cos \theta_{\mathrm{b}} - F_{\mathbf{B}}^{\mathrm{b}} r_{\mathrm{r}} \sin \theta_{\mathrm{b}}\right). \tag{48b}$$

In the preceding analysis, we have derived precise analytical closed-form expressions for the resultant force and torque within the hierarchical helical structure, accounting for deformations in the axial, torsional, bending, and radial directions. From an experimental perspective, the deformation state of the outermost layer, typically characterized by the axial strain $\varepsilon_{\mathrm{r}}$ and torsional strain $\gamma_{\mathrm{r}}$, is readily measurable. Consequently, the proposed bottom-up theoretical framework facilitates a systematic analysis of the hierarchical helical structure's mechanical response, which includes predicting the macroscopic force-displacement relations and microstructure evolution during reversible actuation.

### 3.6 Thermo–mechanical model of TCP with a hierarchical fiber

Following the hierarchical mechanics framework proposed in the preceding sections, we then advance to the TCP muscle as the quaternary structure, where the hierarchical rope above serves as the constituent filament. Unlike the lower-level hierarchies that the constituent filaments/bundles were approximated as helices wound around a straight core with negligible curvature effects to their radius, the rope in a TCP structure possesses a substantial initial curvature due to the mandrel-coiling fabrication process. To accurately predict the actuation performance, especially the coupling between the radial thermal expansion and the axial extension, it is essential to establish a finite strain framework that explicitly accounts for the initial curvature of the rope. Hence, a theoretical model is proposed herein to describe the thermo-mechanical response of the TCP structure.

#### 3.6.1 Geometrical description of the initial curvature in the coiled rope

The geometry of the TCP structure is modeled as a helical coil characterized by helical radius $R_{\mathrm{c}}$ and angle $\theta_{\mathrm{c}}$ of coil. Consequently, the rope, acting as the curvilinear fiber, possesses the intrinsic initial curvature ${\kappa_{\mathrm{r}}}^0$ and initial torsion ${\tau_{\mathrm{r}}}^0$ in the reference (undeformed) configuration. Consistent with the geometric descriptions in Eq. (44), these parameters are strictly determined by the initial helical geometry of the TCP:

$$\kappa_{\mathrm{r}}^{0}=\frac{\cos^{2}\theta_{\mathrm{c}}^{0}}{R_{\mathrm{c}}^{0}},\quad \tau_{\mathrm{r}}^{0}=\frac{\sin\theta_{\mathrm{c}}^{0}\cos\theta_{\mathrm{c}}^{0}}{R_{\mathrm{c}}^{0}}. \tag{49}$$

To analyze the geometric evolution of the coiled rope upon thermal actuation, a local curvilinear coordinate system $\{r_k, \alpha_{\mathrm{r}}, s\}$ is attached to the rope's centerline, where $r_k \in [0, r_{\mathrm{r}}]$ is the local radial coordinate within the rope's cross-section, $\alpha_{\mathrm{r}} \in [0, 2\pi]$ is winding angle of the helical coil, and $s$ is the arc length. The metric tensor $G_{ij}$ of the rope in the reference configuration reflects the non-Euclidean geometry induced by the initial curvature (Liu et al. 2024):

$$[G_{ij}]=\begin{bmatrix} G_{r_k r_k} & G_{r_k \alpha_{\mathrm{r}}} & G_{r_k s} \\ G_{\alpha_{\mathrm{r}} r_k} & G_{\alpha_{\mathrm{r}}\alpha_{\mathrm{r}}} & G_{\alpha_{\mathrm{r}} s} \\ G_{s r_k} & G_{s\alpha_{\mathrm{r}}} & G_{ss} \end{bmatrix}=\begin{bmatrix} 1 & 0 & 0 \\ 0 & r_k^{2} & 0 \\ 0 & 0 & \left(1+\kappa_{\mathrm{c}}^{0} r_k \cos\alpha_{\mathrm{r}}\right)^{2} \end{bmatrix} \tag{50}$$

The term $G_{ij} = 1 + \kappa_{\mathrm{r}}{}^{0} r_k \cos\alpha_{\mathrm{r}}$ represents the ratio between the arc length at any point $(r_k, \alpha_{\mathrm{r}})$ on the rope cross-section and the arc length along the centerline, which features the effect of initial curvature, introducing a geometric non-linearity that distinguishes the present model from the straight-bundle theories used in Section 3.3.

Upon thermo-mechanical actuation, the TCP structure deforms from the reference configuration to the current configuration defined by $\{\bar{r}_k, \bar{\alpha}_{\mathrm{r}}, \bar{s}\}$. Assuming the deformation is uniform along the helical axis, the mapping from the reference to the current configuration is described by the displacement field within the rope as,

$$\bar{r}_k = r_k + u_{\mathrm{radi}}(r_k, \alpha_{\mathrm{r}}), \quad \bar{\alpha}_{\mathrm{r}} = \alpha_{\mathrm{r}} + \phi_{\mathrm{b}} s, \quad \bar{s} = (1+\varepsilon_{\mathrm{r}})s, \tag{51}$$

where $u_{\mathrm{radi}}$ denotes the radial displacement field incorporating thermal expansion and Poisson effects, $\phi_{\mathrm{b}}$ represents the twisting angle per unit length, and $\varepsilon_{\mathrm{r}}$ is the axial strain of the rope's centerline induced by tension.

On the basis of the finite strain theory for curved fibers, the longitudinal Green-Lagrange strain $\varepsilon_{\mathrm{s}}$ at an arbitrary point $(r_k, \alpha_{\mathrm{c}})$ on the rope's cross-section is derived as:

$$\varepsilon_{\mathrm{s}}(r_k, \alpha_{\mathrm{r}}) = \varepsilon_{\mathrm{r}} + \frac{\Delta\kappa_{\mathrm{r}} r_k \cos\alpha_{\mathrm{r}} + u_{\mathrm{radi}}\kappa_{\mathrm{r}}^0 \cos\alpha_{\mathrm{r}}}{1+\kappa_{\mathrm{r}}^0 r_k \cos\alpha_{\mathrm{r}}}, \tag{52}$$

where $\Delta\kappa_{\mathrm{r}} = \kappa_{\mathrm{r}} - {\kappa_{\mathrm{r}}}^0$ is the variation in curvature. Eq. (44) reveals the actuation mechanism of the coiled structure: the term $u_{\mathrm{radi}}{\kappa_{\mathrm{r}}}^0 \cos\alpha_{\mathrm{c}}$ indicates that a thermally induced radial expansion $u_{\mathrm{radi}}$ directly produces a bending strain through the initial curvature ${\kappa_{\mathrm{r}}}^0$, even in the absence of external loading.

Similarly, the shear strain $\gamma_{\alpha_{\mathrm{r}} s}$ induced by the torsional deformation of the rope is expressed as:

$$\gamma_{\alpha_{\mathrm{r}}\mathrm{s}}(r_k, \alpha_{\mathrm{r}}) = \frac{\Delta\tau_{\mathrm{r}} r_k}{1+\kappa_{\mathrm{r}}^0 r_k \cos\alpha_{\mathrm{r}}}. \tag{53}$$

where $\Delta\tau_{\mathrm{r}} = \tau_{\mathrm{r}} - {\tau_{\mathrm{r}}}^0$ is the twist rate derived from the evolution of the coil geometry. The detailed derivation of Eqs. (50-52) is provided in Appendix B.

3.6.2 Effective constitutive relations

Owing to the rope being treated as a homogenized, transversely isotropic continuum, the effective mechanical properties ($E_{\mathrm{r}}^{\mathrm{eff}}$ and $G_{\mathrm{r}}^{\mathrm{eff}}$) can be derived from the tertiary structure analysis in Section 3.5. The constitutive relation for the rope, incorporating the anisotropic thermal expansion, is expressed as:

$$\boldsymbol{\sigma} = \mathbf{C}_{\mathrm{rope}}(\boldsymbol{\varepsilon} - \boldsymbol{\alpha}_{\mathrm{rope}}\Delta T). \tag{54}$$

where $\boldsymbol{\sigma}$ is the stress tensor, $\mathbf{C}_{\mathrm{rope}}$ is the effective stiffness matrix of the rope (The detailed step module is shown in Appendix C), and $\boldsymbol{\alpha}_{\mathrm{rope}}$ is the effective coefficient of thermal expansion (CTE) vector. $\Delta T$ is the temperature increment triggering the actuation.

3.6.3 Equilibrium governing equations of TCP muscle

The internal forces acting on the coiled rope's cross-section can be derived by integrating the constitutive stresses over the cross-sectional area $A_{\mathrm{r}} = \pi r_{\mathrm{r}}^2$:

$$\begin{aligned} F_{\mathrm{T}}^{\text{r-c}} &= \int_{A_{\mathrm{r}}} E_{\mathrm{r}}^{\mathrm{eff}}(\varepsilon_{\mathrm{s}} - \alpha_{\mathrm{r}}\Delta T)dS, \\ M_{\mathrm{B}}^{\text{r-c}} &= \int_{A_{\mathrm{r}}} E_{\mathrm{r}}^{\mathrm{eff}}(\varepsilon_{\mathrm{s}} - \alpha_{\mathrm{r}}\Delta T)r_k \cos\alpha_{\mathrm{r}}\, dS, \\ M_{\mathrm{T}}^{\text{r-c}} &= \int_{A_{\mathrm{r}}} G_{\mathrm{r}}^{\mathrm{eff}}\gamma_{\alpha_{\mathrm{r}}\mathrm{s}} r_k dS. \end{aligned} \tag{55}$$

The term $M_{\mathrm{B}}^{\mathrm{r-c}}$ induced by the axial strain $\varepsilon_{\mathrm{r}}$ and initial curvature $\kappa_{\mathrm{r}}^{0}$ represents the extension-torsion-bending coupling effect of the coiled geometry.

During the actuation of artificial muscles, the TCP is subjected to the external axial load $F_{\mathrm{c}}$ and torque $M_{\mathrm{c}}$. The equilibrium of the coiled structure requires that these external loads balance the internal resultants of the rope. Based on the helical mechanics in the Frenet frame, one can obtain that,

$$\begin{aligned} F_{\mathrm{T}}^{\text{r-c}} &= F_{\mathrm{c}} \sin\theta_{\mathrm{c}}, \\ M_{\mathrm{B}}^{\text{r-c}} &= M_{\mathrm{c}} \cos\theta_{\mathrm{c}} - F_{\mathrm{c}} R_{\mathrm{c}} \sin\theta_{\mathrm{c}}, \\ M_{\mathrm{T}}^{\text{r-c}} &= M_{\mathrm{c}} \sin\theta_{\mathrm{c}} + F_{\mathrm{c}} R_{\mathrm{c}} \cos\theta_{\mathrm{c}}. \end{aligned} \tag{56}$$

Meanwhile, the kinematic compatibility conditions impose strict constraints relating the local deformation variables $\{\varepsilon_{\mathrm{r}}, \tau_{\mathrm{r}}, \kappa_r\}$ to the evolution of the TCP geometry $\{R_{\mathrm{c}}, \theta_{\mathrm{c}}\}$. Specifically, the variation of curvature $\Delta\kappa_{\mathrm{r}}$ and torsion $\Delta\tau_{\mathrm{r}}$ are determined by the helical geometry variations as,

$$\Delta\kappa_{\mathrm{r}} = \frac{\cos^2\theta_{\mathrm{c}}}{R_{\mathrm{c}}} - \frac{\cos^2\theta_{\mathrm{c}}^0}{R_{\mathrm{c}}^0}, \quad \Delta\tau_{\mathrm{r}} = \frac{\sin\theta_{\mathrm{c}}\cos\theta_{\mathrm{c}}}{R_{\mathrm{c}}} - \frac{\sin\theta_{\mathrm{c}}^0\cos\theta_{\mathrm{c}}^0}{R_{\mathrm{c}}^0}. \tag{57}$$

Then, the actuation stroke of the TCP muscle can be defined as,

$$\delta = L_{\mathrm{c}} - L_{\mathrm{c}}^0 = \frac{L_{\mathrm{c}}^0}{\sin\theta_{\mathrm{c}}^0}\sin\theta_{\mathrm{c}} - L_{\mathrm{c}}^0. \tag{58}$$

Consequently, the equilibrium configuration of the TCP for a prescribed thermal load $\Delta T$ is determined by solving the system of Eqs. (56)–(58), thereby establishing a quantitative basis for predicting both actuation stroke and force generation.

## 4. Finite element modeling of a hierarchical helical structure and TCP artificial muscle

### 4.1 Hierarchical Helical Structure

To validate the proposed hierarchical theoretical model, three-dimensional finite element simulations of hierarchical helical structure were performed in Abaqus/Explicit (as shown in Fig. 9). To minimize boundary effects, the model spanned two helical pitches in length. Geometrically, centerlines of individual helices were generated from parametric equations, after which circular cross-sections were swept and circumferentially arrayed to build the solid hierarchical structure.

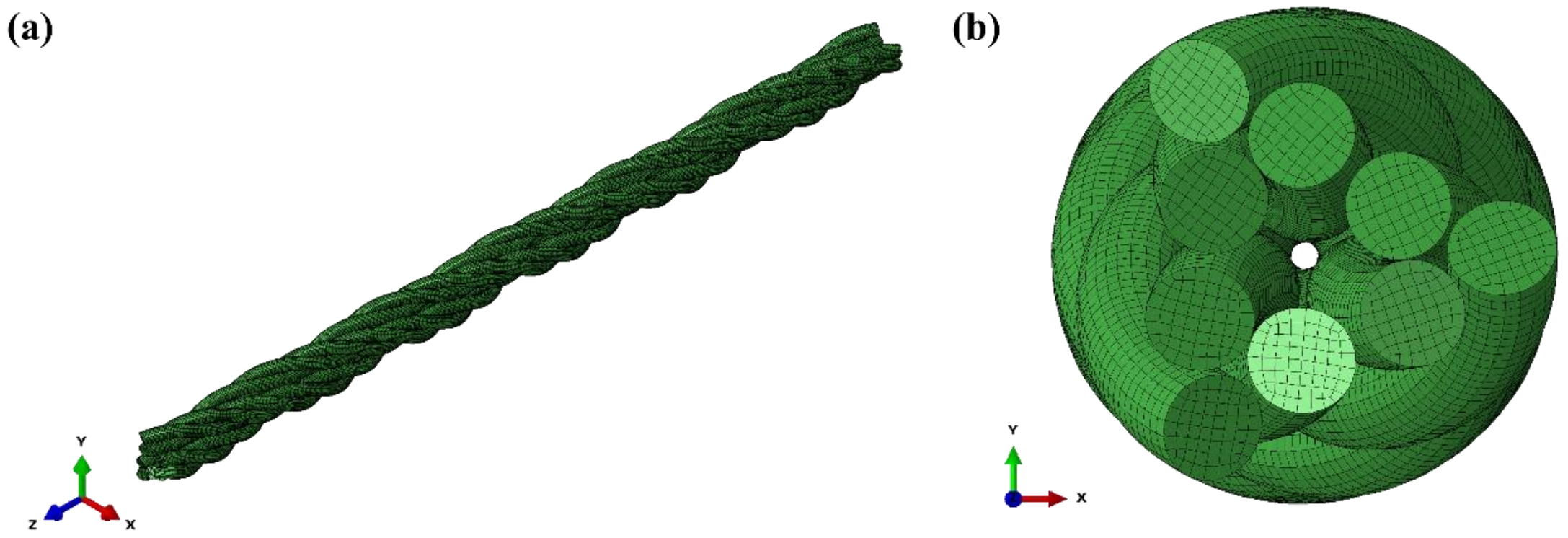

Fig. 9 Finite element mesh of a hierarchical helical structure

Following geometry generation, the domain was discretized using C3D8R eight-node linear reduced-integration solid elements. A mesh-convergence study was conducted to optimize the balance between computational accuracy and cost, identifying a global seed size of 0.045 mm as optimal. Consequently, for a representative 3×3 structure with initial filament and ply helix angles of 15° and a filament radius of $r_{\mathrm{f}} = 0.15$ mm, the final model comprised 242820 elements. This density ensured that deviations relative to finer meshes remained below 5%, satisfying the study's accuracy criteria.

To accurately demonstrate the mechanical anisotropy, the nylon fibers were modeled using a transversely isotropic constitutive law (parameters detailed in Table 1). A critical challenge presented by the complex helical geometry was the continuous variation of principal material directions relative to the global coordinate frame. To address this, a discrete coordinate-field method was adopted (Fig. 10). In this approach, elements within each helical layer were treated as independent sets. By defining the material 1-axis along the vector connecting the centroids of the upper and lower nodes of each element, a unique local coordinate system was assigned. This ensured perfect alignment between the principal axes and the actual helical trajectory.

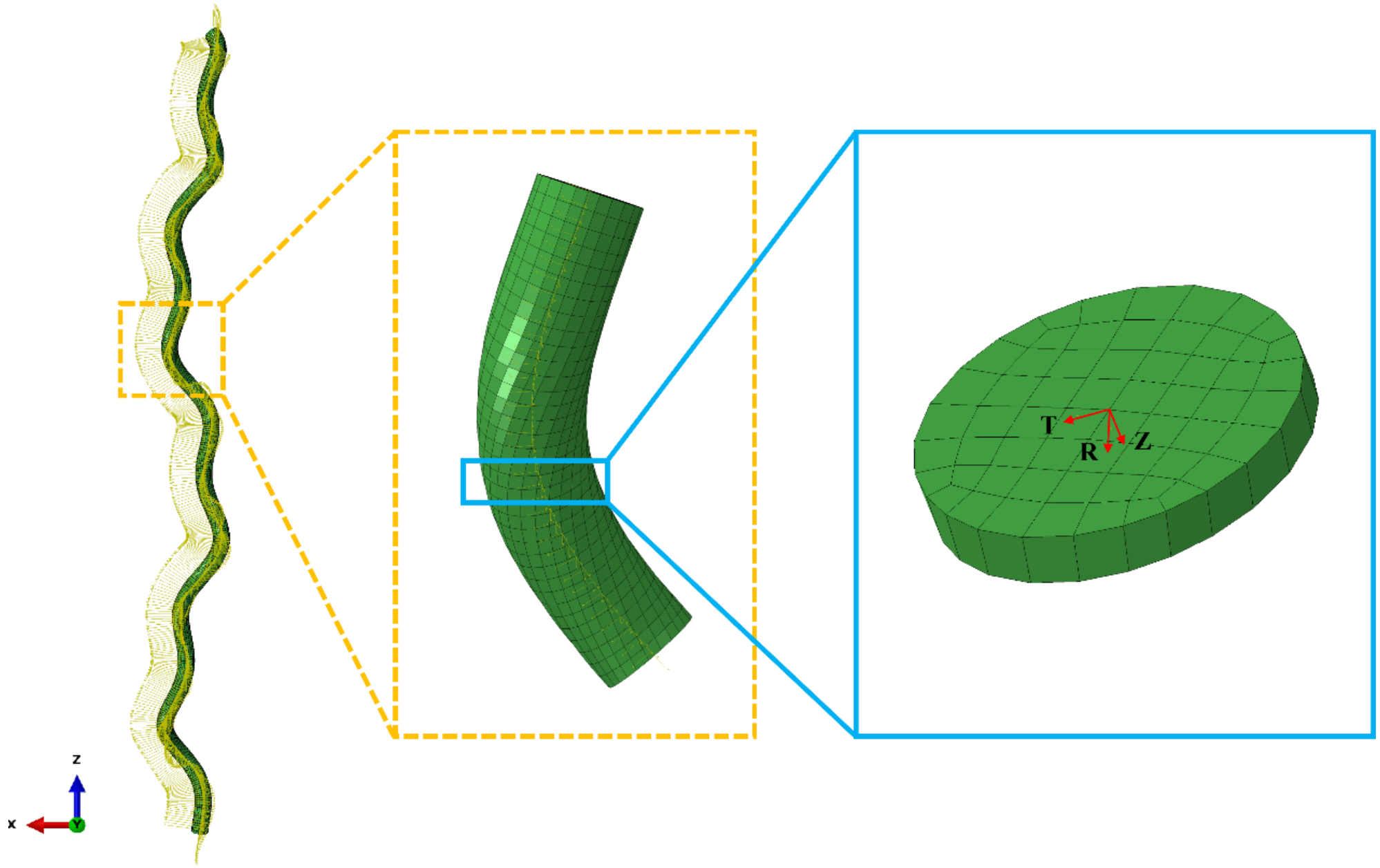

Fig. 10 Local material coordinate system of helical nylon fiber

Boundary conditions and loading were applied via reference points (RPs)

created at both ends of the structure, where all nodes on each end cross-section were kinematically coupled to their corresponding RP. The upper RP was fully constrained ($U1 = U2 = U3 = 0, UR1 = UR2 = UR3 = 0$). At the lower RP, an axial displacement corresponding to 5% engineering strain was prescribed along the fiber axis, while the remaining five degrees of freedom were fixed. The axial reaction force and reaction moment at the lower RP were recorded continuously to analyze the axial deformation and torsion-extension coupling behavior.

Table. 1 The structural and material parameters of artificial muscle

| Parameters | Values | Data sources |
|---|---|---|
| Axial Young's modulus $E_1$ | 2.351 GPa | Measurement |
| Transverse Young's modulus $E_2$=$E_3$ | 1.893 GPa | Measurement |
| Out-of-plane shear modulus $G_{23}$ | 0.760 GPa | Measurement |
| In-plane shear modulus $G_{12}$=$G_{13}$ | 0.653 GPa | Measurement |
| Transverse poisson's ratio $v_{23}$ | 0.450 | Choy and Leung (1985) |
| Axial poisson's ratio $v_{12}$=$v_{13}$ | 0.374 | Measurement |
| Filament radius $r_\mathrm{f}^0$ | 0.15 mm, 0.085 mm | Measurement |
| Helical angle $\theta_\mathrm{f}^0$ | $9.86°, 15.15°, 20.14°$ | Measurement |
| Helical angle $\theta_\mathrm{b}^0$ | $10.23°, 14.88°, 20.30°$ | Measurement |
| Axial CTE $\alpha_1$ | $\alpha_1 = (-5.4T - 101) \times 10^{-6}$ | Hu at al. (2025) |
| Radial CTE $\alpha_2$ | $\alpha_2 = 2.08 \times 10^{-4}$ | Hu at al. (2025) |

## 4.2 Coiled Structure

Complementing the hierarchical analysis, finite element simulations were extended to the single-fiber coiled structure to verify theoretical predictions regarding thermo-mechanical coupling. To minimize end effects and characterize the behavior of central coils, a multi-cycle segment was modeled in Abaqus. The geometry was defined by sweeping a circular cross-section along a centerline determined by the specific helix radius and pitch. Consistent with the previous model, the mesh utilized C3D8R elements (Fig. 11). For a representative case—featuring a fiber radius of 0.38 mm, coil radius of 0.7 mm, bias angle of 45°, and pitch angle of 20°—the model consisted of 53760 elements.

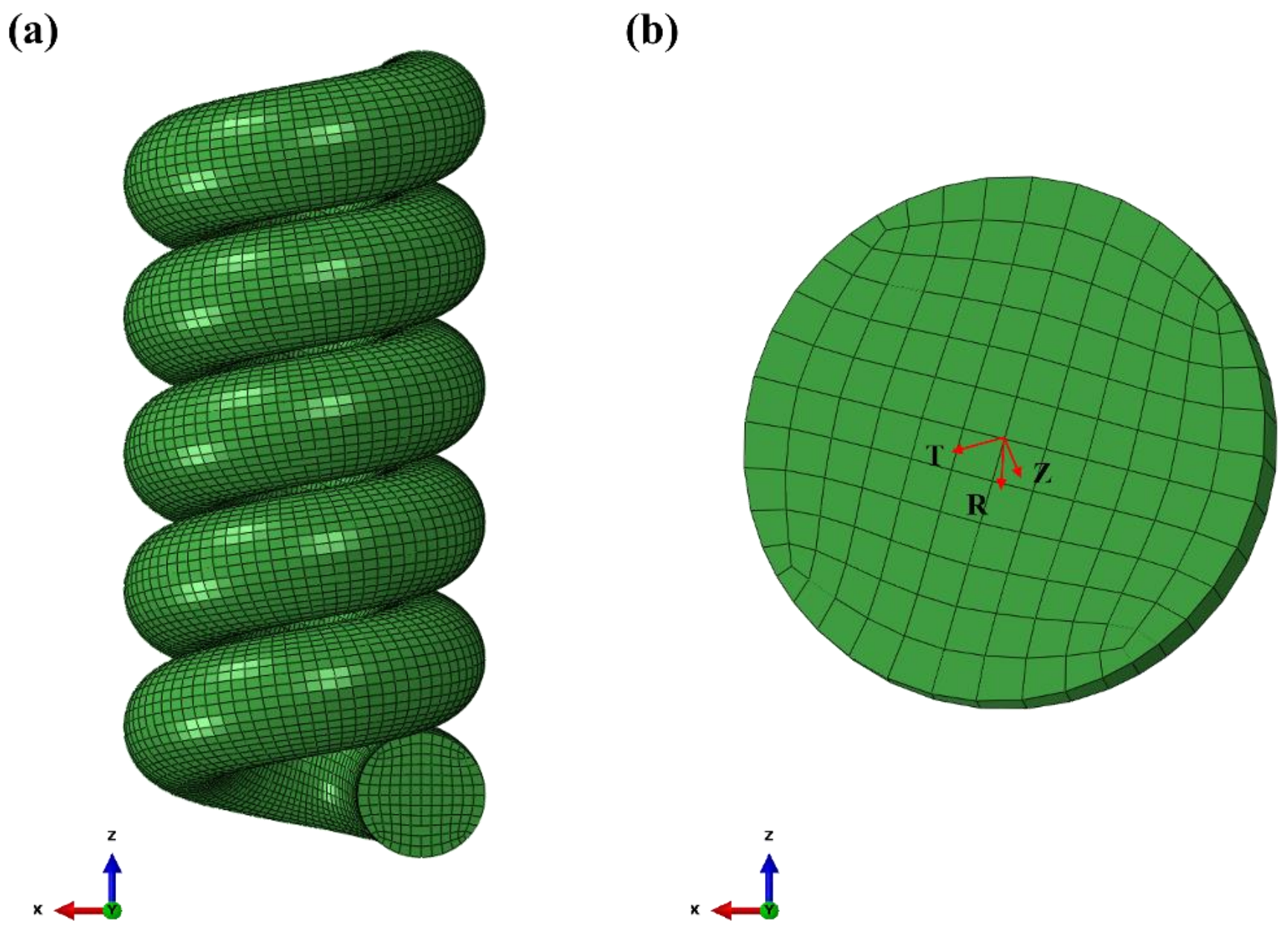

Fig. 11 Finite element mesh of twisted and coiled structure

Material properties were again defined as transversely isotropic, with the addition of a temperature-dependent coefficient of thermal expansion to simulate thermal actuation. Given that the curvature of the coiled structure creates similar orientation challenges to the hierarchical model, the same discrete coordinate-field method was employed. This ensured that the material orientation remained tangent to the coiled centerline throughout the deformation.

The boundary conditions were adapted to simulate actuation under load. RPs were established at both ends; the upper RP was fully fixed, while the lower RP was constrained in transverse and rotational degrees of freedom, permitting only axial translation. The simulation proceeded in two steps: a constant axial load (mimicking a suspended weight) was applied at room temperature until equilibrium, and the temperature was elevated while maintaining the load to induce thermal contraction. The resulting axial displacement and reaction force were recorded for comparison with theoretical predictions.

## 5. Results

In this section, actuation-stroke experiments were first performed on TCP muscles with various structures, including helical and coiled configurations, thereby demonstrating the superior load-bearing capacity of the coiled artificial muscle with the hierarchical structure. Subsequently, the theoretical model for TCP artificial muscles was validated by comparing experimental data and finite element simulations. Finally, within a top-down theoretical framework, the effects of geometric parameters and microstructure evolution on the mechanical properties of the hierarchical helical TCP artificial muscles were systematically investigated. The TCP artificial muscles

5.1 Actuation-stroke experiment

To elucidate the role of hierarchical topology in enhancing axial load-bearing capacity while preserving actuation stroke, the thermomechanical response of TCP muscle was characterized by actuation experiments under cyclic thermal loading. Specimens were fabricated from nylon 6,6 monofilament precursors ($d = 0.3$ mm) following specific torsional insertion and coiling protocols detailed in Section 2.2. Each sample was suspended vertically in a custom test rig equipped with a load cell (resolution: 0.01 N) at the lower end and a fixed clamp at the upper end. Axial loads of 10 g and 100 g were applied to simulate low- and high-stress regimes, respectively. Thermal actuation was induced by hanging the sample from a hook at the upper end and placing it inside a hollow ceramic tube, with weights suspended at the lower end; the artificial muscle in the middle was heated by heating the ceramic tube, with temperatures cycled between 35°C and 120°C at a rate of 2°C/min, corresponding to the glass transition and crystallization domains of nylon 6,6. It is important to note that the artificial muscle should be positioned as close as possible to the ceramic tube without touching it, to avoid excessively high local temperatures that could cause melting. Strain was measured as the relative length change $\varepsilon = \Delta L/L_0$, where $L_0$ is the initial length, using a non-contact laser extensometer (accuracy: 0.1%). Temperature profiles were monitored via embedded thermocouples. Data acquisition occurred at 10 Hz over 1000 s per cycle, with three replicates per configuration to ensure reproducibility. The equilibrium shapes were imaged post-fabrication using optical microscopy to quantify geometric parameters such as helix radius and pitch.

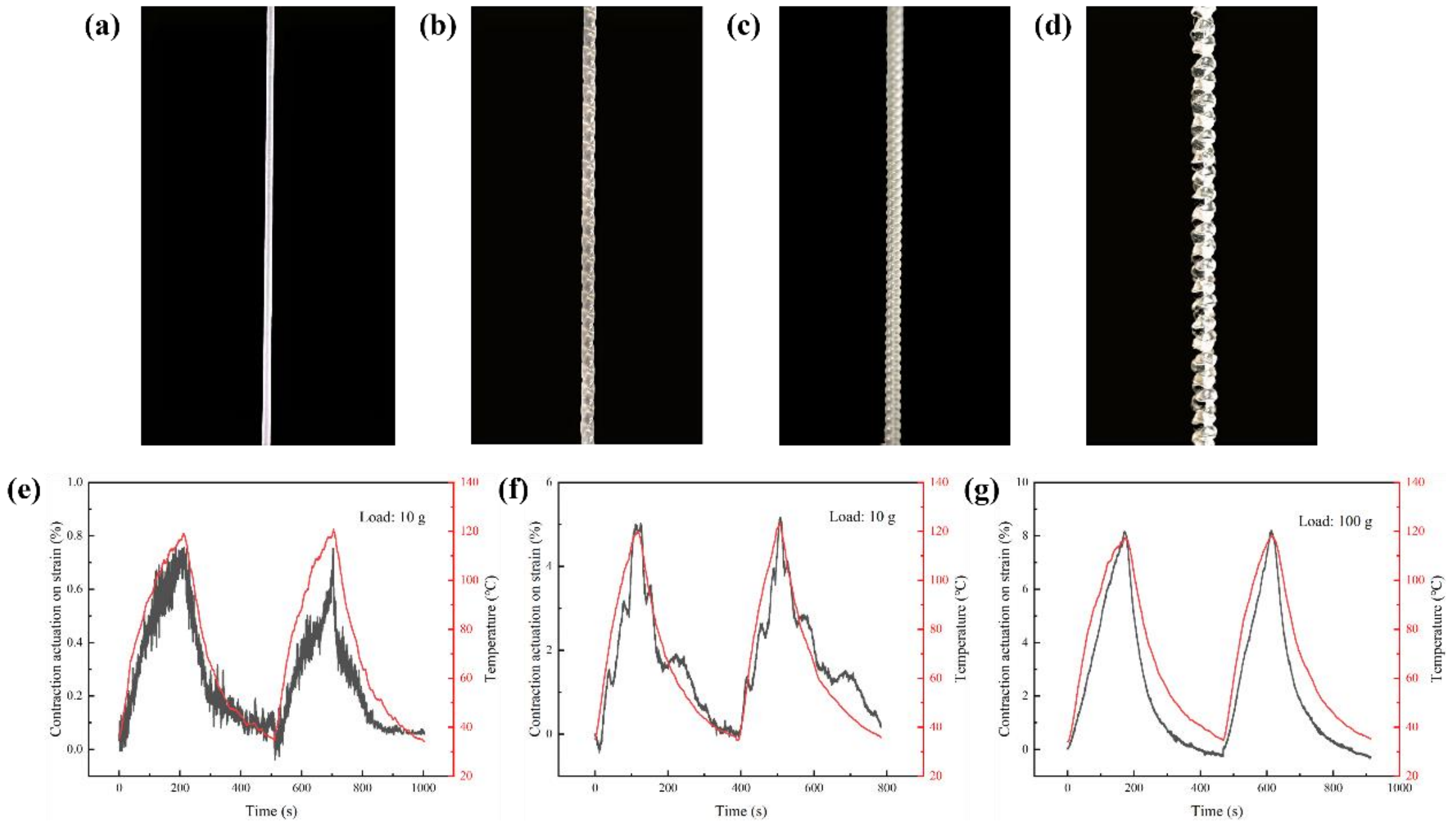

Fig. 12 Morphological evolution and thermomechanical actuation performance of TCP actuators with varying structural hierarchies.. Upper panels illustrate optical images of: (a) the as-received single monofilament precursor; (b) a 2-ply helical structure TCP structure formed by twisting two precursors without coiling; (c) a single-ply coiled TCP structure formed by over-twisting induced coiling; (d) a hierarchical 2-ply coiled TCP structure exhibiting a nested, dual-level helical structure. Lower panels show temporal evolution of contraction actuation strain (black curves, left axis) and surface temperature (red curves, right axis) under cyclic thermal actuation: (e) the 2-ply helical TCP (Load: 10 g) exhibiting low-amplitude actuation; (f) the single-ply coiled TCP (Load: 10 g) showing moderate actuation strain; (g) the hierarchical 2-ply coiled TCP (Load: 100 g), demonstrating high-amplitude actuation strain under a significantly elevated mechanical load. Peak strains approximate 0.8%, 5%, and 8% respectively, with temperature cycles ranging 35–120°C.

To isolate the effects of structural geometry on performance, four distinct topological configurations were prepared: (i) a transversely isotropic nylon precursor prior to twist insertion (Fig. 12a), serving as a pristine monofilament; (ii) a 2-ply helical structure (Fig. 12b) is formed by introducing torsional strain into two parallel precursors; (iii) a single-ply coiled TCP actuator (Fig. 12c), where the insertion of twist beyond the material's buckling threshold induces a transition from a twisted fiber to a helical coil; and (iv) the proposed hierarchical 2-ply coiled TCP actuator (Fig. 12d), characterized by a nested helical structure designed to maximize structural stiffness and actuation work density. The actuation performance of these topologies was characterized by measuring the isotonic contraction strain under distinct loading conditions. In all cases, the actuation strain tracks the thermal profile (red curves) with high fidelity, confirming the mechanism is governed by thermally induced anisotropic expansion and untwisting.

The 2-ply helical TCP (Fig. 12(e)) exhibits a marginal contraction strain of approximately 0.8% under a 10 g load. The absence of a coiled spring-like geometry limits the geometric amplification factor, restricting the actuator's stroke. Conversely, the single-ply coiled TCP (Fig. 12(f)) exhibits a significant performance enhancement, achieving a peak actuation strain of approximately 5% under the same 10 g load, which confirms that the coiled topology is critical for maximizing stroke in TCP actuators. However, the hierarchical 2-ply coiled TCP (Fig. 12(g)) presents superior electromechanical performance. Despite supporting a load an order of magnitude higher (100 g vs. 10 g), the actuator achieves a peak contraction strain of approximately 8%. The dramatic increase in both load-bearing capacity and actuation strain in the hierarchical structure can be attributed to the synergistic effect of the 2-ply and coiled structures. The plying process suppresses torsional instability, allowing the actuator to maintain its coiled geometry under high tension (100 g). Meanwhile, the coiled structure provides the necessary geometric compliance to amplify the thermal expansion of the polymer chains into a macroscopic contraction. Therefore, the coiled TCP with a hierarchical helical structure represents an optimal design for applications requiring high work density and robust stroke capabilities.

5.2 Fundamental mechanical properties and model validation

Owing to the large actuation strains and high load-bearing capacity exhibited by TCP artificial muscles with hierarchical helical structures, a systematic investigation of the effects of geometric parameters and microstructure evolution on their mechanical behavior is necessary. Before undertaking this parametric study, the accuracy and predictive capability of the proposed theoretical model for hierarchical helical structures were validated by a comprehensive comparison against experimental data and FEM simulations.

Figure 13(a) and Figure 14(a) illustrates the mechanical response of a $3 \times 3$ hierarchical helical structure (comprising 3 bundles, each containing 3 filaments) with varying initial helical angles ( $\theta_{\mathrm{b}}^{0} =$ $70°, 75°, 80°$). The theoretical predictions for tensile force $F_T^r$ and torque $M_T^r$ as functions of axial strain $\varepsilon_r$ are plotted alongside experimental measurements and FEM results. The comparison reveals an excellent quantitative agreement among the theoretical model, experiments, and numerical simulations across the entire strain range ($0 \le \varepsilon_r \le 0.02$). As anticipated from the mechanics of helical springs, the structural stiffness exhibits a strong dependence on the initial helical angle. A decrease in $\theta_b^0$ from $80°$ to $70°$ results in a significant increase in the slopes of the tensile force-strain and torque-strain curves, indicating enhanced axial and torsional stiffness.

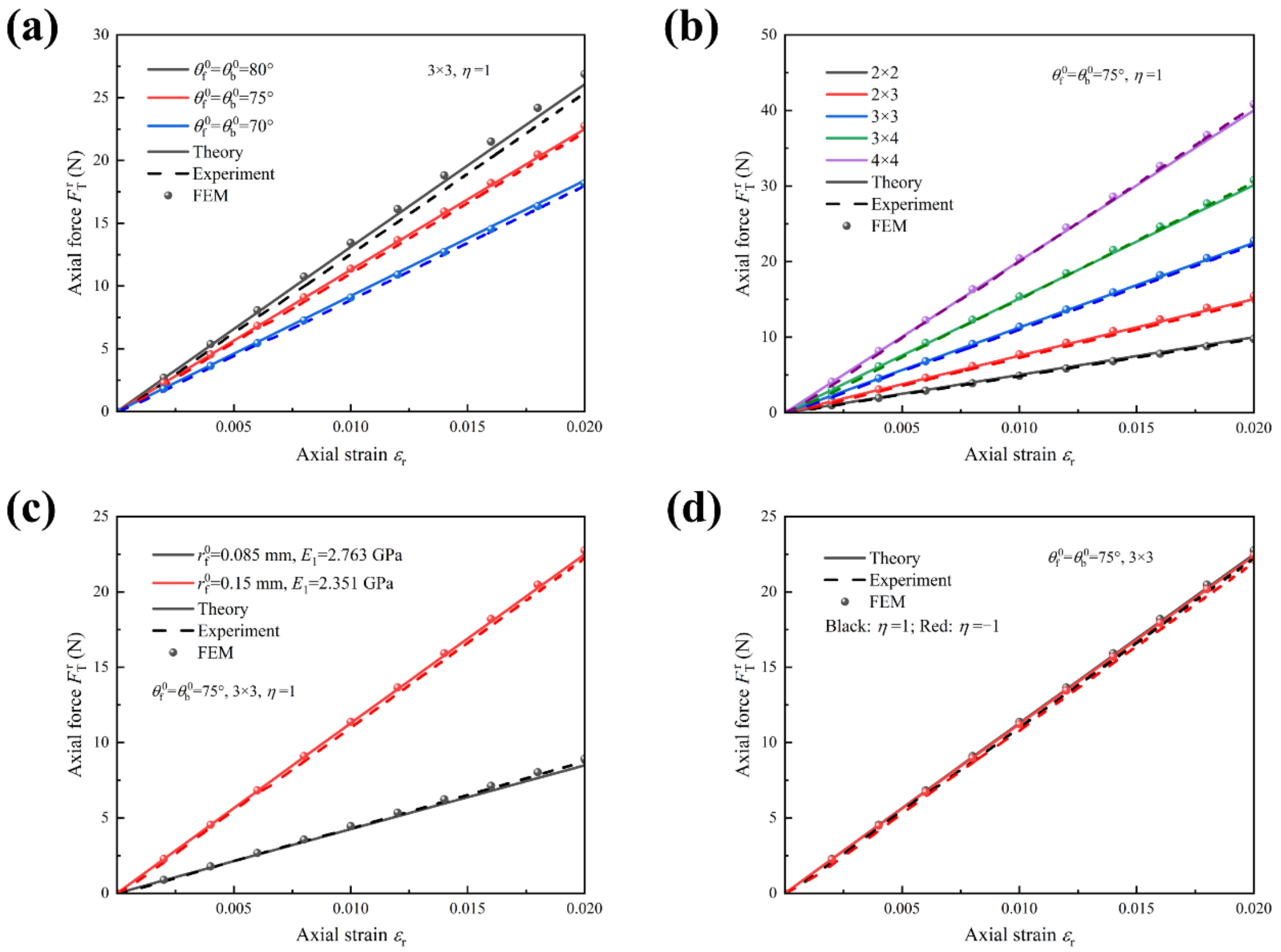

Fig. 13 Comparisons of the axial force–axial strain ($F_T^r - \varepsilon_r$) responses obtained from the theoretical model (solid lines), experimental measurements (dashed lines), and FEM simulations (scatter points). (a) The effect of initial helical angles ($\theta_f^0 = \theta_b^0$) with a $3 \times 3$ structure and $\eta = 1$. (b) The effect of structure sizes (ranging from $2 \times 2$ to $4 \times 4$) at $\theta_{\mathrm{f}}^{0} = \theta_{\mathrm{b}}^{0} = 75°$. (c) The effect of filament radius $r_{\mathrm{f}}^{0}$ and Young's modulus. (d) The effect of the parameter η (comparing $\eta = 1$ and $\eta = -1$) for a $3 \times 3$ structure at $\theta_{\mathrm{f}}^{0} = \theta_{\mathrm{b}}^{0} = 75°$.

The universality of the theoretical framework is further verified in Fig. 13(b) and Fig. 14(b) by examining various hierarchical topologies, ranging from simple $2 \times 2$ arrays to more complex $4 \times 4$ configurations. The tensile force and torque curves exhibit a clear hierarchical scaling law: as the number of filaments and bundles increases (from $2 \times 2$ to $4 \times 4$), the global stiffness of the artificial muscle rises monotonically. This can be attributed to the parallel contribution of a larger number of load-bearing elements. The theoretical model successfully predicts this normalization effect, showing high fidelity to the experimental and FEM results across all tested hierarchical levels.

Figure 13(c) and Figure 14(c) quantify the impact of the constituent filament radius on the global and local mechanical performance. Comparative analyses are performed between $3 \times 3$ structures comprising filaments with radii $r_{\mathrm{f}}^{0} = 0.085\ \mathrm{mm}$ (black) and $r_{\mathrm{f}}^{0} = 0.15\ \mathrm{mm}$ (red). The theoretical model faithfully captures the size-dependent stiffening effects observed in both experiments and simulations. An increase in filament radius leads to a marked enhancement in the tensile force and induced torque at a given axial strain, which is inherently dictated by the increased cross-sectional area and moment of inertia. The excellent agreement between analytical predictions and the discrete data from FEM and experiments evaluates and confirms the model's capability to account for geometric variations in the filament cross-section.

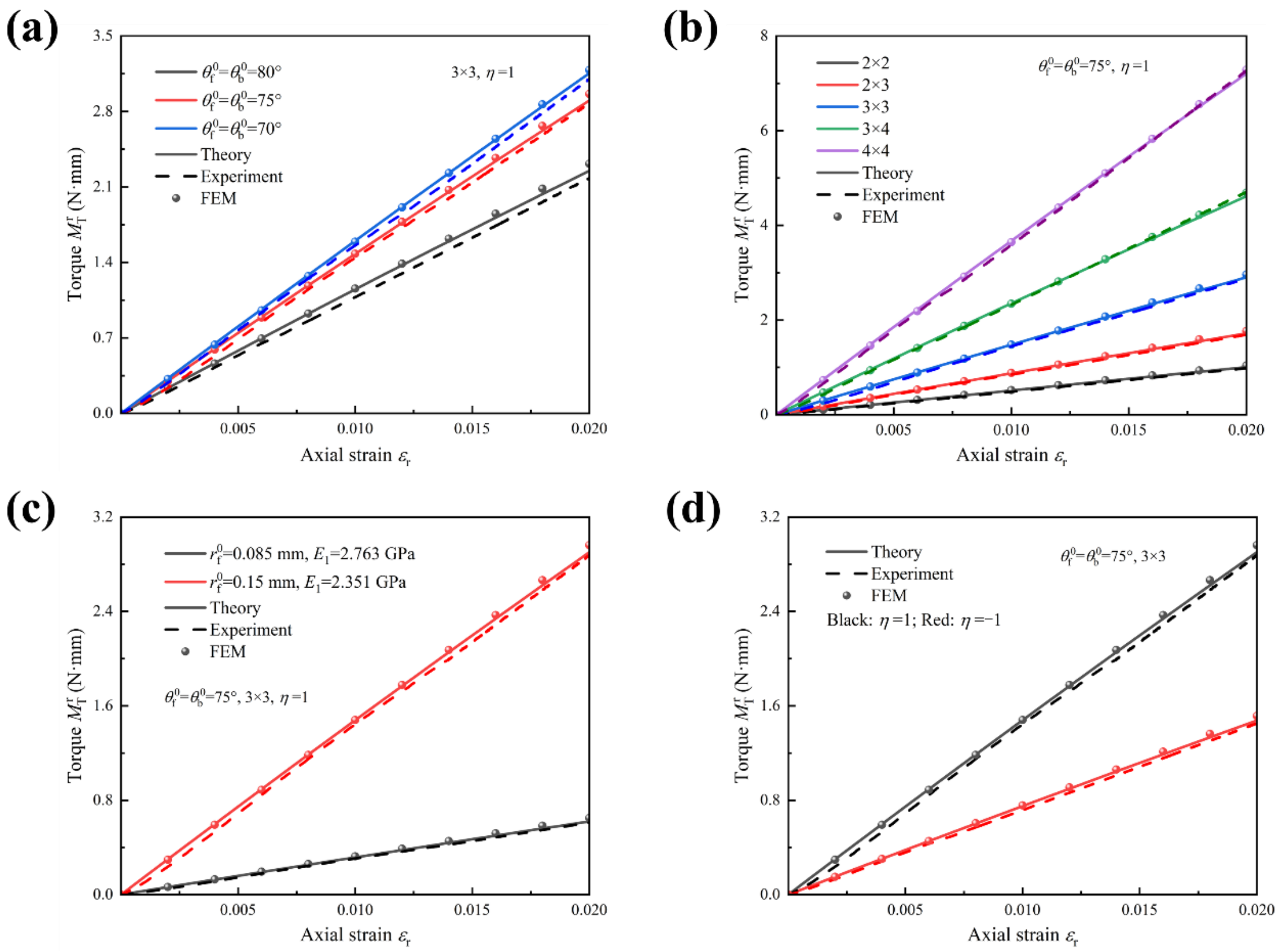

Fig. 14 Comparisons of the torque–axial strain ($M_{\mathrm{T}}^{\mathrm{r}} - \varepsilon_{\mathrm{r}}$) responses obtained from the theoretical model (solid lines), experimental measurements (dashed lines), and FEM simulations (scatter points). (a) The effect of initial helical angles ($\theta_{\mathrm{f}}^{0} = \theta_{\mathrm{b}}^{0}$) with a $3 \times 3$ structure and $\eta = 1$. (b) The effect of structure sizes (ranging from $2 \times 2$ to $4 \times 4$) at $\theta_{\mathrm{f}}^{0} = \theta_{\mathrm{b}}^{0} = 75°$. (c) The effect of filament radius $r_{\mathrm{f}}^{0}$ and Young's modulus $E_1$. (d) The effect of the parameter $\eta$ (comparing $\eta = 1$ and $\eta = -1$) for a $3 \times 3$ structure at $\theta_{\mathrm{f}}^{0} = \theta_{\mathrm{b}}^{0} = 75°$.

The influence of structural chirality on the mechanical behavior of the $3 \times 3$ hierarchical structure is presented in Fig. 13(d) and Fig. 14(d). Two configurations with distinct chirality indices ($\eta = 1$ and $\eta = -1$) are analyzed, while keeping other geometric parameters constant ($\theta_{\mathrm{f}}^{0} = \theta_{\mathrm{b}}^{0} = 75°, r = 0.15$ mm). The results demonstrate a distinct decoupling between the tensile and torsional responses regarding chirality. As shown in the tensile force-strain diagrams, the curves for $\eta = 1$ and $\eta = -1$ virtually overlap, suggesting that the axial stiffness distribution is insensitive to the chirality of the hierarchical assembly. In contrast, the torque-strain response exhibits a marked dependence on chirality; the magnitude of the induced torque varies significantly between the two chiral configurations. This behavior corroborates the theoretical formulation, confirming that while axial extension is governed by the scalar magnitudes of the helical geometry, the torsional coupling is intrinsically linked to the vector nature of the helical handedness. The theoretical curves align perfectly with the FEM and experimental data points for all loading cases.

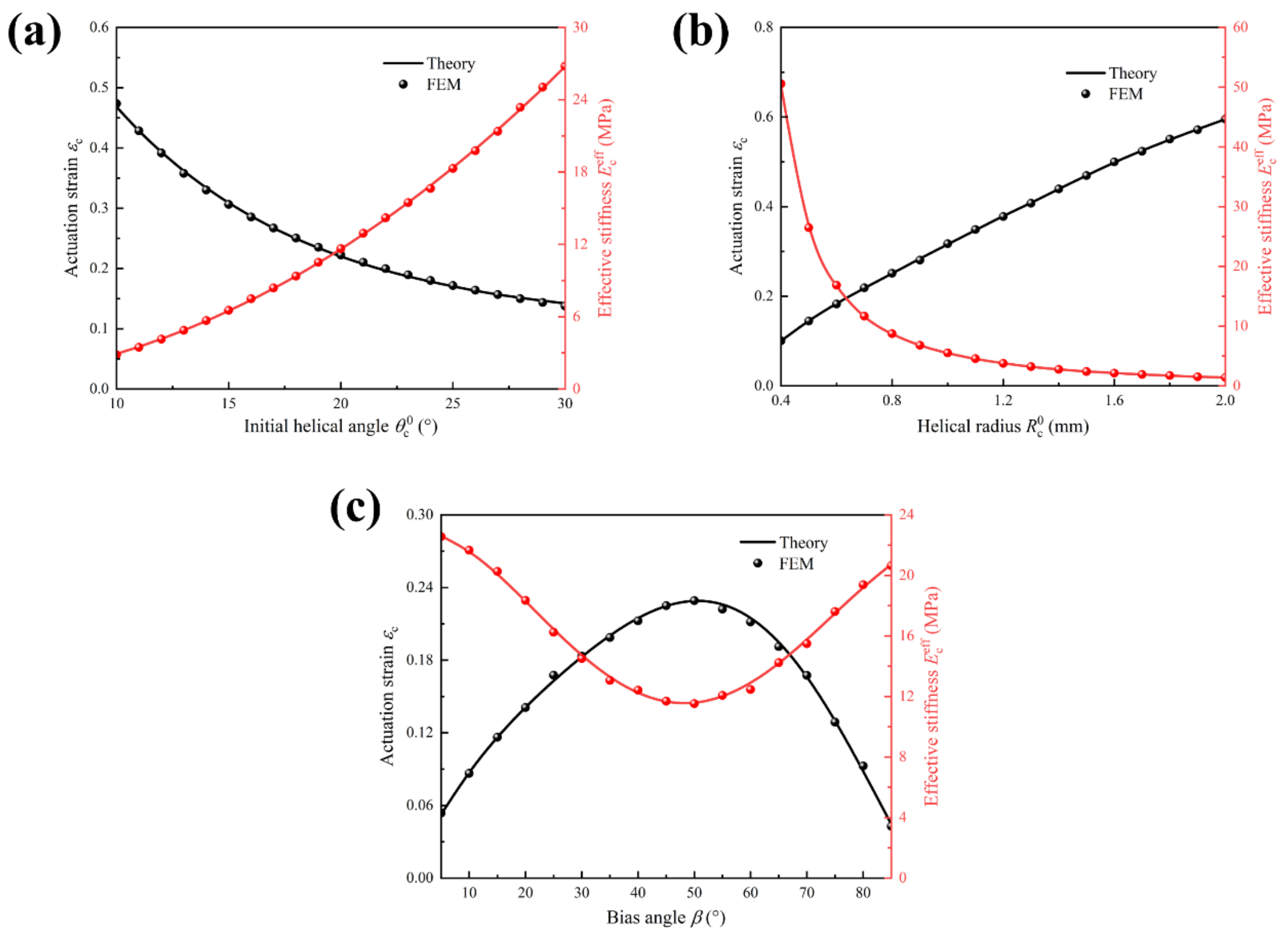

Fig. 15 Influence of geometric parameters on the actuation strain $\varepsilon_c$ (left axis, black) and effective stiffness $E_c^{eff}$ (right axis, red). The solid lines and scatter points represent the theoretical predictions and FEM results, respectively. (a) The effect of the initial helical angle $\theta_c^0$. (b) The effect of the helical radius $R_c^0$. (c) The effect of the bias angle $\varphi$.

Following the validation of the hierarchical helical structure, the predictive fidelity of the thermo-mechanical model for the fully coiled TCP muscle (the quaternary structure) was evaluated. Figure 15 presents the dependence of the macroscopic actuation strain and structural stiffness on three critical geometric parameters of TCP: the initial helical angle ($\theta_c^0$), the helical radius ($R_c^0$), and the fiber bias angle $\varphi$. As depicted in Fig. 15(a), the actuation performance exhibits a strong sensitivity to the initial winding geometry of the coiled structure. An increase in the helical angle $\theta_c^0$ results in a monotonic rise in the effective axial stiffness (red curve), which is physically consistent with the mechanics of helical springs where a larger pitch angle enhances load-bearing resistance. However, the stiffening effect leads to a corresponding decline in the thermally induced actuation strain (black curve), as the geometric amplification factor is reduced at higher pitch angles. Conversely, the influence of the radius $R_c^0$, illustrated in Fig. 15(b), displays an inverse correlation. Larger helical radii significantly reduce the structural stiffness, thereby providing greater geometric compliance that facilitates larger actuation strokes under identical thermal loading conditions. Furthermore, Fig. 15(c) elucidates the role of the microstructural bias angle in determining the actuation efficiency. Unlike the monotonic trends observed for the coil geometry, the actuation strain exhibits a pronounced convex parabolic profile with respect to the bias angle, achieving a peak value at approximately $53°$. This optimal angle arises from the competition between the anisotropic thermal expansion of the fiber and the torsional coupling efficiency within the hierarchical assembly. Meanwhile, the stiffness follows a concave trajectory, reaching a minimum near the strain optimum. Upon all investigated parameter spaces, the theoretical predictions (solid lines) demonstrate exceptional quantitative agreement with the 3D finite element simulation results (discrete points).

The validation results above confirm that the proposed theoretical model accurately characterize the complex coupling between the initial curvature, the hierarchical helical deformations, and the macroscopic thermo-mechanical response of the TCP muscles, establishing a robust basis for the subsequent parametric optimization.

### 5.3 Hierarchical parametric effect of TCP's actuation and stiffness

Following the validation of the theoretical framework against experimental and finite element data, a systematic parametric study is conducted. This section investigates the sensitivity of the TCP muscle's macroscopic actuation characteristics—specifically, actuation strain, effective stiffness, and recovery loads—to variations in thermal loading, hierarchical topology, and microstructural geometry.

#### 5.3.1 Thermal actuation response

Figure 16 illustrates the thermo-mechanical coupling response, tracking the continuous evolution of actuation strain, recovery torque, and recovery bending moment upon heating. The actuation strain decreases monotonically with temperature, attaining a maximum contraction of approximately 22% at 120°C. This contraction is coupled with a quasi-linear increase in recovery torque, which reaches a peak value of 2.9 N·mm, signifying substantial torsional energy release. In contrast, the recovery bending moment exhibits a non-monotonic response; it initially increases to a local maximum near 60°C before declining sharply to −0.10 N·mm as the temperature approaches 120°C.

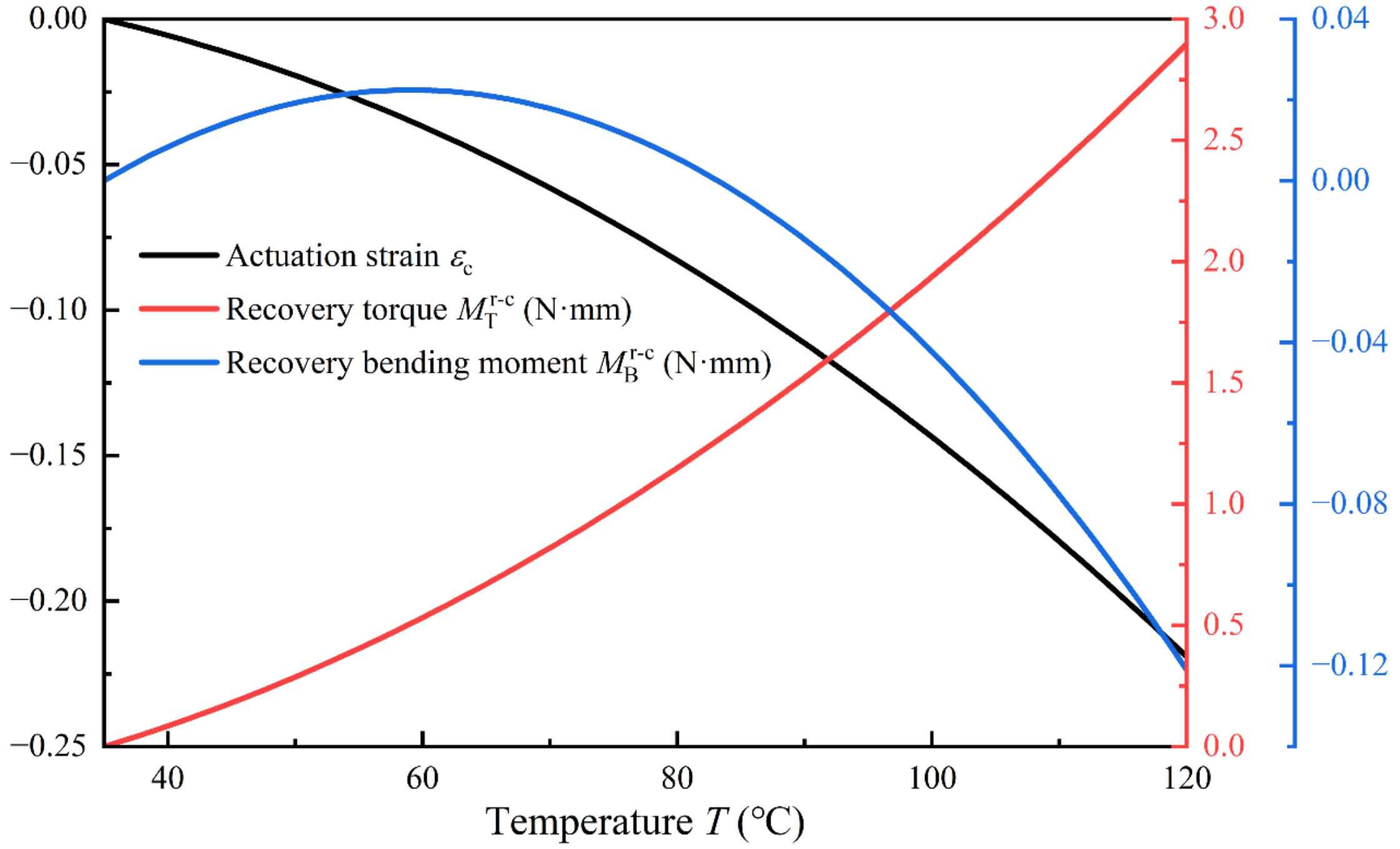

Fig. 16 Theoretical evolution of the actuation strain $\varepsilon_c$ (black line, left axis), recovery torque $M_T^{r\text{-}c}$ (red line, inner right axis), and recovery bending moment $M_B^{r\text{-}c}$ (blue line, outer right axis) with respect to temperature $T$.

### 5.3.2 Effects of hierarchical helical angle

Figure 17 characterizes the sensitivity of actuation metrics to the hierarchical geometry, specifically the primary ($\theta_f^0$) and secondary ($\theta_b^0$) helical angles. Both actuation strain (a) and stiffness (b) display a monotonic increase with $\theta_f^0$, a trend significantly accentuated by the secondary twist. Specifically, increasing $\theta_b^0$ from 50° to 80° elevates the peak strain to 0.24 and nearly doubles the structural stiffness at the upper bound of $\theta_{\mathrm{f}}^{0}$. Recovery torque (c) follows a similar linear positive correlation, culminating in a maximum of 4.0 N·mm. Conversely, the recovery bending moment (d) exhibits a divergent, negative trend; while relatively stable at low $\theta_{\mathrm{b}}^{0}$, it declines sharply to $-0.18$ N·mm at high angles, evidencing a strong coupling between highly twisted hierarchical geometries and bending instability.

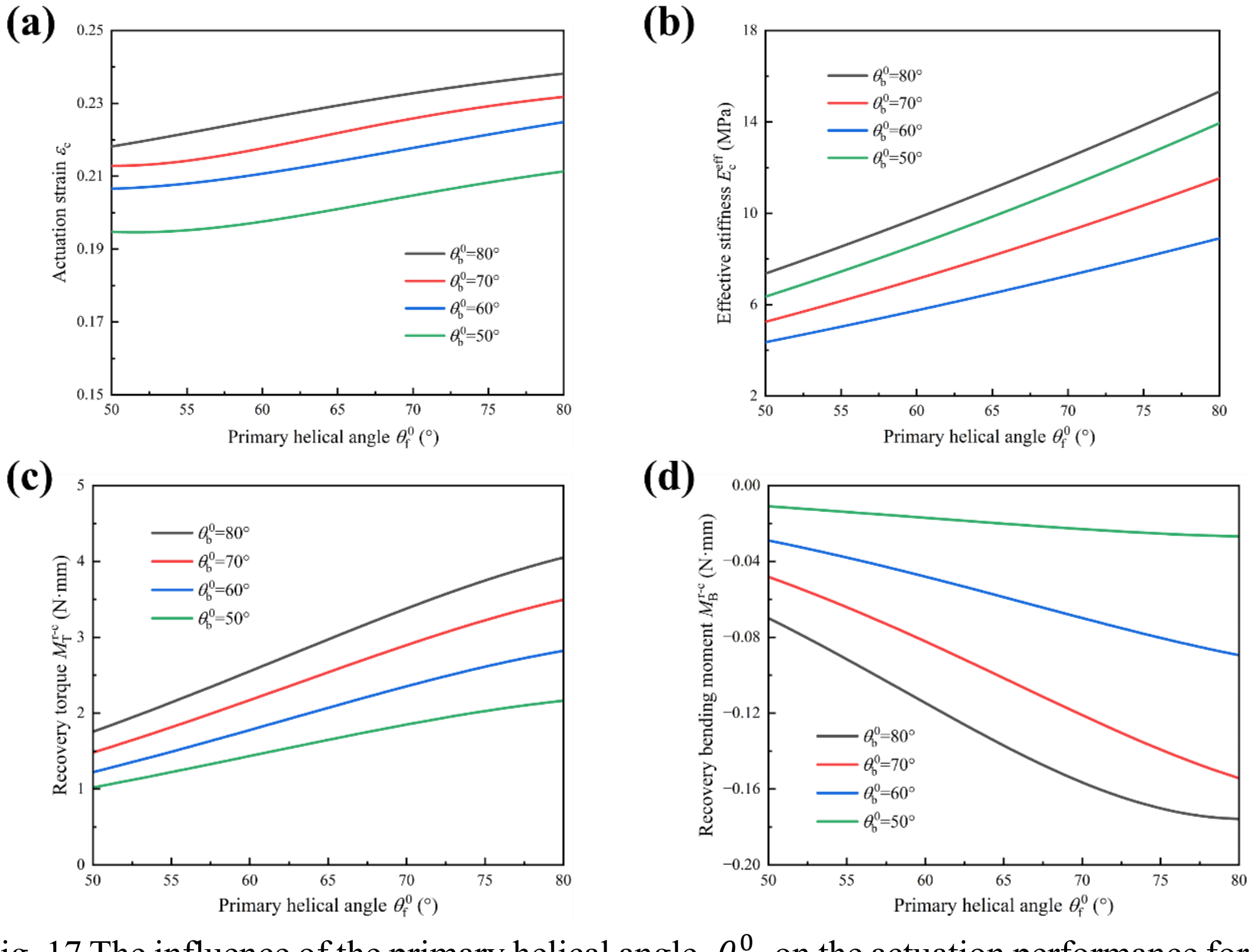

Fig. 17 The influence of the primary helical angle $\theta_{\mathrm{f}}^{0}$ on the actuation performance for different values of $\theta_{\mathrm{b}}^{0}$ (ranging from 50° to 80°). (a) Actuation strain $\varepsilon_{\mathrm{c}}$. (b) Effective stiffness $E_{\mathrm{c}}^{\mathrm{eff}}$. (c) Recovery torque $M_{\mathrm{T}}^{\text{r-c}}$. (d) Recovery bending moment $M_{\mathrm{B}}^{\text{r-c}}$.

### 5.3.3 Effects of filament and bundle court in the hierarchy

The scaling laws governing the interplay between structural complexity and mechanical output are examined in Fig. 18, focusing on fiber strand counts at the primary ($n$) and secondary ($m$) levels. Actuation strain (a) undergoes monotonic decay with increasing $m$, a trend intensified by higher $n$. Conversely, stiffness (b) follows an asymptotic trajectory, saturating at levels inversely proportional to $n$, implying a compromise between bundle size and modulus. Recovery torque (c) exhibits a pronounced increase, reaching $600$ N·mm for the $n = 15$ configuration. This gain comes at the cost of stability, as the recovery bending moment (d) diverges to $-100$ N·mm, demonstrating severe bending coupling in high-mass hierarchical structures.

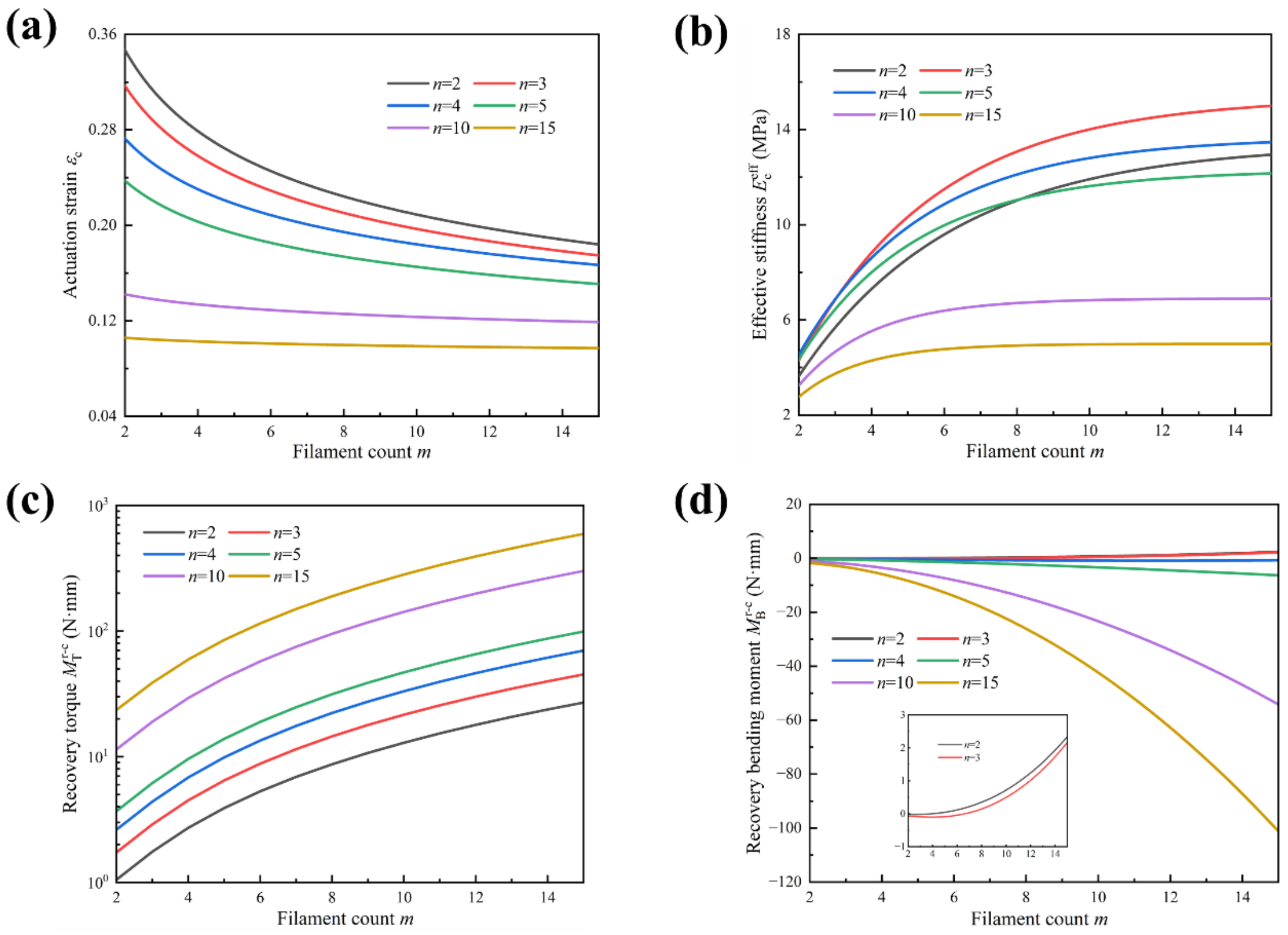

Fig. 18 Theoretical predictions of the actuation performance as a function of the filament count $m$ for different values of $n$ (ranging from 2 to 15). (a) Actuation strain $\varepsilon_c$ plotted on a logarithmic scale. (b) Effective stiffness $E_c^{eff}$. (c) Recovery torque $M_T^{r\text{-}c}$ plotted on a linear scale. (d) Recovery bending moment $M_B^{r\text{-}c}$.

### 5.3.4 Effects of size effects

To clarify the role of fiber cross-sectional dimensions, Figure 19

delineates response variations with respect to the primary fiber radius ($r_f^0$). A distinct trade-off governs actuation performance: as $r_f^0$ increases from $0.10$ to $0.25$ mm, actuation strain (a) decays monotonically from $0.33$ to $0.12$, whereas structural stiffness exhibits a nonlinear, convex increase, peaking at $42$ MPa. Parallel to stiffness, the recovery torque (b) scales substantially, reaching $13.5$ N·mm. Conversely, the recovery bending moment diverges to $-1.2$ N·mm, indicating that while thicker geometries enhance load-bearing capacity, they simultaneously amplify bending coupling effects.

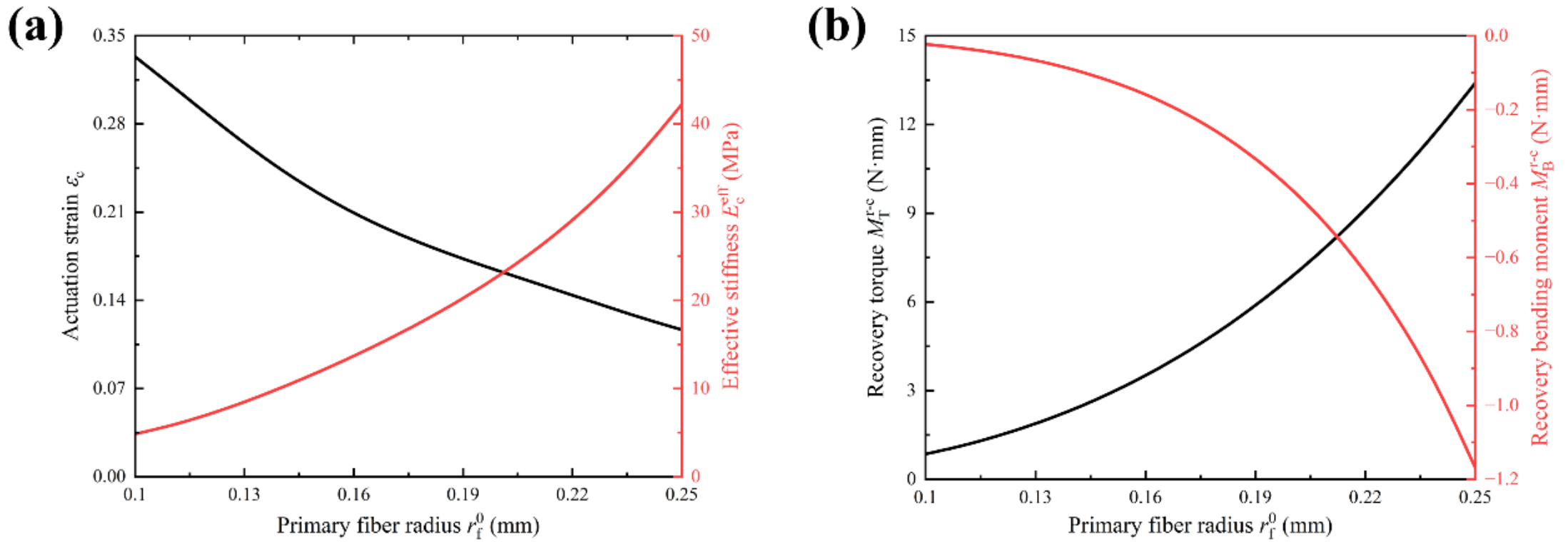

Fig. 19 Influence of the primary fiber radius $r_f^0$ on the actuation performance. (a) The variations of actuation strain $\varepsilon_c$ (black curve, left axis) and effective stiffness $E_c^{eff}$ (red curve, right axis). (b) The variations of recovery torque $M_T^{r\text{-}c}$ (black curve, left axis) and recovery bending moment $M_B^{r\text{-}c}$ (red curve, right axis).

### 5.3.5 Effects of structural chirality

Figure 20 presents the effect of topological chirality, denoted by the index $\eta$, revealing how the specific chiral arrangement ($\eta = 1$ vs. $\eta = -1$) modulates thermo-mechanical behavior. While actuation strain (a) is marginally enhanced in the $\eta = -1$ configuration, structural stiffness (a) is maximized for $\eta = 1$, indicating a competition between deformability and rigidity. Recovery torque (b) displays notable insensitivity to chirality, with both $\eta = \pm 1$ curves overlapping and peaking at 3.5 $\mathrm{N \cdot mm}$. Conversely, the recovery bending moment (b) diverges; the $\eta = 1$ geometry undergoes a steeper decline in moment, signifying a stronger coupling with bending instability compared to the $\eta = -1$ case.

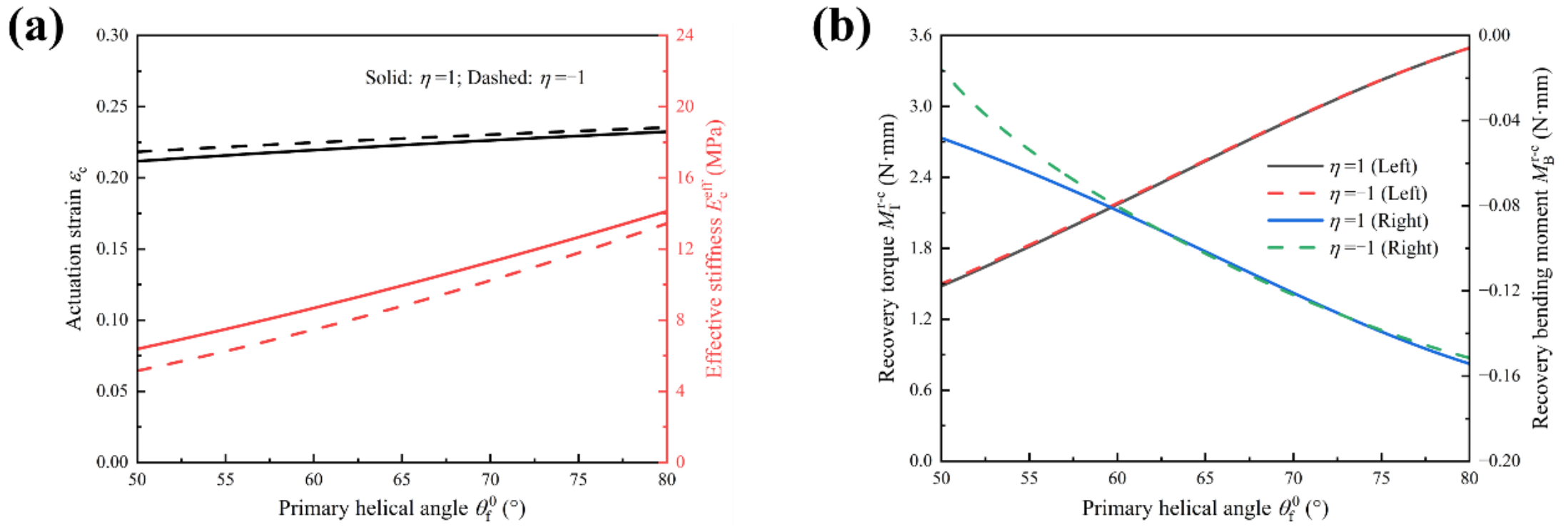

Fig. 20 Comparison of actuation performance for different chirality parameters η (solid lines for $\eta = 1$, dashed lines for $\eta = -1$) as a function of the primary helical angle $\theta_{\mathrm{f}}^{0}$. (a) The variations of actuation strain $\varepsilon_{\mathrm{c}}$ (black lines, left axis) and effective stiffness $E_{\mathrm{c}}^{\mathrm{eff}}$ (red lines, right axis). (b) The variations of recovery torque $M_{\mathrm{T}}^{\mathrm{r\text{-}c}}$ (left axis) and recovery bending moment $M_{\mathrm{B}}^{\mathrm{r\text{-}c}}$ (right axis).

## 5.4 Actuation stress and energy density

Although kinematic response and static stiffness characterize deformation mechanics, the practical utility of hierarchical TCP actuators is ultimately determined by their work capacity. This section extends the parametric analysis to include isometric actuation stress and volumetric energy density, with the latter computed as the area under the stress-strain curve over the actuation stroke as,

$$W = \sigma_{\mathrm{c}} \varepsilon_{\mathrm{c}}, \tag{59}$$

$$\sigma_{\mathrm{c}} = E_{\mathrm{c}}^{\mathrm{eff}} \varepsilon_{\mathrm{c}}, \tag{60}$$

$$\varepsilon_c = \frac{\delta(T_1) - \delta(T_0)}{L_c^0 \sin\theta_c^0}, \tag{61}$$

$$E_c^{eff} = \frac{F L_c^0 \sin^2\theta_c^0}{\pi r_r^2 \delta(T_0)}. \tag{62}$$

5.4.1 Geometric dependence on helical angles

The mechanical work capacity, quantified by isometric actuation stress and volumetric energy density, is analyzed in Fig. 21 as a function of hierarchical twist parameters ($\theta_f^0$ and $\theta_b^0$). Both metrics demonstrate a robust monotonic dependence on the hierarchical twist. The actuation stress (a) scales linearly with $\theta_f^0$, a trend significantly amplified by higher $\theta_b^0$ values, culminating in a peak stress of 3.6 MPa. Similarly, the energy density (b) exhibits a convex increasing trend, rising from 0.18 MPa to 0.87 MPa across the investigated domain. These results suggest that highly twisted hierarchical geometries are essential for maximizing the actuator's mechanical work potential.

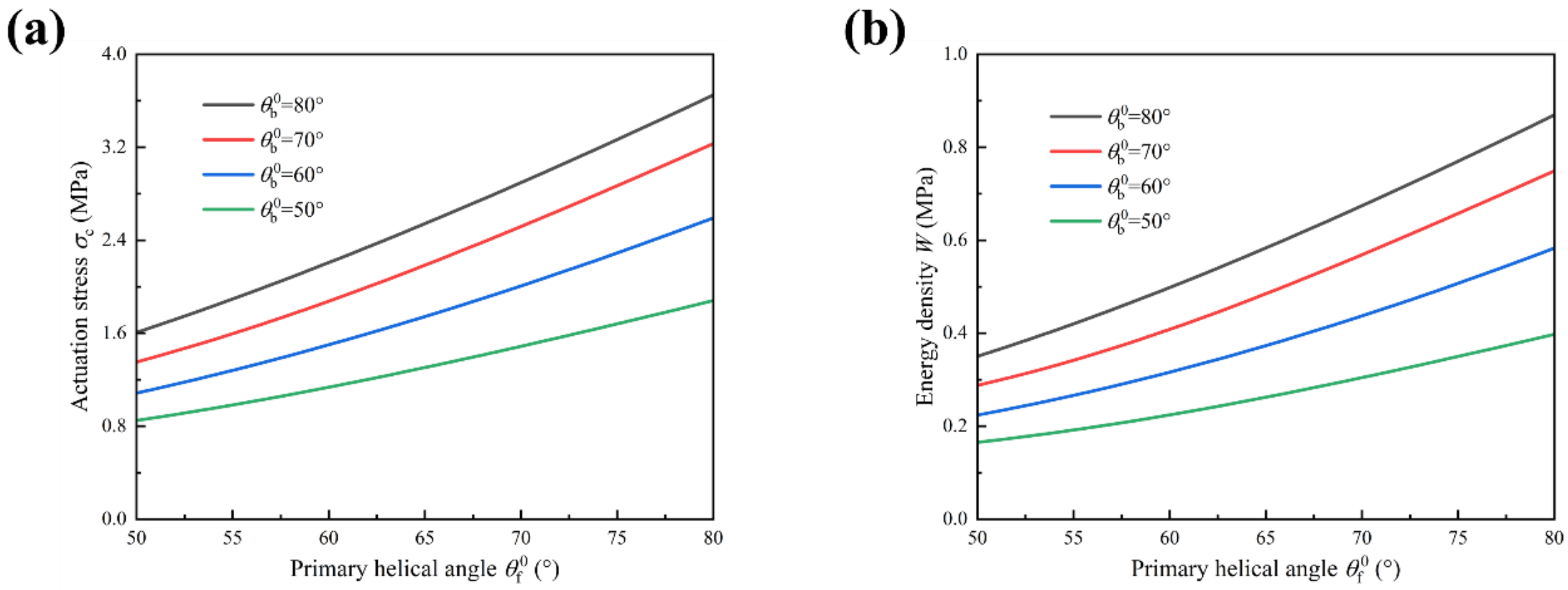

Fig. 21 Theoretical predictions of the actuation performance as a function of the primary helical angle $\theta_f^0$ for different values of $\theta_b^0$ (ranging from $50°$ to $80°$). (a) Actuation stress $\sigma_c$. (b) Energy density $W$.

5.4.2 Optimization of hierarchical arrangement

In contrast to the monotonic angular trends, actuation performance reveals a complex dependency on the topological strand distribution $(n, m)$, as illustrated in Fig. 22. Both actuation stress (a) and energy density (b) exhibit a non-monotonic dependence on $m$, revealing a distinct geometric optimum. The configuration with $n = 3$ yields maximum performance, achieving a peak stress of 2.8 MPa at $m \approx 9$ and a peak energy density of 0.6 MPa at $m \approx 6$. Deviating from this optimum by increasing the primary bundle size ($n \geq 5$) results in a sharp degradation of work capacity, with the $n = 15$ case reduced to below 0.5 MPa. These data delineate a critical topological window for maximizing specific mechanical output.

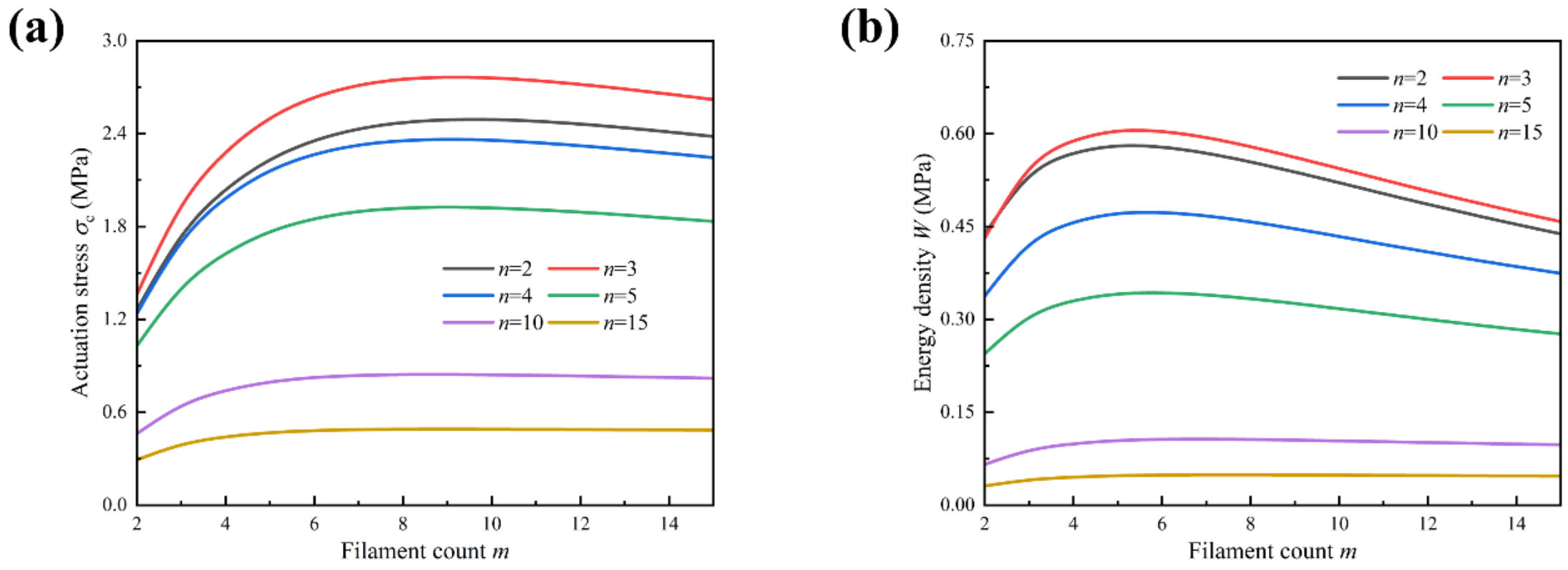

Fig. 22 Theoretical predictions of the actuation performance as a function of the filament count $m$ for different values of $n$ (ranging from 2 to 15). (a) Actuation stress $\sigma_c$. (b) Energy density $W$.

5.4.3 Size-dependent strengthening mechanism

Analysis of the primary fiber radius ($r_f^0$) reveals a divergence in scaling behaviors between force generation and energetic efficiency, as shown in Fig. 23. A distinct decoupling characterizes the mechanical response:

actuation stress (a) scales quasi-linearly with $r_f^0$, tripling from 1.6 MPa to nearly 5.0 MPa. Conversely, volumetric energy density (b) exhibits notable invariance, plateauing near 0.60 MPa. This indicates that while geometric scaling enhances load-bearing capacity, specific work output remains an intrinsic, size-independent property of the hierarchical structure.

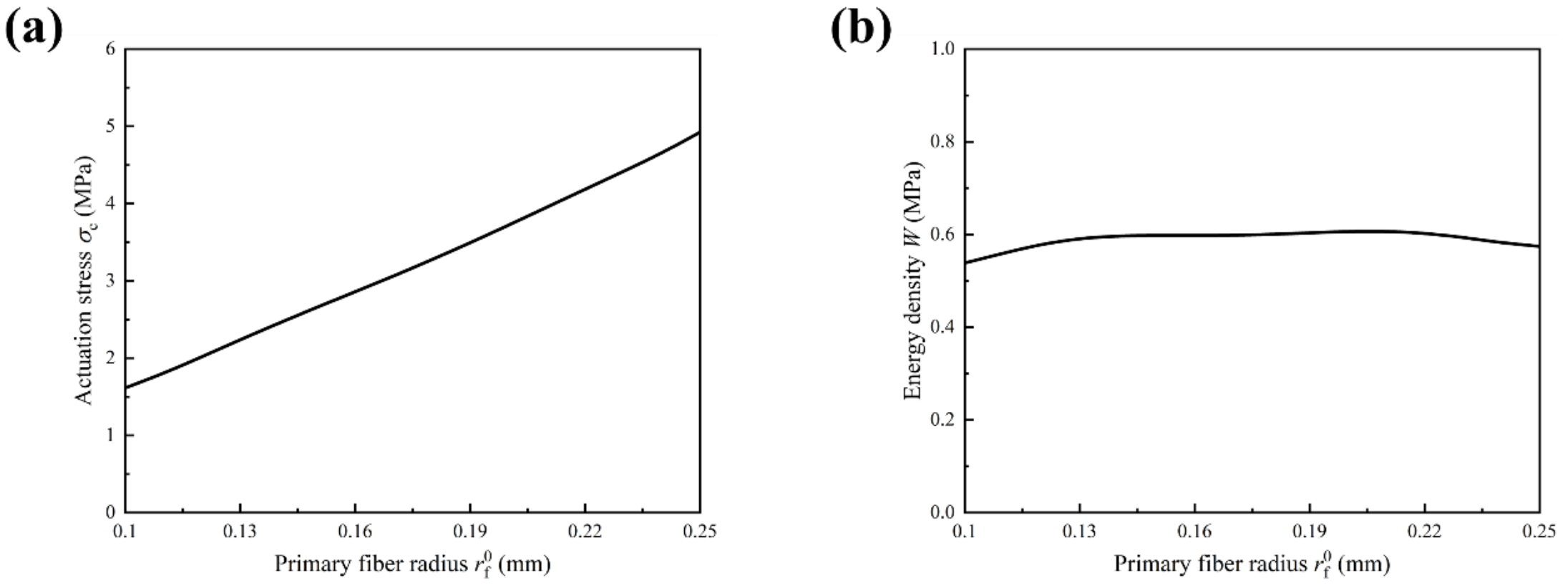

Fig. 23 Theoretical predictions of the actuation performance as a function of the primary fiber radius $r_f^0$. (a) Actuation stress $\sigma_c$. (b) Energy density $W$.

### 5.4.4 Chirality-induced coupling efficiency

Finally, the modulation of work capacity by the topological chirality index ($\eta$) is plotted in Fig. 24, highlighting the performance disparity between homochiral and heterochiral configurations. Both actuation stress (a) and energy density (b) increase monotonically with the primary helical angle, yet exhibit a distinct sensitivity to the chiral topology. The $\eta = 1$ configuration consistently yields higher mechanical outputs than the $\eta = -1$ case. Specifically, at $\theta_f^0 = 80°$, the actuation stress for $\eta = 1$ reaches 3.3 MPa, surpassing the 3.1 MPa observed for $\eta = -1$. Similarly, volumetric energy density favors the $\eta = 1$ structure, peaking at 0.78 MPa. This confirms that the homochiral arrangement provides a quantifiable enhancement to the overall thermo-mechanical performance.

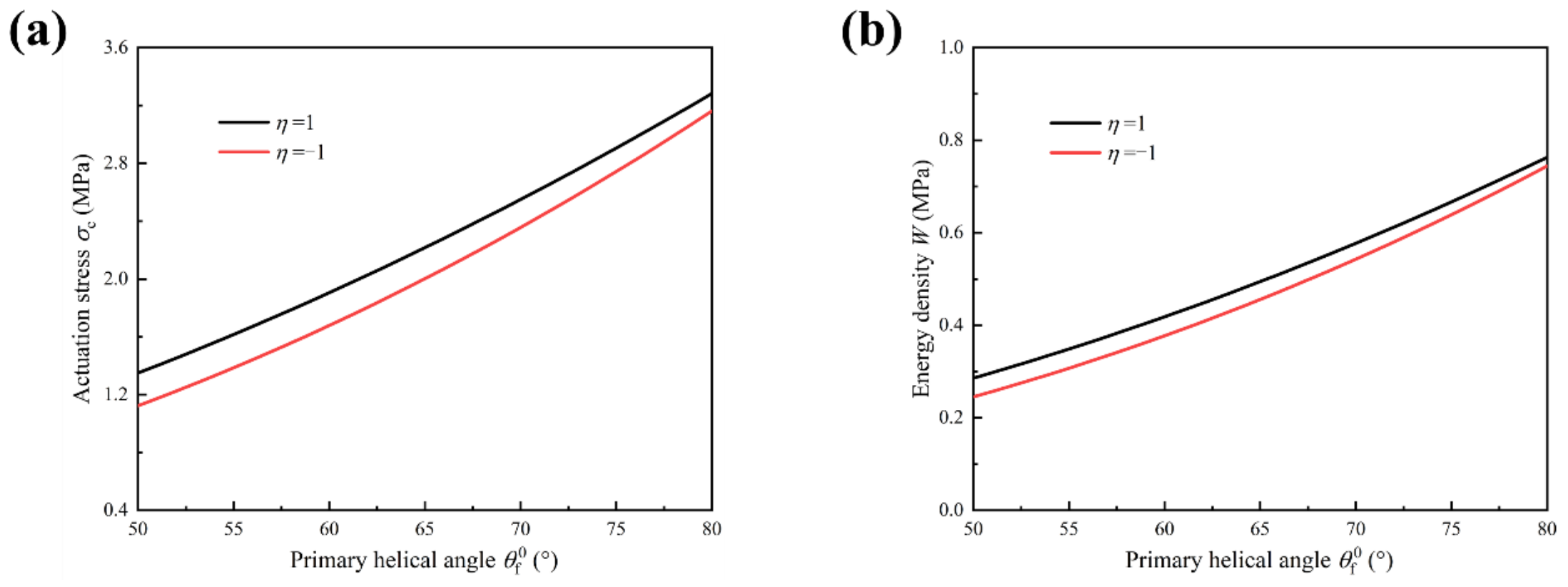

Fig. 24 Comparisons of the actuation performance with different chirality parameters $\eta$ (black lines for $\eta = 1$, red lines for $\eta = -1$) as a function of the primary helical angle $\theta_f^0$. (a) Actuation stress $\sigma_c$. (b) Energy density $W$.

## 6. Discussion

The primary contribution of this study is the resolution of a persistent trade-off in TCP artificial muscles: the historical inability to elevate load capacity without drastically sacrificing actuation stroke. Unlike single-fiber baselines that typically compromise kinematic range to support higher loads (Haines et al., 2014; Yuan et al., 2019), our hierarchical topological design effectively decouples force generation from deformation limits. We achieved an isometric actuation stress of ~5.0 MPa—a threefold increase over the baseline—while maintaining a biological-like stroke of ~22%. Crucially, this performance is maximized within a specific topological window (homochiral, $n = 3$), suggesting that the hierarchical structure functions not merely as a bundle, but as a mechanically coherent meta-structure. This successful decoupling sets the stage for understanding the underlying physical mechanisms that allow such a rigid structure to maintain high deformability.

## 6.1 Mechanism of synergy on stiffness and chirality

The physical basis for this decoupling lies in a counter-intuitive "stiffness-stroke synergy" revealed by our parametric analysis. While classical mechanics dictates that increased stiffness impedes displacement (Wahl, 1963), our results demonstrate that in TCPs, higher twist densities amplify the thermal untwisting torque exponentially, thereby overcoming the linear rise in structural rigidity. Furthermore, this synergistic mechanism is strictly dependent on chiral topology. The performance gap between homochiral ($\eta = 1$) and heterochiral ($\eta = -1$) configurations elucidates a critical "chiral matching" effect: in homochiral topologies, radial thermal expansion works constructively with the untwisting moment, whereas heterochiral forms likely induce antagonistic shear stresses (Lamuta et al., 2018). However, while optimizing the helical angle and chirality enhances performance, increasing the topological complexity of the bundle itself reveals a critical boundary.

## 6.2 Topological Limits of geometric jamming

Despite the benefits of hierarchical scaling, our data delineate a strict upper bound governed by "geometric jamming." The non-monotonic dependence of actuation stress on bundle count, peaking at $n = 3$ and degrading significantly for $n \geq 5$, implies a transition from stable load distribution to frictional saturation. We postulate that low bundle counts ($n = 3$) adopt a stable triangular packing that permits radial "breathing" during thermal cycling. Conversely, denser configurations ($n \geq 5$) suffer from circumferential confinement, where excessive inter-filamentary friction dissipates thermal work as heat rather than mechanical output. This finding aligns with wire rope mechanics and establishes that simply adding more fibers is a diminishing return strategy. Instead, the path to higher power lies in geometric scaling rather than topological densification.

### 6.3 Scale invariance and engineering scalability

In contrast to the limitations imposed by bundle count ($n$), the volumetric energy density exhibits remarkable scale invariance with respect to fiber radius ($r$). Our analysis shows that energy density remains constant (~0.60 MPa) even as the fiber radius is scaled up. While the peak stress falls within the typical range of polymer actuators (1–10 MPa) (Chen et al., 2024), the constancy of energy density distinguishes this structure from surface-dominated actuators (e.g., CNT yarns) that typically suffer performance dilution with increased size (Mirvakili and Hunter, 2018). This implies that the hierarchical TCP allows for the "linear amplification" of force: the actuator can be geometrically scaled to achieve arbitrary absolute force outputs for heavy-duty applications (e.g., exoskeletons) without inherent efficiency losses, effectively bridging the gap between micro-robotic prototypes and industrial-grade actuation.

### 6.4 The competition of limitations and stability

The realization of such high volumetric energy density, however, introduces intrinsic stability challenges. The divergence of the recovery bending moment at high helical angles (Fig. 17d) signals a strong propensity for lateral buckling, or "snarling," which acts as the physical price for high-torque density. While this necessitates external guide mechanisms in practical deployments (Saharan and Tadesse, 2018), it confirms the structure's high-energy state. Additionally, the increased thermal mass of hierarchical bundles inevitably prolongs cooling cycles compared to monofilaments. Future iterations must address this thermal hysteresis, potentially by converting the interstitial void space of the hierarchy—currently a structural feature—into a conduit for active fluidic cooling.

## 7. Conclusion

Twisted and coiled polymer artificial muscles exhibit a programmable hierarchical structure that governs their macroscopic actuation authority, offering a versatile platform for artificial muscle design. The present study has established a comprehensive mechanics-based framework for the design and modeling of hierarchical TCP actuators, explicitly aimed at resolving the intrinsic dichotomy between load-bearing capacity and actuation stroke. By systematically restructuring monofilament precursors into rationally designed multi-level helices, we demonstrated that the mechanical limitations of thermal actuators are not solely dictated by material properties but are significantly governed by topological structure. The central achievement of this work is the successful decoupling of force generation from kinematic constraints. Through an optimized hierarchical topology, we achieved an isometric actuation stress of approximately 5.0 MPa, that is a threefold enhancement over single-fiber baselines, while preserving a biological-like contraction stroke of ~22%. These findings refute the conventional assumption that high-load polymer actuators must necessarily sacrifice dynamic range, confirming instead that mechanical output can be amplified through geometric ordering without compromising the intrinsic actuation mechanism.

Our theoretical and experimental analyses elucidate two fundamental physical principles that dictate the efficacy of these meta-structures. First, we identified a strict topological threshold at a primary bundle count of $n = 3$. This configuration represents a critical equilibrium between load sharing and "geometric jamming"; beyond this threshold ($n \geq 5$), the structural benefits are negated by excessive inter-filamentary friction and circumferential confinement, which dissipate thermal energy rather than generating work. Second, the study substantiates the necessity of chiral matching ($\eta = 1$), revealing a "stiffness-stroke synergy" where the radial thermal expansion of homochiral filaments constructively amplifies the untwisting torque. These nonlinear behaviors were accurately demonstrated by our multi-scale constitutive model, validating its utility as a robust predictive tool for determining the effective stiffness and actuation limits of complex fibrous assemblies.

From an engineering perspective, the most significant implication of this research is the scale invariance of the volumetric energy density (~0.60 MPa). Unlike surface-dominated actuators that typically suffer from performance dilution upon upscaling, the hierarchical TCP structure permits the linear amplification of absolute force output through geometric scaling (increasing fiber radius) without compromising specific efficiency. This characteristic effectively bridges the gap between micro-robotic prototypes and high-load industrial applications, such as powered exoskeletons. Collectively, these results signify a paradigm shift in artificial muscle design, moving from material synthesis toward structural programming, where predictable mechanical performance is encoded directly into the geometry. Future efforts will focus on addressing the thermal hysteresis inherent to these high-mass structures, potentially through the integration of active fluidic cooling within the hierarchical porosity.

## References

Aziz, S., Spinks, G.M., 2020. Torsional artificial muscles. Mater. Horiz. 7, 667–693.

Chen, Y.P., Hu, J.J., Xie, Y.Y., Liu, L., Liu, D.B., 2024. Effect of temperature softening on the actuation performance of twisted and coiled polymer muscles. Sens. Actuators A Phys. 374, 115444.

Cherubini, A., Moretti, G., Vertechy, R., Fontana, M., 2015. Experimental characterization of thermally-activated artificial muscles based on coiled nylon fishing lines. AIP Adv. 5(6), 067158.

Choy, C., Leung, W., 1985. Elastic moduli of ultradrawn polyethylene. J. Polym. Sci. Polym. Phys. Ed. 23, 1759–1780.

Chu, H.T., Hu X.H., Wang Z., Mu J.K., Li N., Zhou X.S., Fang S.L., Haines C.S., Park J.W., Qin S., Yuan N.Y., Xu J., Tawfick S., Kim H., Conlin P., Cho M., Cho K., Oh J., Nielsen S., Alberto K.A., Razal J.M., Foroughi J., Spinks G.M., Kim S.J., Ding J.N., Leng J.S., 2021. Baughman R.H. Unipolar stroke, electroosmotic pump carbon nanotube yarn muscles. Science 371(6528), 494–498.

Costello, G.A., 1997. Theory of wire rope. Springer, New York.

Espinosa, H.D., Filleter, T., Naraghi M., 2012. Multiscale Experimental Mechanics of Hierarchical Carbon-Based Materials. Adv. Mater. 24(21), 2805–2823.

Foroughi, J., Spinks, G.M., Wallace, G.G., Oh, J., Kozlov, M.E., Fang, S.,

Wang, T., Huang, A.R., Kaner, R.B., Baughman, R.H., 2011. Torsional carbon nanotube artificial muscles. Science 334, 494–497.

Kongahage, D., Spinks, G.M., Foroughi, J., 2021. Twisted and coiled multi-ply yarns artificial muscles. Sens. Actuators A Phys. 318, 112490.

Gao, P.X., Yuan, X.J., Ren, M., Dong, L.Z., Di J.T., 2025. Self-Sensing Archimedean Spiral Artificial Muscle Fibers with Bidirectional Actuation for Electromagnetic Wave/Light Modulation and Mechanical Display. Adv. Funct. Mater. e19002.

Gao, Y., Li, B., Wang, J.S., Feng, X.Q., 2021. Fracture toughness analysis of helical fiber-reinforced biocomposites. J. Mech. Phys. Solids 146, 104206.

Gao, Z.W., Guo, J.J., Zhang, Y.H., Zhou, Z.W., Zhang, C.N., Li, H., Chen, B., Wang, J.Z., 2024. Multilayer modeling framework for analyzing thermo-mechanical properties and responses of twisted and coiled polymer actuators. Smart Mater. and Struct. 33, 045031.

Gotti, C., Sensini, A., Zucchelli, A., Carloni, R., Focarete, M.L., 2020. Hierarchical fibrous structures for muscle - inspired soft - actuators: A review. Appl. Mater. Today 20, 100772.

Haines, C.S., Lima, M.D., Li, N., Spinks, G.M., Foroughi, J., Madden, J.D.W., Kim, S.H., Fang, S.L., de Andrade, M. J., Göktepe, F., Göktepe, Ö., Mirvakili, S.M., Naficy, S., Lepró, X., Oh, J.Y., Kozlov, M.E., Kim, S.J., Xu, X.R., Swedlove, B.J., Wallace, G.G., Baughman, R.H., 2014. Artificial muscles from fishing line and sewing thread. Science 343, 868–872.

Han, Y.C., Yong, H.D., Zhou, Y.H., 2024. A multi-scale mechanical model of multilevel helical structures with filament damage. Int. J. Mech. Sci. 290, 112666.

He, W.J., Ge, Z.L., Kong, W.Y., Ye, L., Zhang, Z., Zhao, X.W., Lin, W., Wang, G.L., 2022. Construction of twisted/coiled poly (lactic acid) fiber-based artificial muscles and stable actuating mechanism. ACS Sustainable Chem. & Eng. 10(46), 15186–15198.

Hu, J., Liu, L., Liu, H., Liu D.B., 2025. Thermal Torsion Effect of Twisted Polymer Actuators. Acta Mech. Solida Sin. 38: 320–330.

Swartz, A.M., Ruiz, D.R.H., Feigenbaum, H.P., Shafer, M.W., Browder, C.C., 2018. Experimental characterization and model predictions for twisted polymer actuators in free torsion. Smart Mater. Struct. 27(11), 114002.

Higueras-Ruiz, D.R., Shafer, M.W., Feigenbaum, H.P., 2021. Cavatappi artificial muscles from drawing, twisting, and coiling polymer tubes. Sci. Robot. 6, eabd5383.

Karami, F., Tadesse, Y., 2017. Modeling of twisted and coiled polymer (TCP) muscle based on phenomenological approach. Smart Mater. Struct. 26(12), 125010.

Lamuta, C., Messelot, S., Tawfick, S., 2018. Theory of the tensile actuation of fiber reinforced coiled muscles. Smart Mater. Struct. 27(5), 055018.

Lang, T.H., Yang, L.X., Yang, S.J., Sheng, N., Zhang, Y.Y., Song, X.F., Guo, Y., Fang, S.L., Mu, J.K., Baughman, R.H., 2024. Emerging innovations in electrically powered artificial muscle fibers. Natl. Sci. Rev. 11(10) nwae232.

Leng, X.Q., Hu, X.Y., Zhao, W.B., An, B.G., Zhou, X., Liu, Z.F., 2021. Recent advances in twisted-fiber artificial muscles. Advanced Intelligent Systems 3(5), 2000185.

Lima, M.D., Li, N., de Andrade, M.J., Fang, S.L., Oh, J., Spinks, G.M., Kozlov, M.E., Haines, C.S., Suh, D., Foroughi, J., Kim, S.J., Chen, Y.S., Ware, T., Shin, M.K., Machado, L.D., Fonseca, A.F., Madden, J.D.W., Voit, W.E., Galvao, D.S., Baughman, R.H., 2012. Electrically, chemically, and photonically powered torsional and tensile actuation of hybrid carbon nanotube yarn muscles. Science 338(6109), 928–932.

Liu, D.B, Zheng, S.M., He, Y.M., 2018. Effect of friction on the mechanical behavior of wire rope with hierarchical helical structures. Math. Mech. Solids 24(7), 2154-2180.

Liu, L., Liu, H., Zhang, Z.Y., Liu, D.B., 2024. A new model of thermo-mechanical actuation for twisted and coiled polymer muscles with initial curvature. Smart Mater. Struct. 33(6): 065022.

Meng, Q.H., Gao, Y., Shi X.H., Feng, X.Q., 2022. Three-dimensional crack bridging model of biological materials with twisted Bouligand structures. J. Mech. Phys. Solids 159, 104729.

Meng, D., Xie, J.Z., Waterhouse, G.I.N., Zhang, K., Zhao, Q.H., Wang, S., Qiu, S., Chen, K.J., Li, J.X., Ma, C.Z., Pan, Y., Xu, J., 2020. Biodegradable Poly (butylene adipate-co-terephthalate) composites reinforced with bio-based nanochitin: preparation, enhanced mechanical and thermal properties. J. Appl. Polym. Sci. 137(12), 48485.

Mirvakili, S.M., Hunter, I.W., 2018. Artificial muscles: Mechanisms, applications, and challenges. Adv. Mater. 30, 1704407.

Mu, J.K., Wang, G., Yan, H.P., Li, H.Y., Wang, X.M., Gao, E.L., Hou, C.Y., Pham, A.T.C., Wu, L.J., Zhang, Q.H., Li, Y.G., Xu, Z.P., Guo, Y., Reichmanis, E., Wang, H.Z., Zhu, M.F., 2018. Molecular-channel driven actuator with considerations for multiple configurations and color switching. Nat. Commun. 9, 590.

Mu, J.K., Jung de Andrade, M., Fang, S., Wang, X., Gao, E.L., Li, N., Kim, S.H., Wang, H., Hou, C., Zhang, Q., Zhu, M.F., Qian, D., Lu, H., Kongahage, D., Talebian, S., Foroughi, J., Spinks, G., Kim, H., Ware, T., Sim, H.J., Lee, D.Y., Jang, Y., Kim, S.J., Baughman, R.H., 2019. Sheath-run artificial muscles. Science 365, 150–155.

Peng, Y.Y., Sun, F.X., Xiao, C.Q., Iqbal, M.I., Sun, Z.G., Guo, M.R., Gao, W.D., Hu, X.R., 2021. Hierarchically Structured and Scalable Artificial Muscles for Smart Textiles. ACS Appl. Mater. Interfaces 13(45), 54386–54395.

Saharan, L., de Andrade, M.J., Saleem, W., Baughman, R.H., Tadesse, Y.,

2018. iGrab: hand orthosis powered by twisted and coiled polymer muscles. Smart Mater. Struct. 26, 105048.

Semochkin, A.N., 2016. A device for producing artificial muscles from nylon fishing line with a heater wire. IEEE International Symposium on Assembly and Manufacturing 61, 26–30.

Singh, G., Varshney, V., Sundararaghavan, V., 2024. Understanding creep in vitrimers: insights from molecular dynamics simulations. Polymer 313(15), 127667.

Sun, J.F., Zhao, J.G., 2022. Physics-based modeling of twisted-and-coiled actuators using Cosserat rod theory. IEEE T. Robot. 38(2), 779–796.

Sun J., Zhang S.J., Deng J., Li, J., Wang, D.H., Liu, J.K., Liu Y.X., 2025. Recent advances in twisted and coiled artificial muscles and their applications. SmartBot 9(2), 12005.

Sutton, L., Moein, H., Rafiee, A., Madden, J.D.W., Menon, C., 2016. Design of an assistive wrist orthosis using conductive nylon actuators. 6th IEEE International Conference on Biomedical Robotics and Biomechatronics 1074–1079.

Tsai, S.M., Wang, Q., Hur, O., Bartlett, M.D., King, W.P., Tawfick, S., 2025. High cycle performance of twisted and coiled polymer actuators. Sens. Actuators A Phys. 381, 116041.

Wahl, A.M., 1963. Mechanical Springs, second ed. McGraw-Hill, New York.

Weissman, Z., Ashcroft, B., Nguyen, P., Sun J.F., 2025. Efficient Pneumatic Twisted-and-Coiled Actuators Through Dual Enforced Anisotropy. IEEE/ASME T. Mech. 30(4), 2946–2954.

Wang, S.J., Xiao, Y., Xu, Z.P., 2022. Energy-conversion efficiency and power output of twisted-filament artificial muscles. Extreme Mech. Lett. 50, 101531.

Xiao, R., Tian, C.S., 2019. A constitutive model for strain hardening behavior of predeformed amorphous polymers: Incorporating dissipative dynamics of molecular orientation. J. Mech. Phys. Solids 125, 472–487.

Xiao, Y., Huang, Z.X., Wang, S.N., 2014. An elastic rod model to evaluate effects of ionic concentration on equilibrium configuration of DNA in salt solution. J. Biol. Phys. 40, 179–192.

Xiao, Y., Luo, Z., Li C., 2022. Synergistic effect of axial-torsional-radial deformation on the multi-strand helical filament artificial muscles. Appl. Math. Model. 109, 760–774.

Xiao, Y., Luo, Z., Li C., 2023. Mechanical response of twisted multifilament artificial muscles upon thermal actuation. Appl. Math. Model. 118, 502–517.

Xu, C.X., Jiang, Z.L., Wang, B.X, Chen, J.P., Sun, T., Fu, F.F., Wang, C.S., Wang, H.P., 2024. Biospinning of hierarchical fibers for a self-sensing actuator. Chem. Eng. J. 485, 150014.

Xu, C.X., Jiang, Z.L., Zhong, T.T., Chen, C., Ren, W.T., Sun, T., Fu F.F.,

2023. Multi-strand Fibers with Hierarchical Helical Structures Driven by Water or Moisture for Soft Actuators. ACS Omega 8(2), 2243–2252.

Yang, Q.X., Li, G.Q., 2016. A top-down multi-scale modeling for actuation response of polymeric artificial muscles. J. Mech. Phys. Solids 92, 237–259.

Yuan, J.K., Neri W., Zakri, C., Merzeau, P., Kratz, K., Lendlein A., Poulin P., 2019. Shape memory nanocomposite fibers for untethered high-energy microengines. Science 365(6449), 155–158.

Zhang, H., Yang, G., Shen, W., Zhang, H., Zheng, T., Zhang, C., Chen, T., 2024. A compound twisted and coiled actuators with payload-insensitive untwisting characteristics. Sens. Actuators A Phys. 374, 115407.

Zhang, J., Sheng, J., O'Neill, C.T., Walsh, C.J., Wood, R.J. Ryu, J.H., Desai, J.P., Yip, M.C., 2019. Robotic artificial muscles: Current progress and future perspectives. IEEE Trans. Robot. 35(3), 761–781.

Zhao, Z.L., Zhao, H.P., Wang, J.S., Zhang, Z., Feng, X.Q., 2014. Mechanical properties of carbon nanotube ropes with hierarchical helical structures. J. Mech. Phys. Solids 71, 64–83.

Zheng, S.M., Liu D.B., He Y.M., 2018. The influence of fiber migration on the mechanical properties of yarns with hierarchical helical structures. J. Strain Anal. Eng. 53(2), 88–105.

Zheng, T.Y., Zhang, X.J., Wang, M., Li, M.H., Zhang, C.W., Zhang, M.L., 2025. Performance measurement of twisted and coiled polymer with its driven bionic wrist mechanism via temperature self-sensing. Measurement 247, 116810.

Zhu, Z.D., Di, J.T., Liu, X.Y., Qin J.Q., Cheng, P., 2022. Coiled polymer fibers for artificial muscle and more applications. Matter 5(4), 1092–1103.

Appendix A. The bending stiffness for the helical bundle

This appendix outlines the derivation of the geometric projection term required for the effective bending stiffness in Eq. (37). Starting from the filament position vector $r(\tilde{s})$ defined in Eq. (32), the unit tangent vector $t_i$ is obtained by differentiating with respect to the bundle's phase angle $\alpha_b$:

$$\mathbf{t}_i = \frac{d\mathbf{r}}{d\alpha_\mathrm{b}}\left(\frac{ds}{d\alpha_\mathrm{b}}\right)^{-1}, \tag{A1}$$

The metric factor $ds/d\alpha_\mathrm{b}$, relating the filament arc length to the bundle deformation, is derived from the geometric constraints as,

$$\frac{ds}{d\alpha_\mathrm{b}} = \frac{1}{\Lambda} = \sqrt{(\rho - r_\mathrm{b}\cos\beta_\mathrm{f})^2 + (\rho\tan\theta_\mathrm{f})^2}, \tag{A2}$$

Substituting the Frenet frame derivatives of the bundle into Eq. (A.1), the explicit components of the filament's tangent vector are determined,

$$\begin{aligned}\mathbf{t}_i = & \frac{-\rho\sin\alpha_\mathrm{b} + \rho\tan\theta_\mathrm{f}\cdot\cos\alpha_\mathrm{b}\cdot\sin\beta_\mathrm{f} + r_\mathrm{b}\sin\alpha_\mathrm{b}\cdot\cos\beta_\mathrm{f}}{\Lambda}\mathbf{e}_1 \\ & + \frac{\rho\cos\alpha_\mathrm{b} + \rho\tan\theta_\mathrm{f}\cdot\sin\alpha_\mathrm{b}\cdot\sin\beta_\mathrm{f} - r_\mathrm{b}\cos\alpha_\mathrm{b}\cdot\cos\beta_\mathrm{f}}{\Lambda}\mathbf{e}_2 \\ & + \frac{\rho\tan\theta_\mathrm{f}\cos\beta_\mathrm{f}}{\Lambda}\mathbf{e}_3.\end{aligned} \tag{A3}$$

Then, the local curvature vector is given by $\kappa_\mathrm{f} = dt_i/ds$. To calculate the contribution of the filament's bending moment to the bundle's binormal axis, we evaluate the projection $\kappa_\mathrm{f} b_i \cdot e_3$, where $b_i = t_i \times n_i$ is the filament's binormal vector. Utilizing the vector identity for curvature $\kappa_\mathrm{f} b_i = t_i \times \frac{dt_i}{ds}$, the projection simplifies to:

$$\kappa_{\mathrm{f}}\mathbf{b}_i \cdot \mathbf{e}_3 = \left( \mathbf{t}_i \times \frac{d\mathbf{t}_i}{d\alpha_{\mathrm{b}}} \frac{1}{\Lambda} \right) \cdot \mathbf{e}_3, \tag{A4}$$

Executing the cross product and simplifying the trigonometric terms yields the specific geometric nonlinearity term $\Psi$ as,

$$\begin{aligned} \Psi &= \Lambda^3 (\kappa_{\mathrm{f}}\mathbf{b}_i \cdot \mathbf{e}_3) \\ &= (r_{\mathrm{b}}^2 - \rho^2 \tan^2\theta_{\mathrm{f}}) \cos^2\beta_{\mathrm{f}} - \rho \left( r_{\mathrm{b}} + \frac{\rho^2 \tan^2\theta_{\mathrm{f}}}{r_{\mathrm{b}}} \right) \cos\beta_{\mathrm{f}} + \rho^2 (1 + 2\tan^2\theta_{\mathrm{f}}). \end{aligned} \tag{A5}$$

Appendix B. Kinematic relations for the curved rope

This appendix details the derivation of the metric tensors and the finite strain components for the TCP structure, accounting for the initial curvature effects of the coiled rope. The rope is modeled as a curvilinear fiber with a circular cross-section of radius $r_{\mathrm{r}}$ and an initial curvature $\kappa_{\mathrm{r}}^{(0)}$.

In the reference configuration, a local curvilinear coordinate system $\{x_1, x_2, x_3\} = \{r_k, \alpha_{\mathrm{r}}, s\}$ is attached to the centerline of the rope. Here, $r_{\mathrm{k}} \in [0, r_{\mathrm{r}}]$ is the radial coordinate, $\alpha_{\mathrm{r}} \in [0,2\pi]$ is the winding angle of the helical coil, and $s$ denotes the arc length along the centerline. The position vector $\mathbf{x}$ of an arbitrary material point is defined by the Frenet frame $\{\mathbf{T}, \mathbf{N}, \mathbf{B}\}$ of the helical centerline:

$$\mathbf{x}(r_k, \alpha_{\mathrm{r}}, s) = \mathbf{r}_{\mathrm{c}}(s) + r_k \cos\alpha_{\mathrm{r}} \mathbf{N}(s) + r_k \sin\alpha_{\mathrm{r}} \mathbf{B}(s), \tag{A6}$$

The squared differential arc length in the reference configuration is given by $ds^2 = G_{ij} dx_i dx_j$. Due to the intrinsic curvature, the arc length of the rope located at $(r_k, \alpha_{\mathrm{r}})$ differs from the one of the centerlines by a factor of the Jacobian determinant $J_{Jaco} = 1 + \kappa_r^{(0)} r_k \cos\alpha_r$. Consequently, the covariant metric tensor $G_{ij}$ is derived as:

$$[G_{ij}] = \frac{\partial \mathbf{x}}{\partial x_i} \cdot \frac{\partial \mathbf{x}}{\partial x_j} = \mathrm{diag}\left(1,\ \ r_k^2,\ \ (1+\kappa_{\mathrm{r}}^{(0)} r_k \cos\alpha_{\mathrm{r}})^2\right). \tag{A7}$$

Eq. (A7) corresponds to Eq. (42), where the non-Euclidean geometry induced by the initial curvature is explicitly identified by the component $G_{33}$.

Upon thermo-mechanical actuation, the rope in a current configuration described by coordinates $\{\bar{x}_1, \bar{x}_2, \bar{x}_3\} = \{\bar{r}_k, \bar{\alpha}_{\mathrm{r}}, \bar{s}\}$ . Assuming the deformation is uniform along the helical axis, the metric tensor $g_{ij}$ in the current configuration retains a form analogous to Eq. (A7), but incorporates the variational curvature $\kappa_{\mathrm{r}} = \kappa_{\mathrm{r}}^{(0)} + \Delta\kappa_{\mathrm{r}}$ and the deformed radius:

$$g_{33} = (1 + \kappa_{\mathrm{r}} \bar{r}_k \cos\bar{\alpha}_{\mathrm{r}})^2, \tag{A8}$$

Then, the longitudinal stretch ratio $\lambda_{\mathrm{s}}$ at a generic material point is defined as the ratio of the differential arc length in the deformed configuration to that in the reference configuration:

$$\lambda_{\mathrm{s}} = \sqrt{\frac{g_{33}}{G_{33}}} \frac{d\bar{s}}{ds} = \frac{1 + \kappa_{\mathrm{r}}(r_k + u_{\mathrm{radi}})\cos\alpha_{\mathrm{r}}}{1 + \kappa_{\mathrm{r}}^{(0)} r_k \cos\alpha_{\mathrm{r}}}(1 + \varepsilon_{\mathrm{r}}). \tag{A9}$$

where it is assumed that the variation in the winding angle of the helical coil does not significantly alter the cosine projection in the linearized regime (i.e., $\cos\bar{\alpha}_{\mathrm{r}} \approx \cos\alpha_r$ ). Substituting $\kappa_{\mathrm{r}} = \kappa_{\mathrm{r}}^{(0)} + \Delta\kappa_{\mathrm{r}}$ into Eq. (A9) and neglecting higher-order terms (such as the product of curvature variation and radial displacement $\Delta\kappa_{\mathrm{r}} u_{\mathrm{radi}}$), the linearized longitudinal Green-Lagrange strain $\varepsilon_{\mathrm{s}} = \lambda_{\mathrm{s}} - 1$ is obtained:

$$\varepsilon_{\mathrm{s}} \approx \varepsilon_{\mathrm{r}} + \frac{\Delta\kappa_{\mathrm{r}} r_k \cos\alpha_{\mathrm{r}} + u_{\mathrm{radi}} \kappa_{\mathrm{r}}^{(0)} \cos\alpha_{\mathrm{r}}}{1 + \kappa_{\mathrm{r}}^{(0)} r_k \cos\alpha_{\mathrm{r}}}, \tag{A10}$$

Similarly, the shear strain $\gamma_{\alpha_r \mathrm{s}}$ arising from the torsional deformation is determined by the twist rate relative to the curved metric.

$$\gamma_{\alpha_{\mathrm{r}}\mathrm{s}} = \frac{\Delta\tau_{\mathrm{r}} r_k}{\sqrt{G_{33}}} = \frac{\Delta\tau_{\mathrm{r}} r_k}{1 + \kappa_{\mathrm{r}}^{(0)} r_k \cos\alpha_{\mathrm{r}}}. \tag{A11}$$

## Appendix C. Effective stiffness matrix for helical anisotropy

The derivation of the effective stiffness matrix $\mathbf{C}_{\mathrm{rope}}$, which appears in the constitutive relations (Eq. 46), is provided in this appendix. The precursor fiber (rope) is modeled as a solid cylinder comprising concentric layers. Each layer is assumed to be a homogeneous, transversely isotropic continuum in its local material frame, characterized by a helical bias angle $\varphi$ that varies along the radial direction.

We define a local material coordinate system $\{1', 2', 3'\}$, where the $1'$-axis aligns with the microfibril (fiber) direction, the $2'$-axis is transverse to the fiber within the cylindrical surface, and the $3'$-axis aligns with the radial direction of the macro-structure. Under the assumption of transverse isotropy, the constitutive relation in the material frame is given by:

$$\boldsymbol{\sigma}' = \mathbf{Q}\boldsymbol{\varepsilon}'. \tag{A12}$$

where $\mathbf{Q}$ is the stiffness matrix in the local frame. The non-zero components of $\mathbf{Q}$, expressed in terms of the five independent engineering constants $(E_1, E_2, G_{12}, \nu_{12}, \nu_{23})$, are derived from the inversion of the compliance matrix $S$:

$$\begin{aligned} &Q_{11} = \frac{E_1(1-\nu_{23}^2)}{\Omega}, \quad Q_{22} = \frac{E_2(1-\nu_{12}\nu_{21})}{\Omega}, \quad Q_{12} = \frac{E_2(\nu_{12}+\nu_{23}\nu_{12})}{\Omega}, \\ &Q_{23} = \frac{E_2(\nu_{23}+\nu_{12}\nu_{21})}{\Omega}, \quad Q_{44} = G_{23}, \quad Q_{55} = G_{12}, \quad Q_{66} = G_{12}. \end{aligned} \tag{A13}$$

Here $\Omega = 1 - 2\nu_{12}\nu_{21} - \nu_{23}^2 - 2\nu_{12}\nu_{21}\nu_{23}$, and $\nu_{21} = \nu_{12}(E_2/E_1)$. It should be note that indices 4, 5, and 6 correspond to the $2'$-$3'$, $1'$-$3'$, and $1'$-$2'$ shear planes, respectively.

To bridge the micro-scale material properties to the macro-scale structural mechanics, a coordinate transformation from the local material frame $\{1',2',3'\}$ to the global curvilinear coordinate system $\{r_k, \alpha_{\mathrm{r}}, s\}$ (e.g. $\{1,2,3\}$) is employed.

The helical geometry introduces a rotation about the radial axis ($r_k$ or axis 1). The bias angle $\varphi$ is defined as the angle between the microfibril axis ($1'$) and the global longitudinal axis ($s$). Let $c = cos\varphi$ and $s = \sin\varphi$. The transformation of the fourth-order stiffness tensor follows the transformation laws for off-axis composites.

Then, the effective stiffness components $\mathbf{C}_{\mathrm{rope}} = \left[C_{ij}\right]$ in the global system are expressed as follows:

$$
\begin{cases}
C_{11} = Q_{33}, \\
C_{22} = Q_{11}s^4 + 2(Q_{12} + 2Q_{66})s^2c^2 + Q_{22}c^4, \\
\bar{C}_{33} = Q_{11}c^4 + 2(Q_{12} + 2Q_{66})s^2c^2 + Q_{22}s^4, \\
C_{12} = Q_{13}s^2 + Q_{23}c^2, \\
C_{13} = Q_{13}c^2 + Q_{23}s^2, \\
C_{23} = (Q_{11} + Q_{22} - 4Q_{66})s^2c^2 + Q_{12}(s^4 + c^4), \\
C_{16} = (Q_{13} - Q_{23})sc, \\
C_{26} = (Q_{11} - Q_{12} - 2Q_{66})s^3c - (Q_{22} - Q_{12} - 2Q_{66})sc^3, \\
C_{36} = (Q_{11} - Q_{12} - 2Q_{66})sc^3 - (Q_{22} - Q_{12} - 2Q_{66})s^3c, \\
C_{66} = (Q_{11} + Q_{22} - 2Q_{12} - 2Q_{66})s^2c^2 + Q_{66}(s^4 + c^4).
\end{cases}
\tag{A14}
$$

# Figures

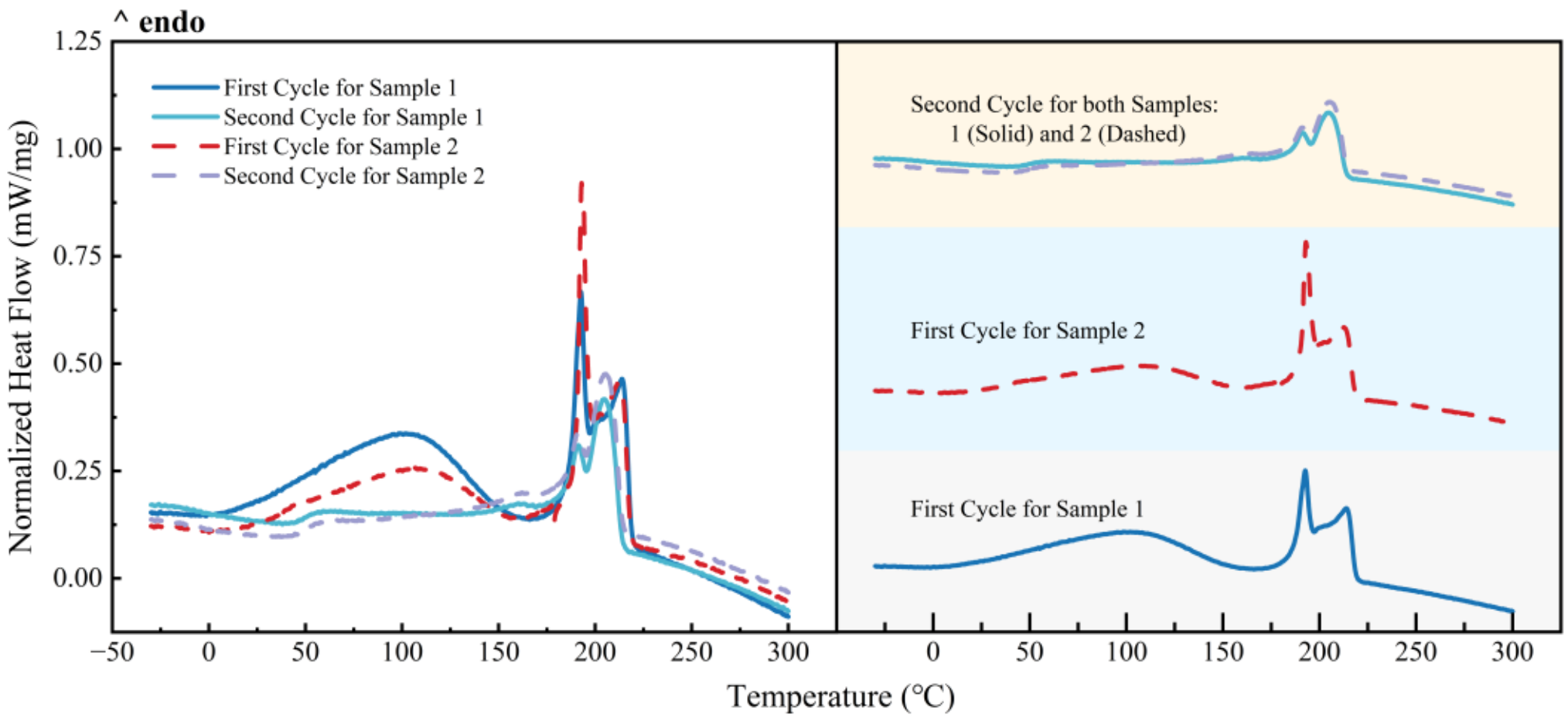

Fig. 1 The results for Sample 1 and Sample 2 by DSC.

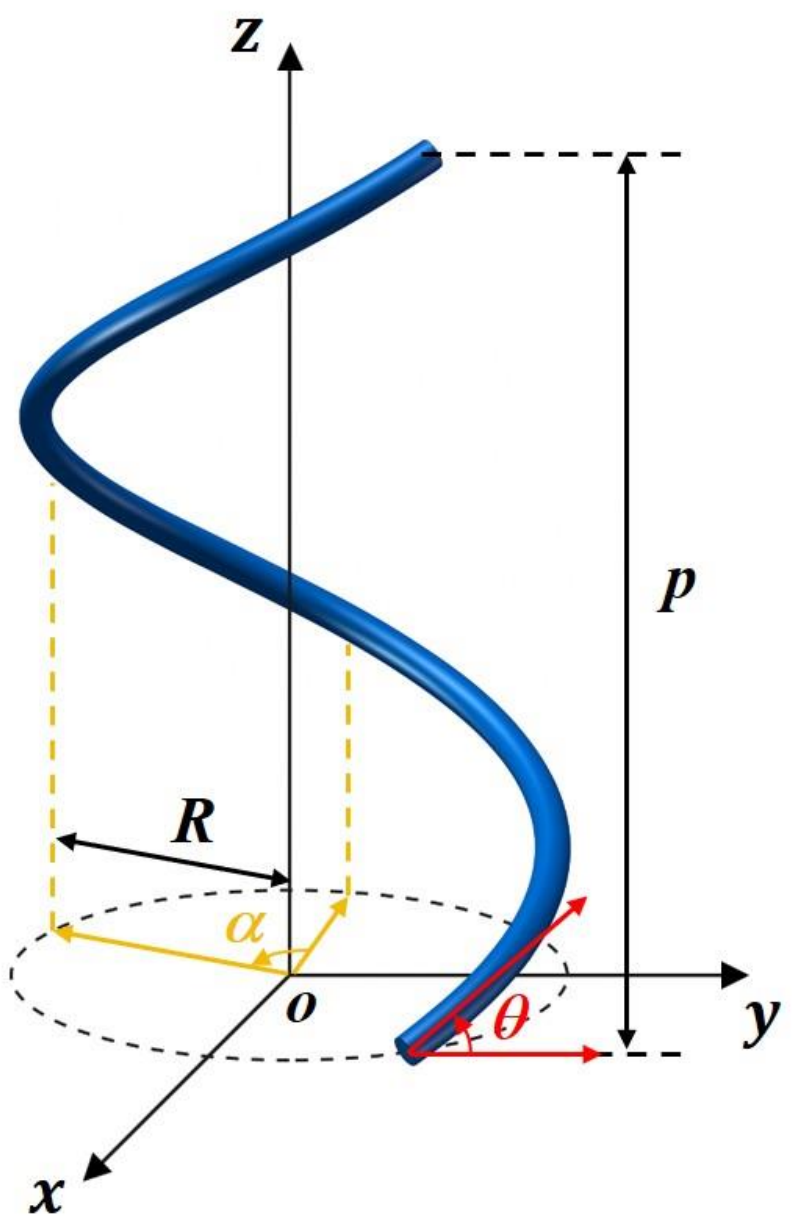

Fig. 2 Geometrical descriptions of a helical curve.

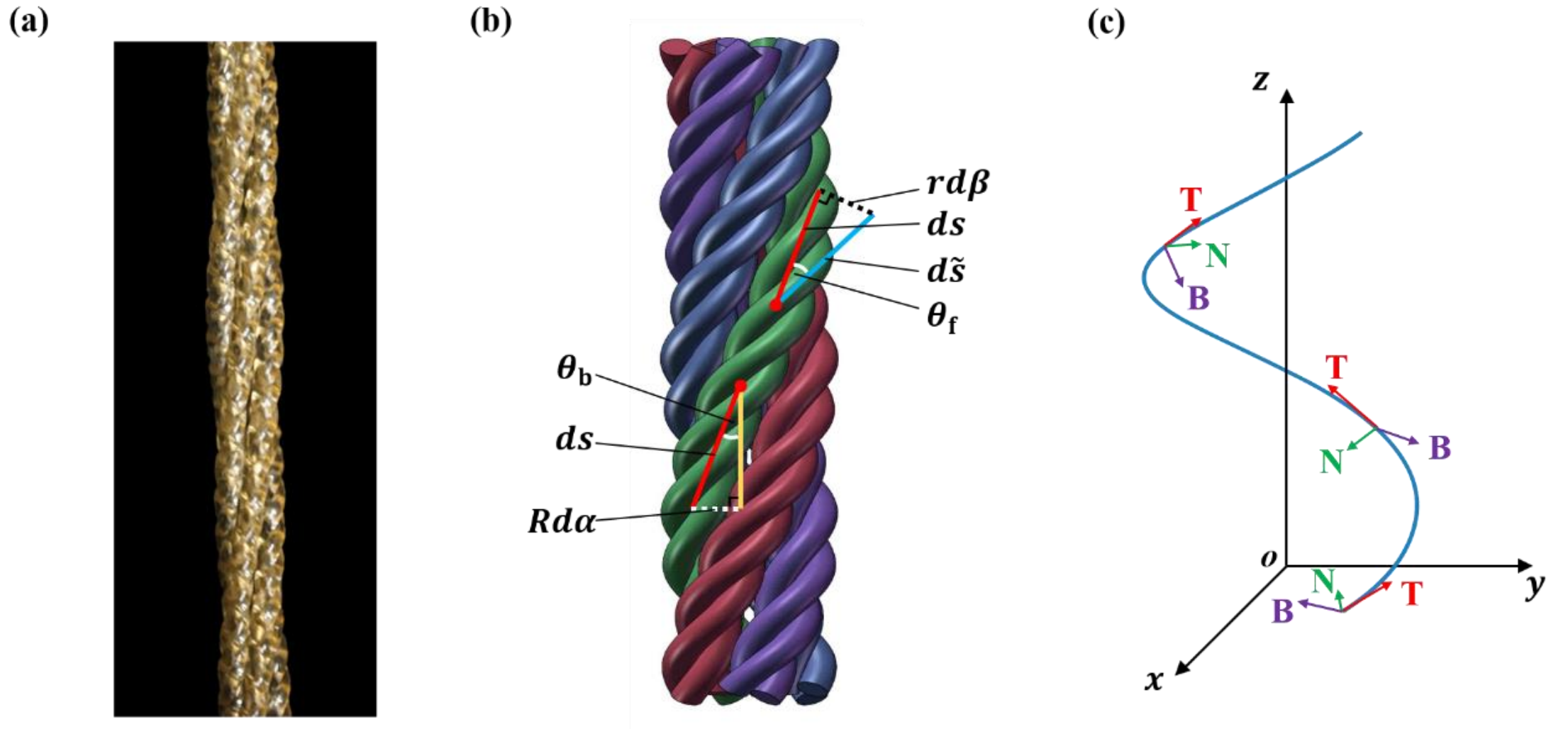

Fig. 3 Optical image (a) and geometrical descriptions (b) of a hierarchical fiber in artificial muscles with hierarchical helical structure (3×4); (c) local right-handed Frenet coordinate frame.

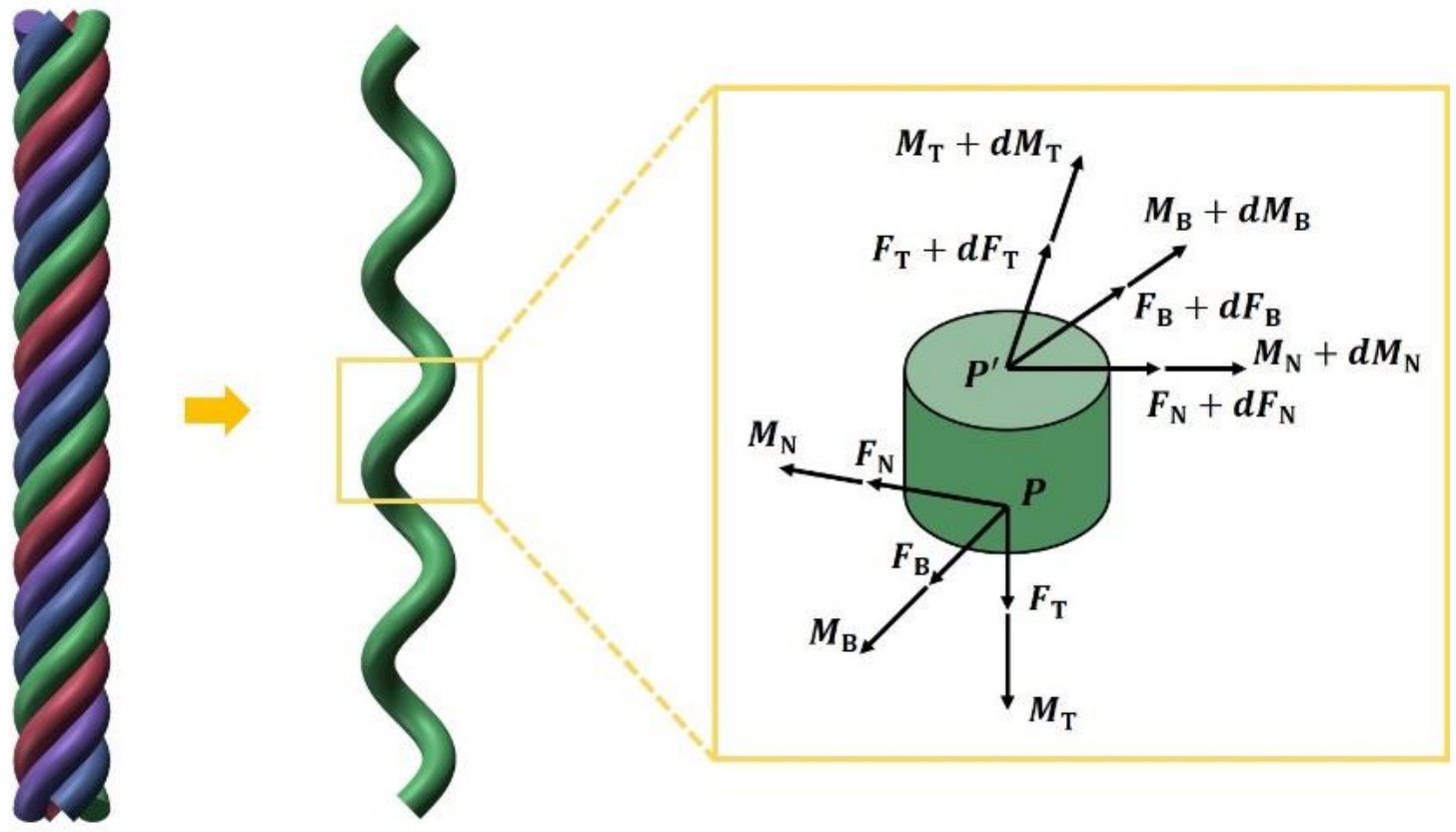

Fig. 4 Model of helical structures. (a) a straight bundle consisting of m filaments and (b) internal forces of a helical filament.

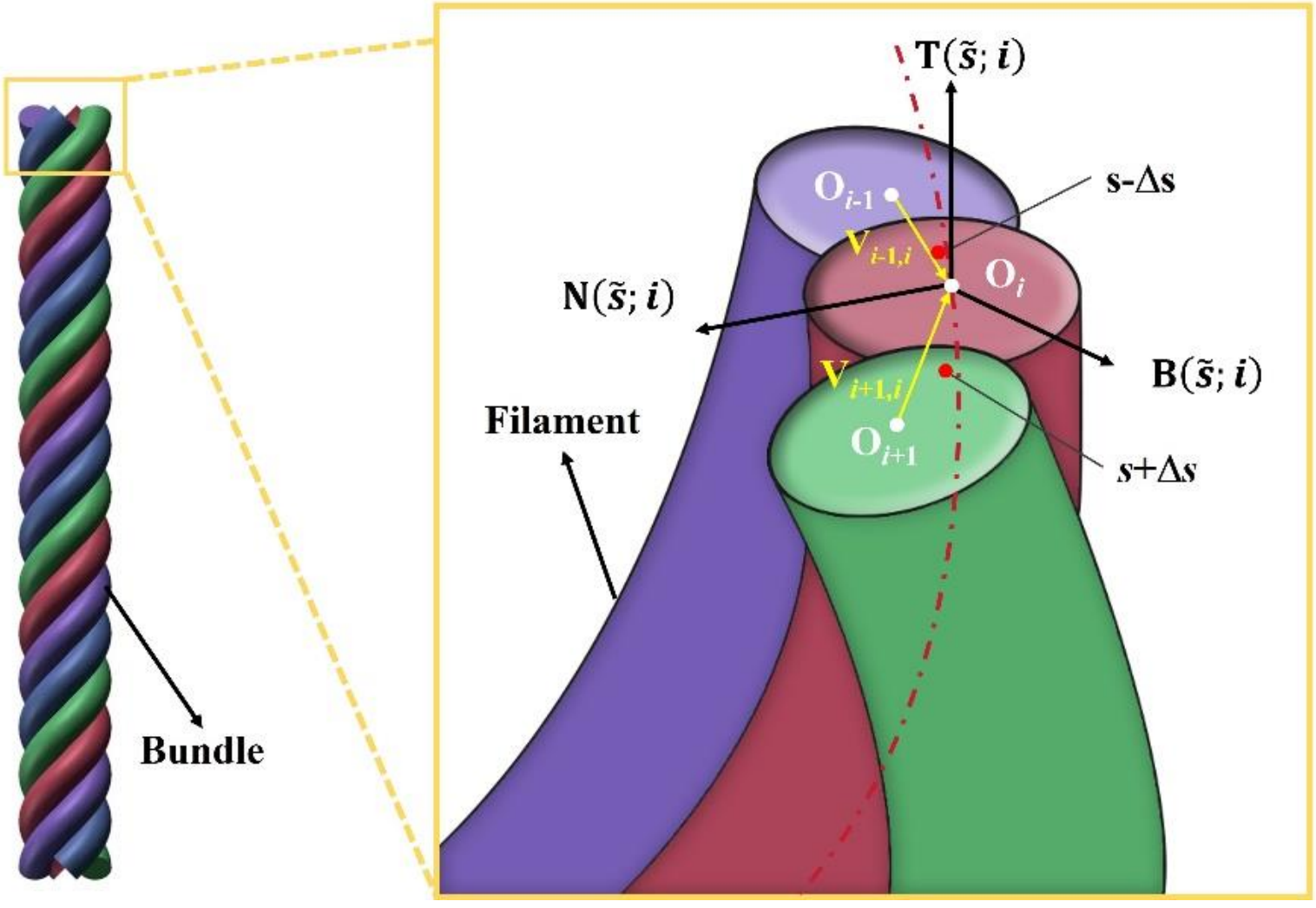

Fig. 5 The force analysis of the three adjacent filaments in a straight bundle.

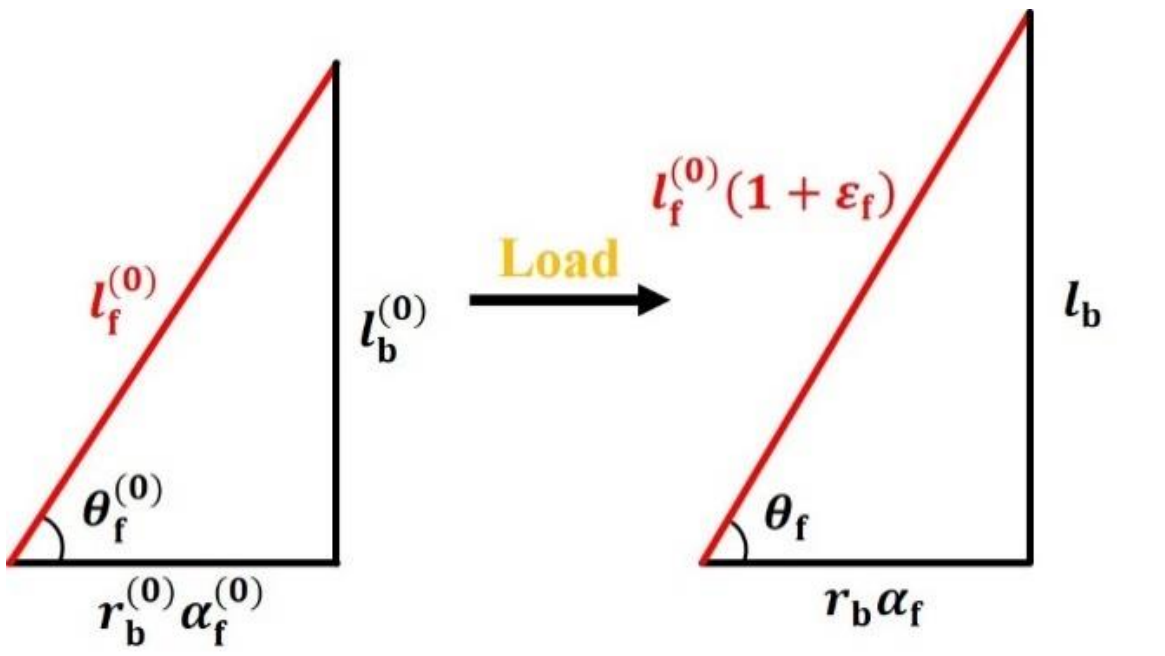

Fig. 6 The geometric relations between the helical filament and straight bundle in the undeformed and the deformed configurations.

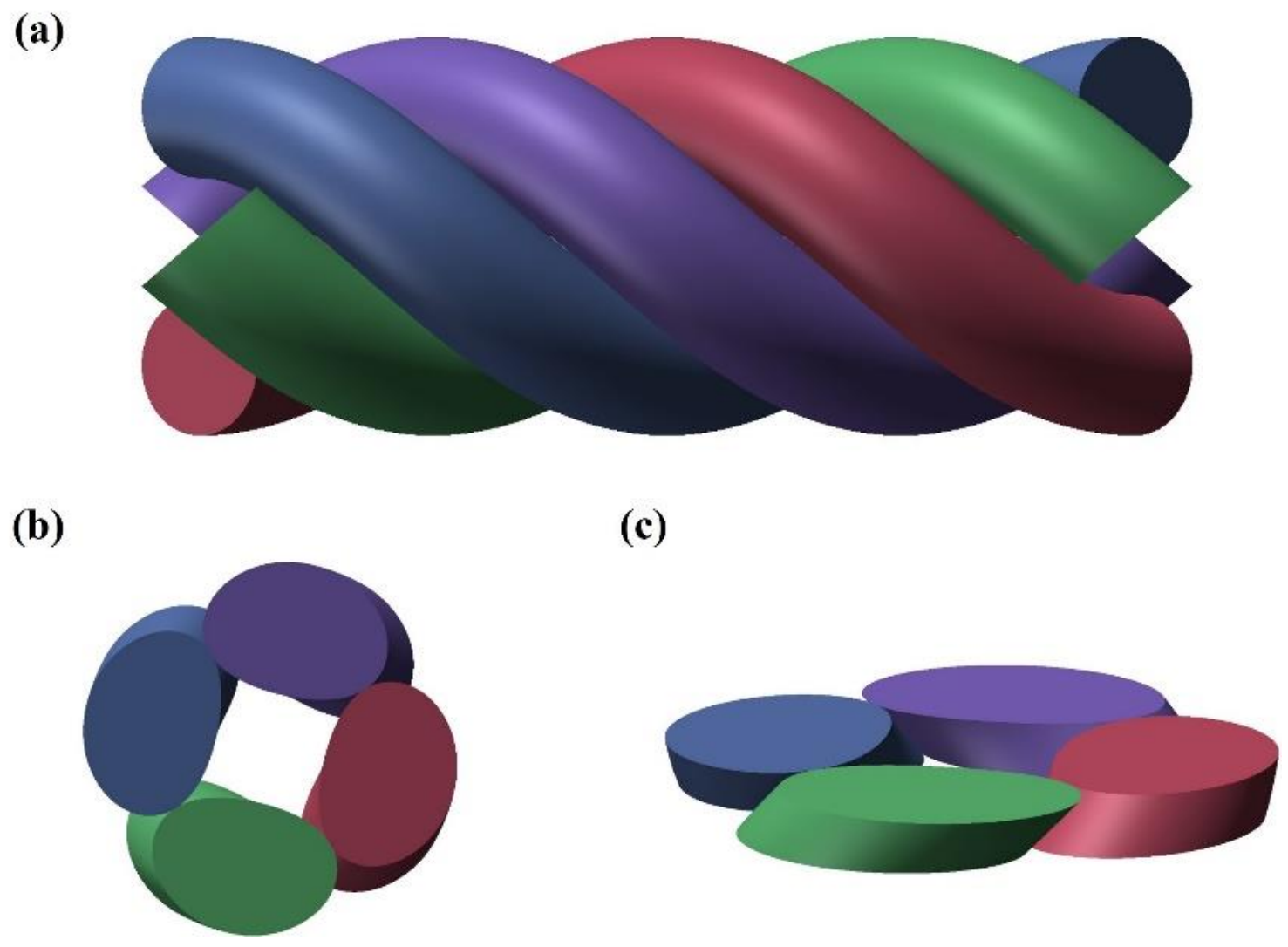

Fig. 7 The lateral and front views of a straight bundle with 4-filaments

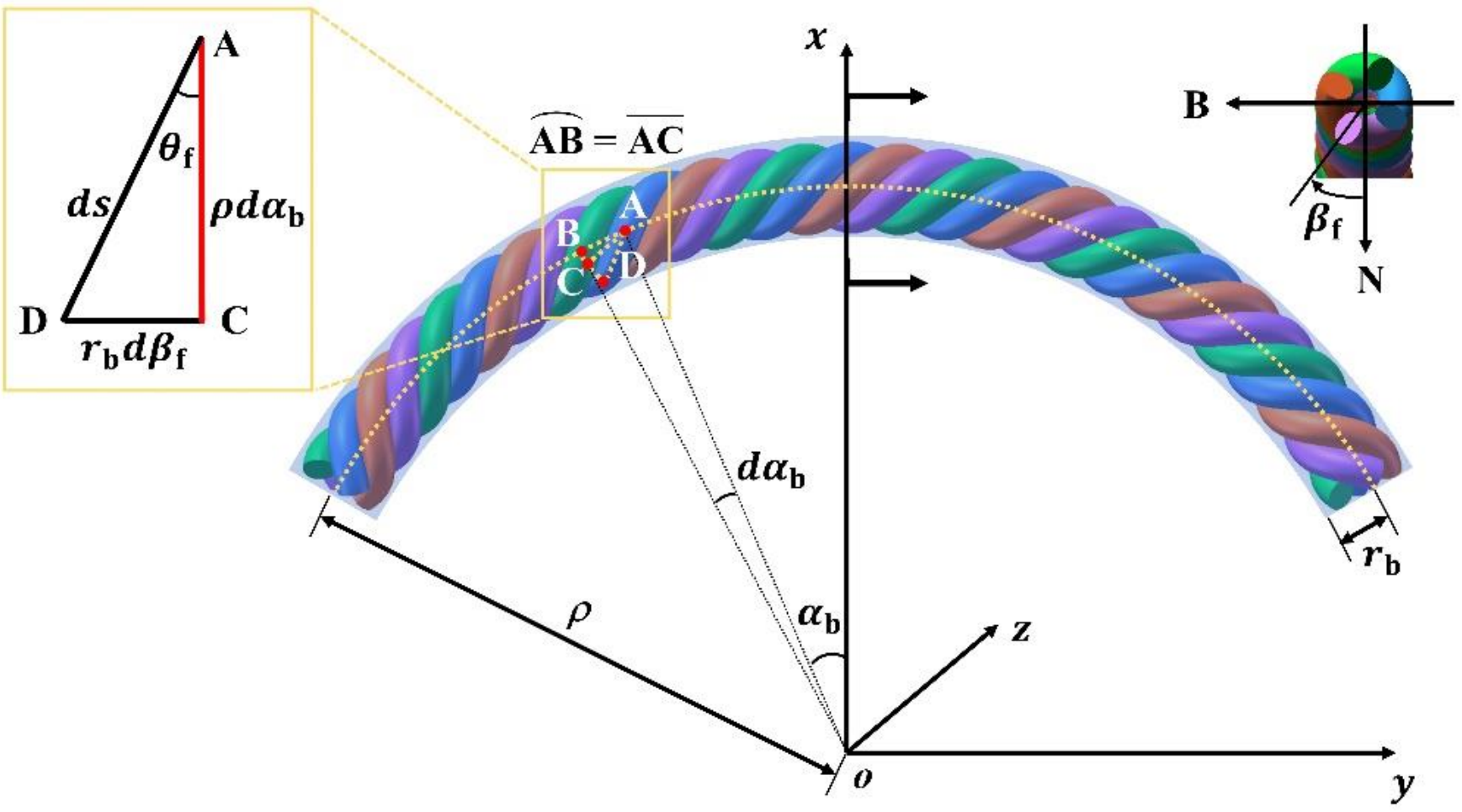

Fig. 8 The schematic diagram of geometric relations for a bending bundle

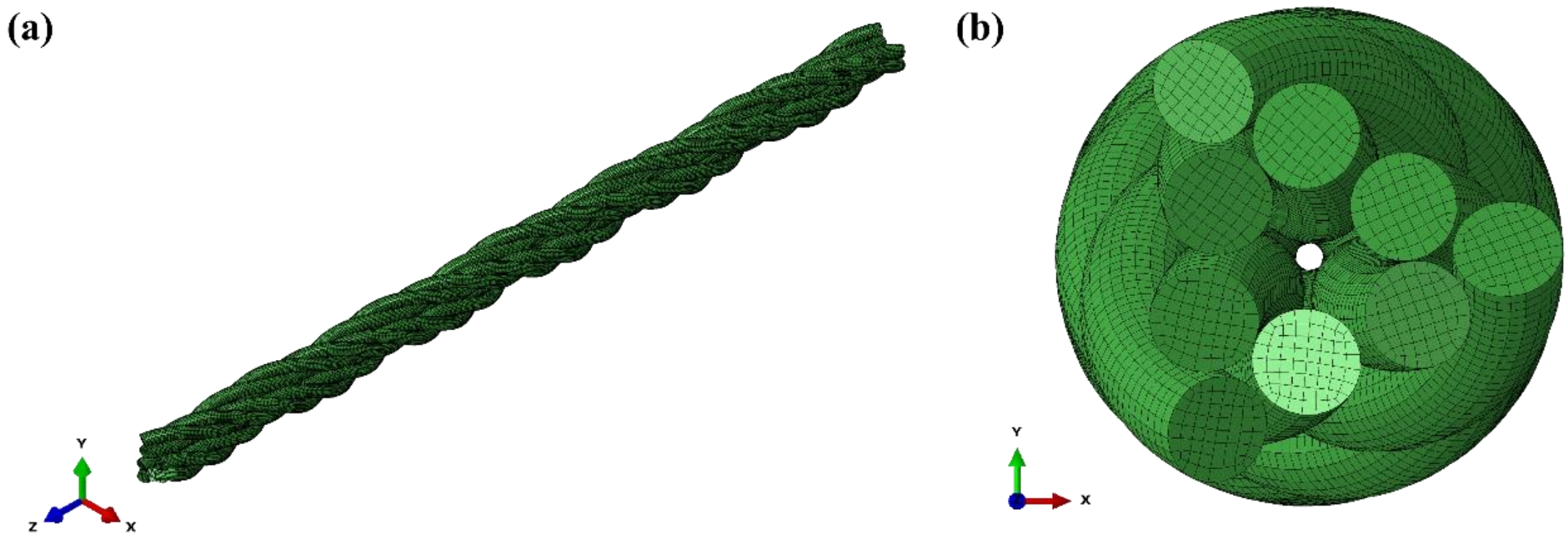

Fig. 9 Finite element mesh of a hierarchical helical structure

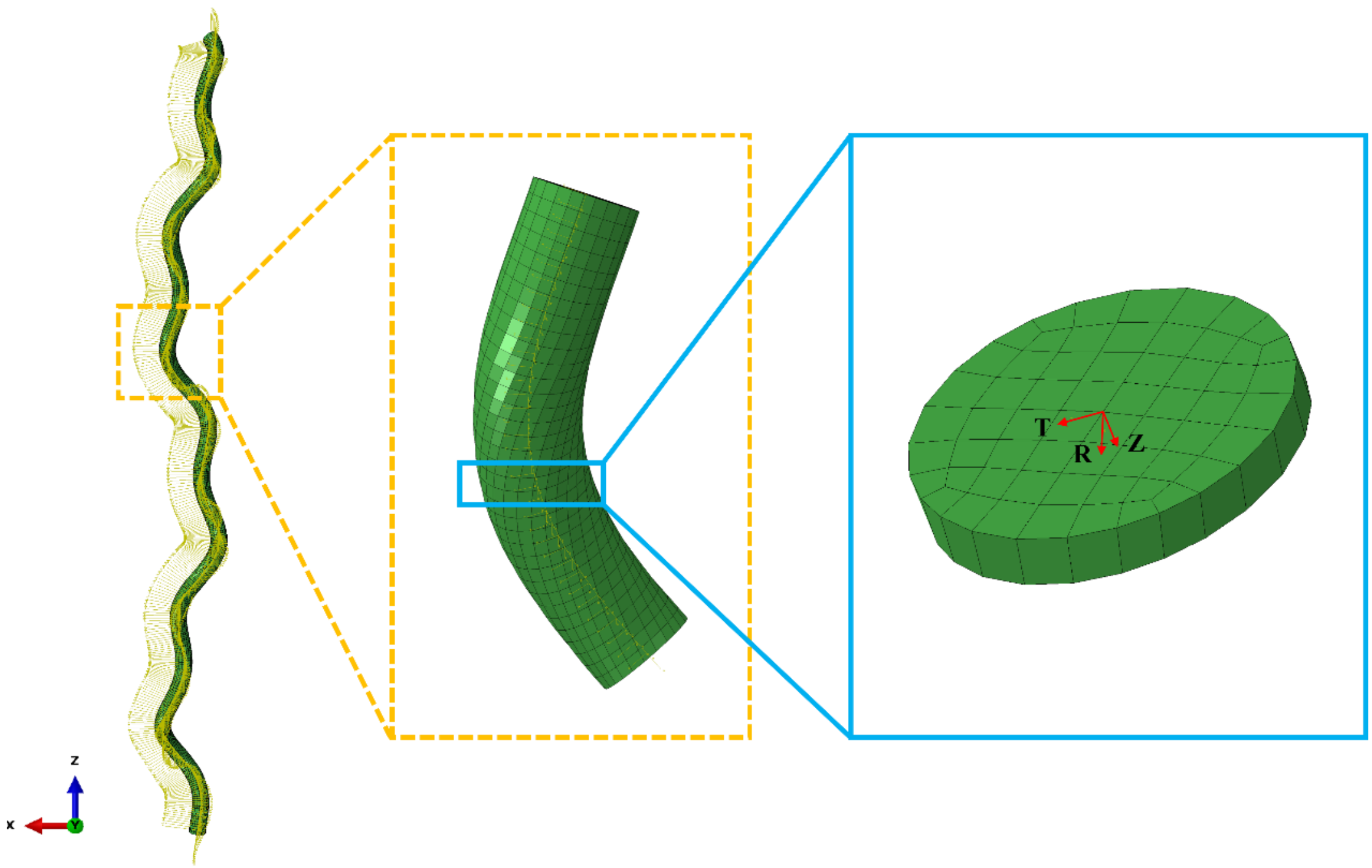

Fig. 10 Local material coordinate system of helical nylon fiber

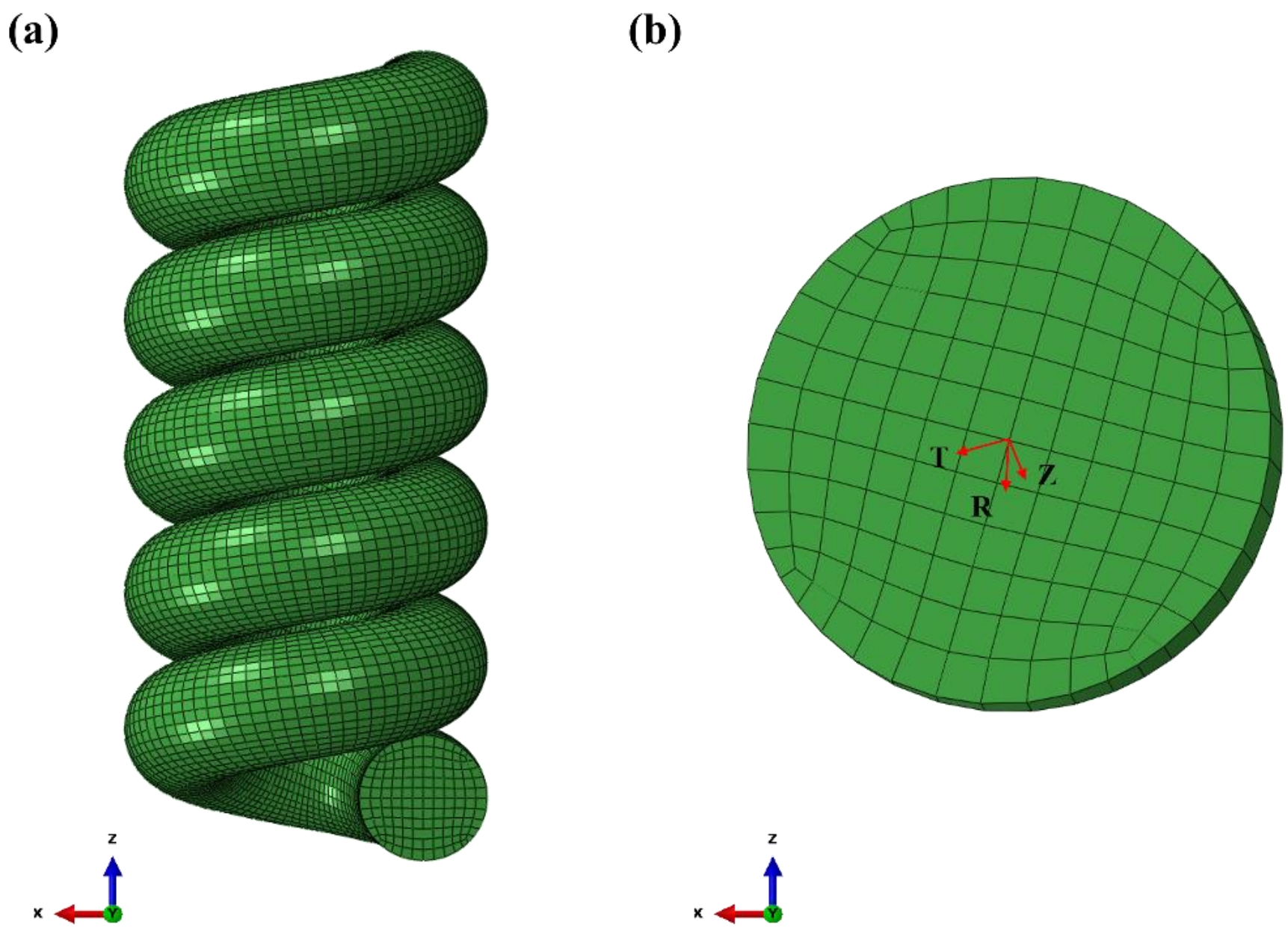

Fig. 11 Finite element mesh of twisted and coiled structure

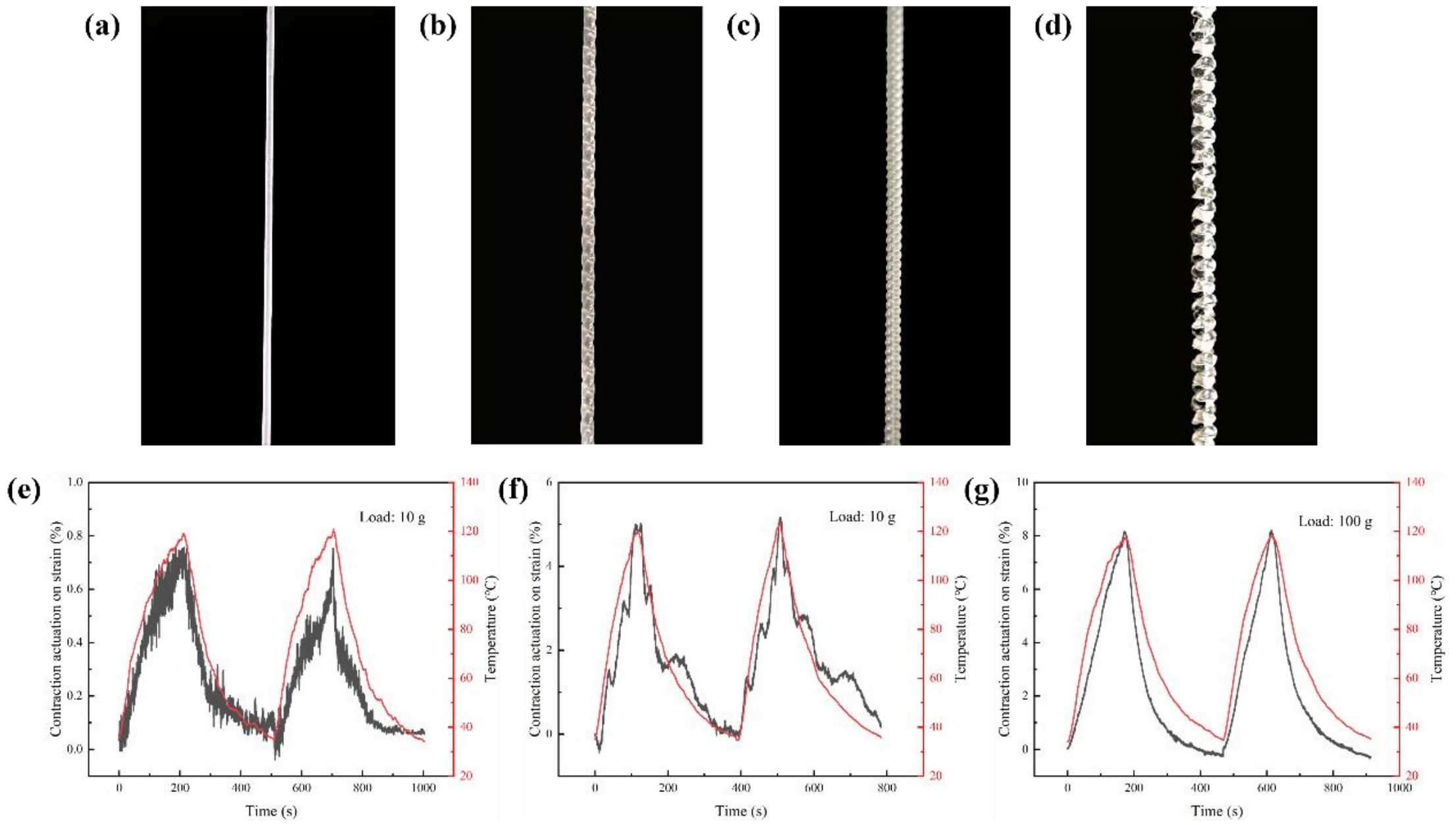

Fig. 12 Morphological evolution and thermomechanical actuation performance of TCP actuators with varying structural hierarchies. Upper panels illustrate optical images of: (a) the as-received single monofilament precursor; (b) a 2-ply helical structure TCP structure formed by twisting two precursors without coiling; (c) a single-ply coiled TCP structure formed by over-twisting induced coiling; (d) a hierarchical 2-ply coiled TCP structure exhibiting a nested, dual-level helical structure. Lower panels show temporal evolution of contraction actuation strain (black curves, left axis) and surface temperature (red curves, right axis) under cyclic thermal actuation: (e) the 2-ply helical TCP (Load: 10 g) exhibiting low-amplitude actuation; (f) the single-ply coiled TCP (Load: 10 g) showing moderate actuation strain; (g) the hierarchical 2-ply coiled TCP (Load: 100 g), demonstrating high-amplitude actuation strain under a significantly elevated mechanical load. Peak strains approximate 0.8%, 5%, and 8% respectively, with temperature cycles ranging 35–120°C.

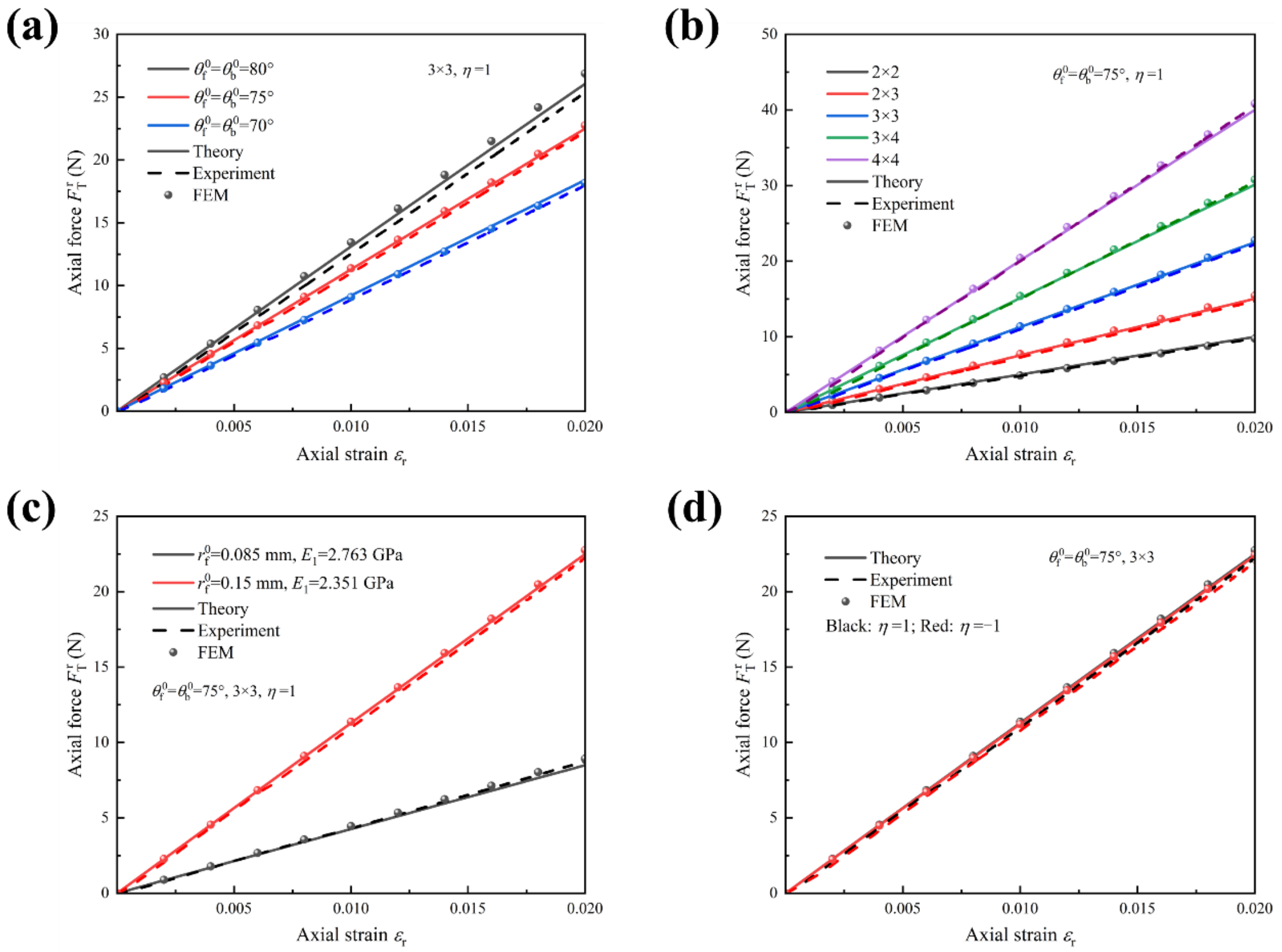

Fig. 13 Comparisons of the axial force–axial strain ($F_T^r - \varepsilon_r$) responses obtained from the theoretical model (solid lines), experimental measurements (dashed lines), and FEM simulations (scatter points). (a) The effect of initial helical angles ($\theta_f^0 = \theta_b^0$) with a $3 \times 3$ structure and $\eta = 1$. (b) The effect of structure sizes (ranging from $2 \times 2$ to $4 \times 4$) at $\theta_f^0 = \theta_b^0 = 75°$. (c) The effect of filament radius $r_f^0$ and Young's modulus. (d) The effect of the parameter $\eta$ (comparing $\eta = 1$ and $\eta = -1$) for a $3 \times 3$ structure at $\theta_f^0 = \theta_b^0 = 75°$.

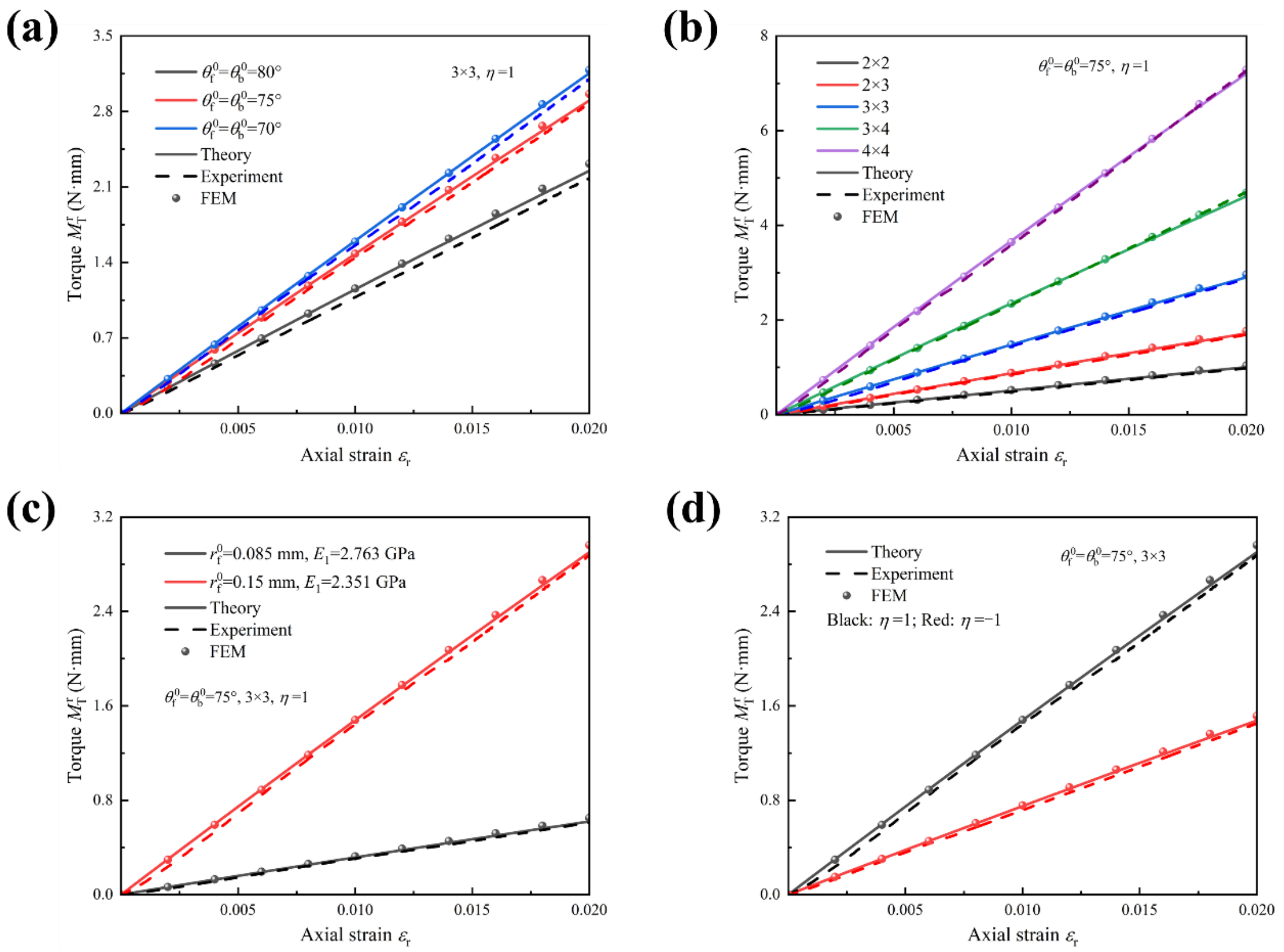

Fig. 14 Comparisons of the torque–axial strain ($M_{\mathrm{T}}^{\mathrm{r}} - \varepsilon_{\mathrm{r}}$) responses obtained from the theoretical model (solid lines), experimental measurements (dashed lines), and FEM simulations (scatter points). (a) The effect of initial helical angles ($\theta_{\mathrm{f}}^{0} = \theta_{\mathrm{b}}^{0}$) with a $3 \times 3$ structure and $\eta = 1$. (b) The effect of structure sizes (ranging from $2 \times 2$ to $4 \times 4$) at $\theta_{\mathrm{f}}^{0} = \theta_{\mathrm{b}}^{0} = 75°$. (c) The effect of filament radius $r_{\mathrm{f}}^{0}$ and Young's modulus $E_1$. (d) The effect of the parameter $\eta$ (comparing $\eta = 1$ and $\eta = -1$) for a $3 \times 3$ structure at $\theta_{\mathrm{f}}^{0} = \theta_{\mathrm{b}}^{0} = 75°$.

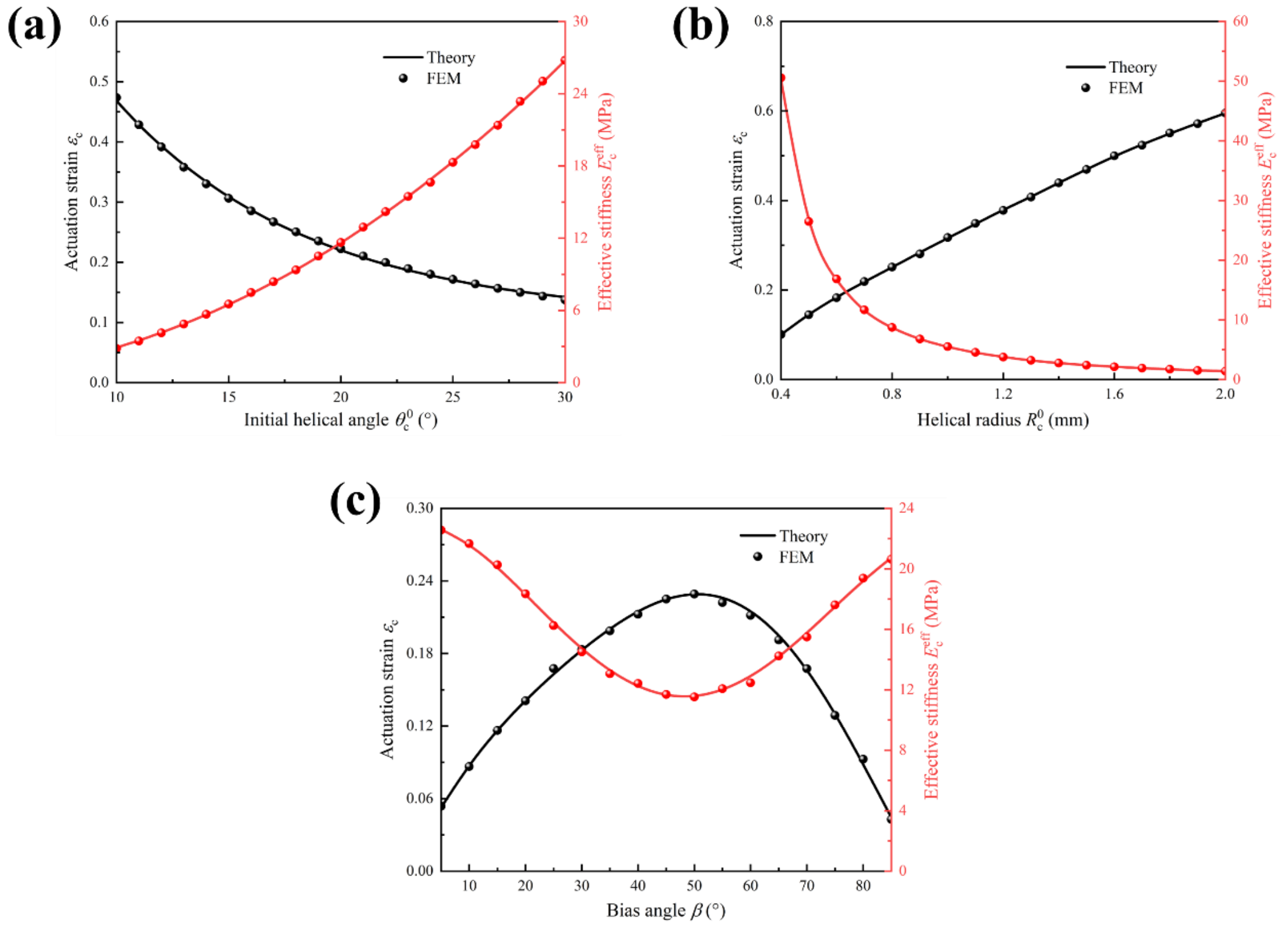

Fig. 15 Influence of geometric parameters on the actuation strain $\varepsilon_c$ (left axis, black) and effective stiffness $E_c^{eff}$ (right axis, red). The solid lines and scatter points represent the theoretical predictions and FEM results, respectively. (a) The effect of the initial helical angle $\theta_c^0$. (b) The effect of the helical radius $R_c^0$. (c) The effect of the bias angle $\varphi$.

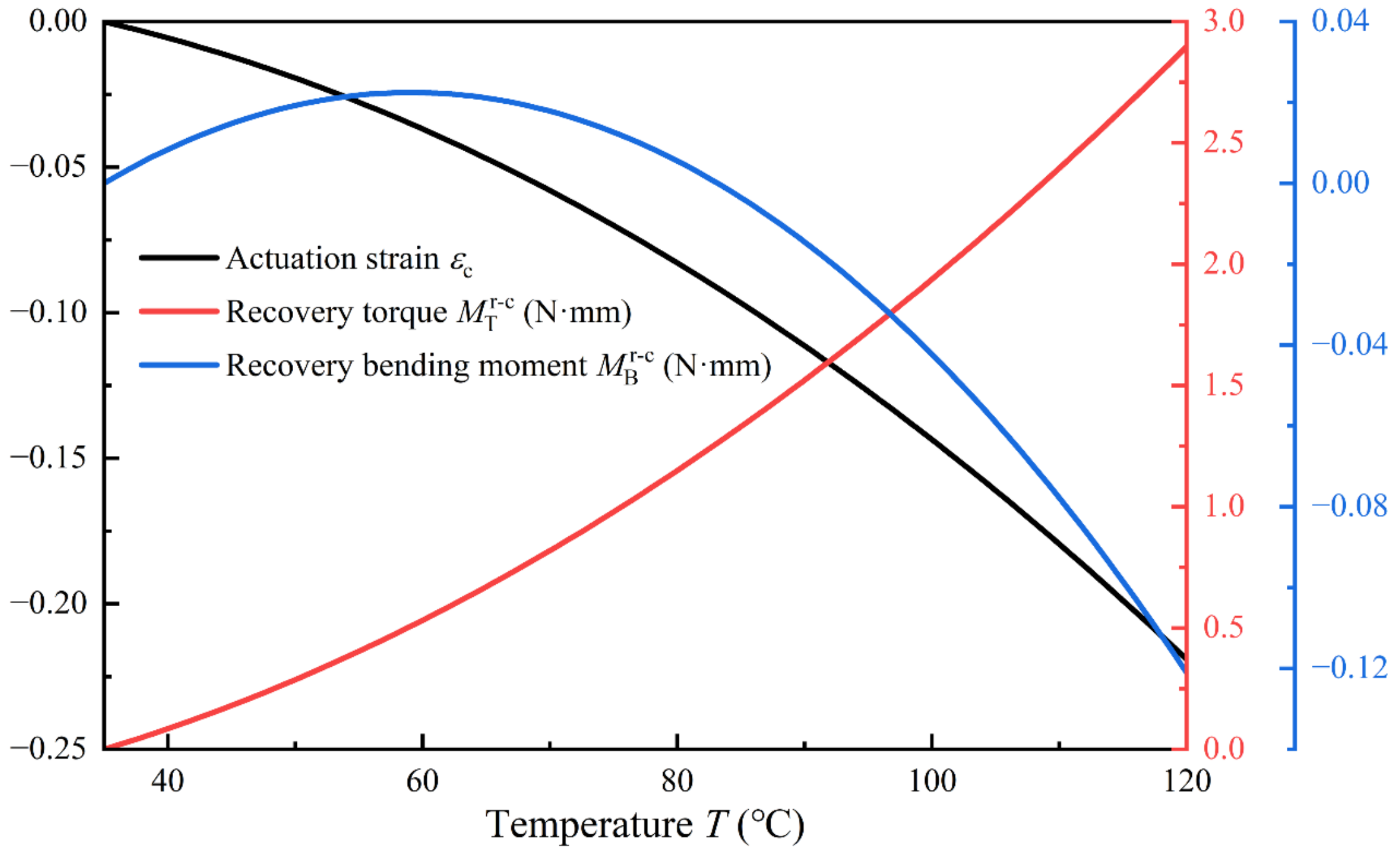

Fig. 16 Theoretical evolution of the actuation strain $\varepsilon_c$ (black line, left axis), recovery torque $M_T^{r\text{-}c}$ (red line, inner right axis), and recovery bending moment $M_B^{r\text{-}c}$ (blue line, outer right axis) with respect to temperature $T$.

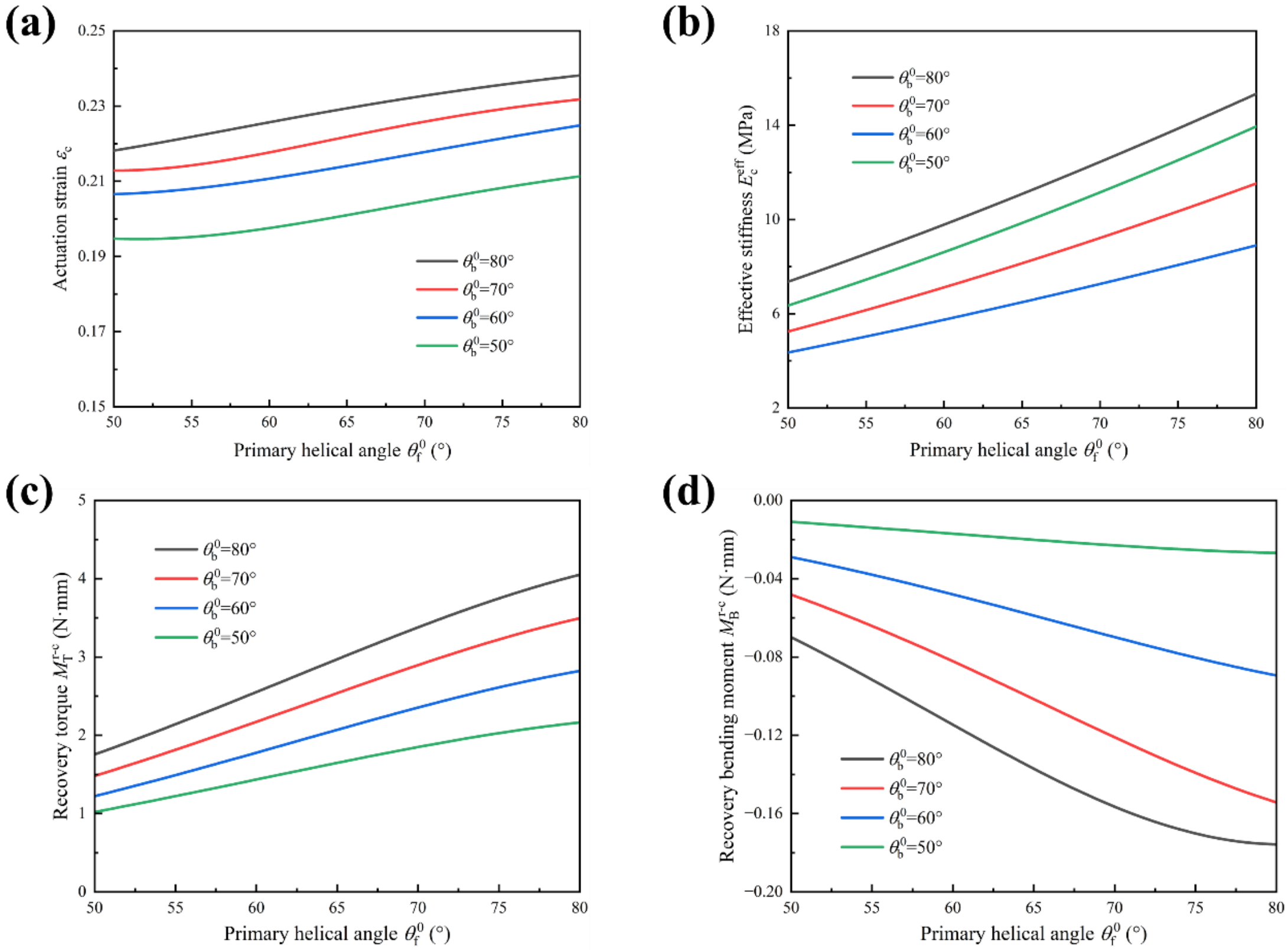

Fig. 17 The influence of the primary helical angle $\theta_f^0$ on the actuation performance for different values of $\theta_b^0$ (ranging from 50° to 80°). (a) Actuation strain $\varepsilon_c$. (b) Effective stiffness $E_c^{eff}$. (c) Recovery torque $M_T^{r\text{-}c}$. (d) Recovery bending moment $M_B^{r\text{-}c}$.

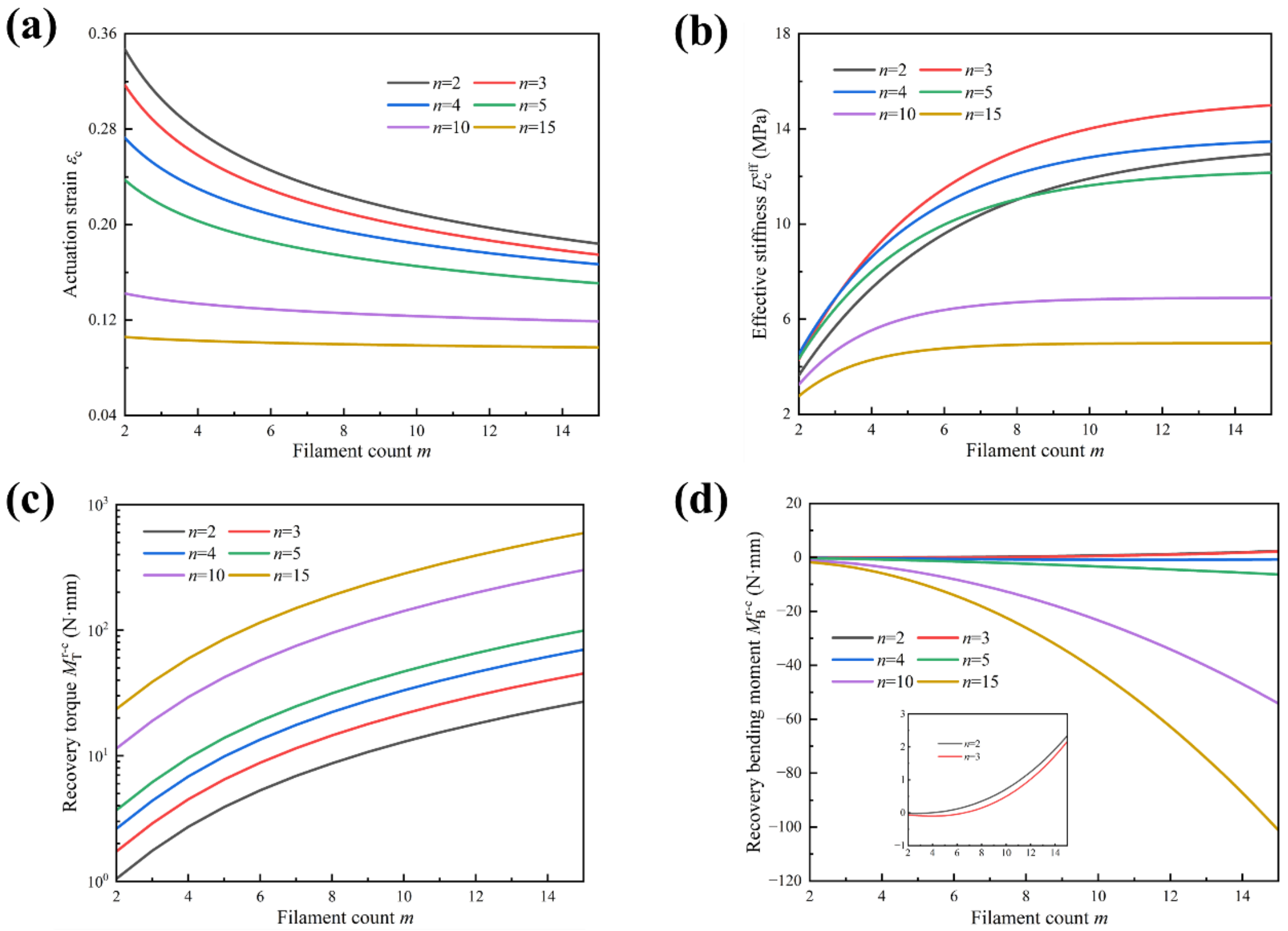

Fig. 18 Theoretical predictions of the actuation performance as a function of the filament count $m$ for different values of $n$ (ranging from 2 to 15). (a) Actuation strain $\varepsilon_{\mathrm{c}}$ plotted on a logarithmic scale. (b) Effective stiffness $E_{\mathrm{c}}^{\mathrm{eff}}$. (c) Recovery torque $M_{\mathrm{T}}^{\text{r-c}}$ plotted on a linear scale. (d) Recovery bending moment $M_{\mathrm{B}}^{\text{r-c}}$.

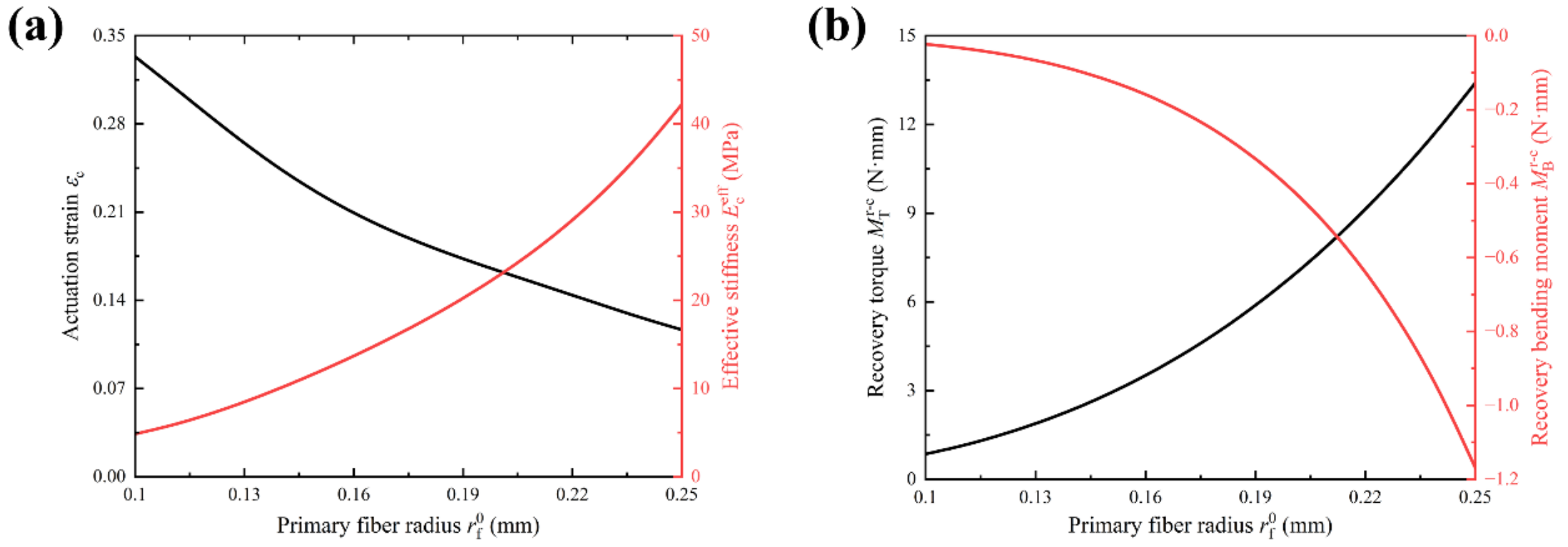

Fig. 19 Influence of the primary fiber radius $r_{\mathrm{f}}^{0}$ on the actuation performance. (a) The variations of actuation strain $\varepsilon_{\mathrm{c}}$ (black curve, left axis) and effective stiffness $E_{\mathrm{c}}^{\mathrm{eff}}$ (red curve, right axis). (b) The variations of recovery torque $M_{\mathrm{T}}^{\text{r-c}}$ (black curve, left axis) and recovery bending moment $M_{\mathrm{B}}^{\text{r-c}}$ (red curve, right axis).

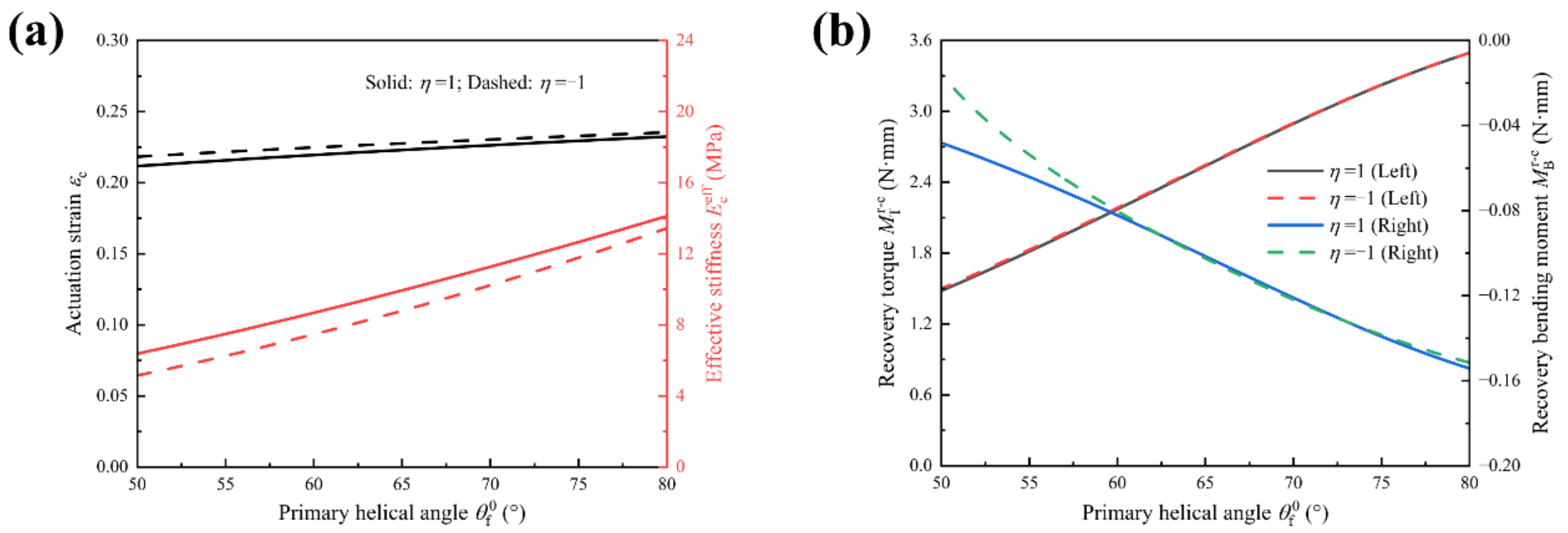

Fig. 20 Comparison of actuation performance for different chirality parameters $\eta$ (solid lines for $\eta = 1$, dashed lines for $\eta = -1$) as a function of the primary helical angle $\theta_f^0$. (a) The variations of actuation strain $\varepsilon_c$ (black lines, left axis) and effective stiffness $E_c^{eff}$ (red lines, right axis). (b) The variations of recovery torque $M_T^{r\text{-}c}$ (left axis) and recovery bending moment $M_B^{r\text{-}c}$ (right axis).

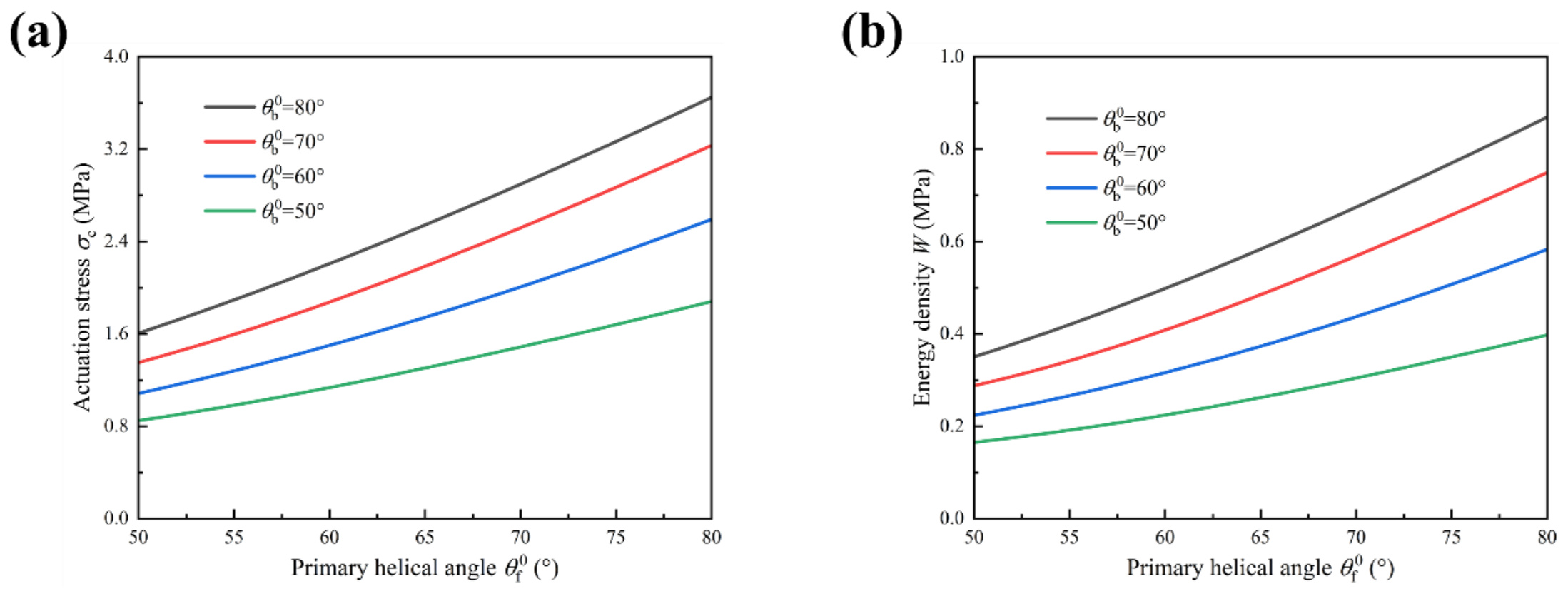

Fig. 21 Theoretical predictions of the actuation performance as a function of the primary helical angle $\theta_f^0$ for different values of $\theta_b^0$ (ranging from 50° to 80°). (a) Actuation stress $\sigma_c$. (b) Energy density $W$.

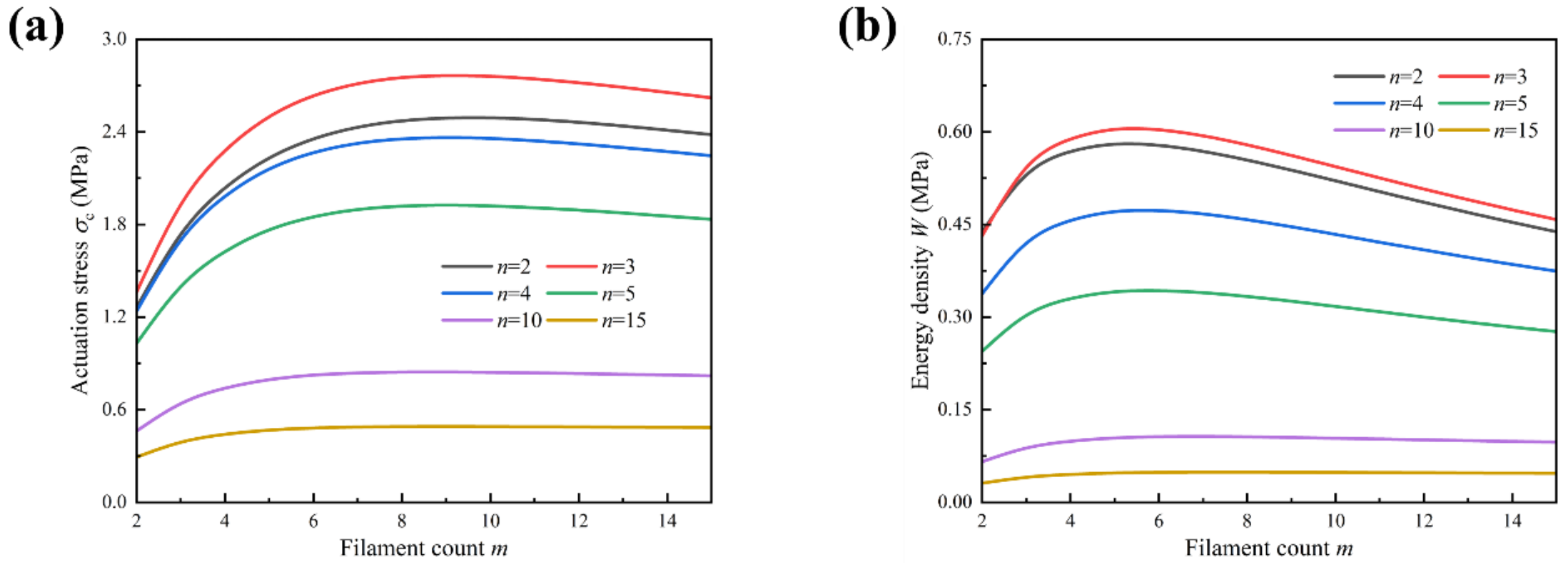

Fig. 22 Theoretical predictions of the actuation performance as a function of the filament count $m$ for different values of $n$ (ranging from 2 to 15). (a) Actuation stress $\sigma_c$. (b) Energy density $W$.

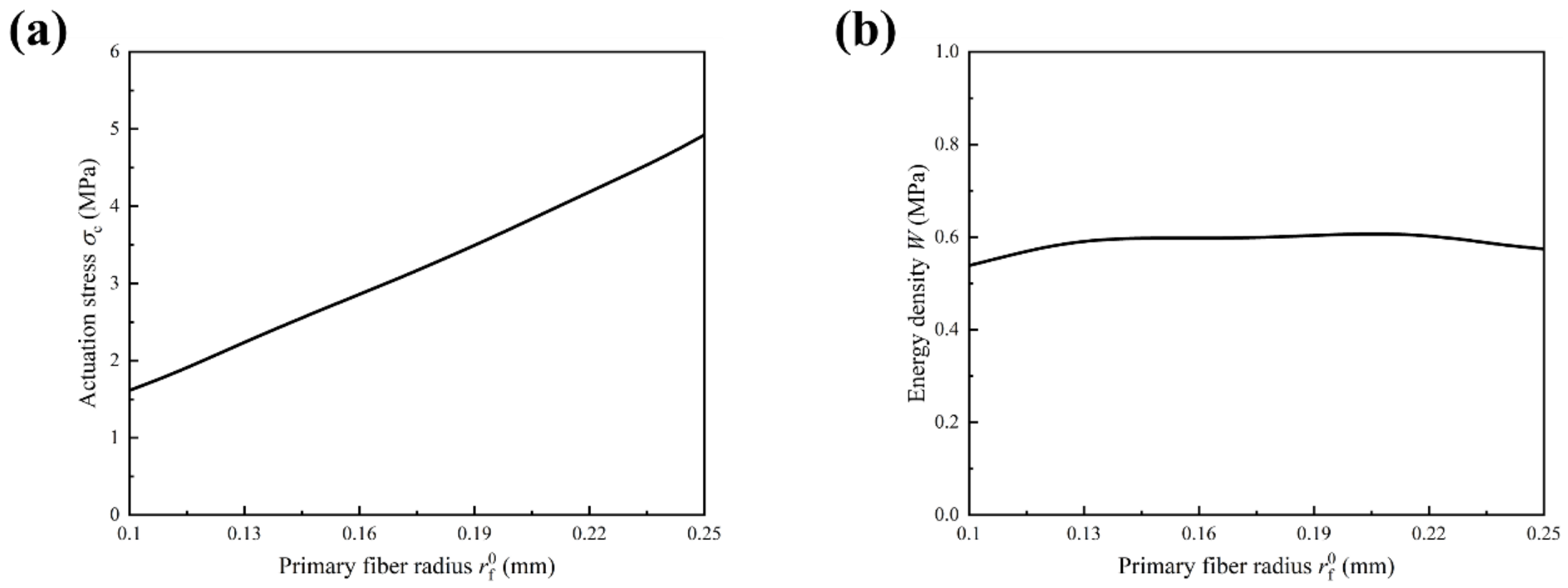

Fig. 23 Theoretical predictions of the actuation performance as a function of the primary fiber radius $r_f^0$. (a) Actuation stress $\sigma_c$. (b) Energy density $W$.

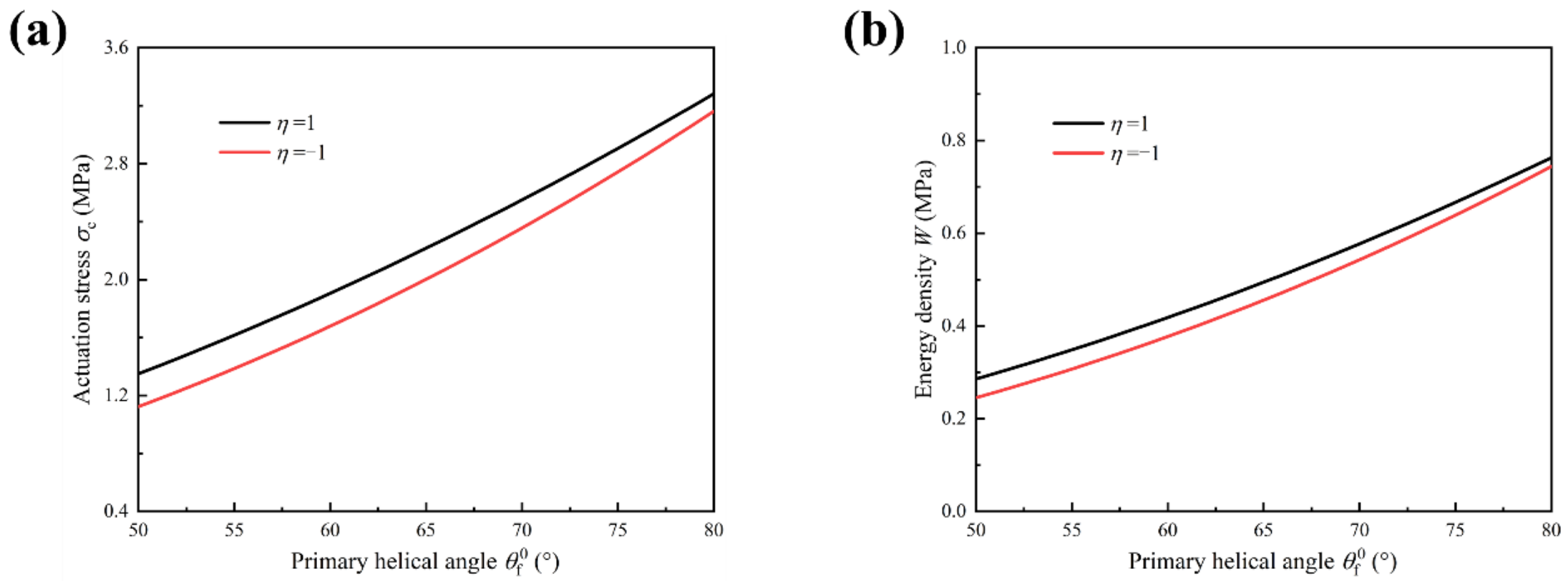

Fig. 24 Comparisons of the actuation performance with different chirality parameters $\eta$ (black lines for $\eta = 1$, red lines for $\eta = -1$) as a function of the primary helical angle $\theta_f^0$. (a) Actuation stress $\sigma_c$. (b) Energy density $W$.